\documentclass[aps,preprint,floatfix]{revtex4}
\usepackage{graphicx,color}
\usepackage{psfrag}
\usepackage{amssymb}
\usepackage{eso-pic}
\usepackage[dvips,letterpaper]{geometry}

\begin{document}

%
%
%
%
\def\oti{{\otimes}}
\def\lb{ \left[ }
\def\rb{ \right]  }
\def\tilde{\widetilde}
\def\bar{\overline}
\def\hat{\widehat}
\def\*{\star}
\def\[{\left[}
\def\]{\right]}
\def\({\left(}		\def\BL{\Bigr(}
\def\){\right)}		\def\BR{\Bigr)}
	\def\BBL{\lb}
	\def\BBR{\rb}
%
%
\def\zb{{\bar{z} }}
\def\zbar{{\bar{z} }}
\def\frac#1#2{{#1 \over #2}}
\def\inv#1{{1 \over #1}}
\def\half{{1 \over 2}}
\def\d{\partial}
\def\der#1{{\partial \over \partial #1}}
\def\dd#1#2{{\partial #1 \over \partial #2}}
\def\vev#1{\langle #1 \rangle}
\def\ket#1{ | #1 \rangle}
\def\rvac{\hbox{$\vert 0\rangle$}}
\def\lvac{\hbox{$\langle 0 \vert $}}
\def\2pi{\hbox{$2\pi i$}}
\def\e#1{{\rm e}^{^{\textstyle #1}}}
\def\grad#1{\,\nabla\!_{{#1}}\,}
\def\dsl{\raise.15ex\hbox{/}\kern-.57em\partial}
\def\Dsl{\,\raise.15ex\hbox{/}\mkern-.13.5mu D}
%
%
\def\ga{\gamma}		\def\Ga{\Gamma}
\def\be{\beta}
\def\al{\alpha}
\def\ep{\epsilon}
\def\vep{\varepsilon}
\def\la{\lambda}	\def\La{\Lambda}
\def\de{\delta}		\def\De{\Delta}
\def\om{\omega}		\def\Om{\Omega}
\def\sig{\sigma}	\def\Sig{\Sigma}
\def\vphi{\varphi}
%
%
\def\CA{{\cal A}}	\def\CB{{\cal B}}	\def\CC{{\cal C}}
\def\CD{{\cal D}}	\def\CE{{\cal E}}	\def\CF{{\cal F}}
\def\CG{{\cal G}}	\def\CH{{\cal H}}	\def\CI{{\cal J}}
\def\CJ{{\cal J}}	\def\CK{{\cal K}}	\def\CL{{\cal L}}
\def\CM{{\cal M}}	\def\CN{{\cal N}}	\def\CO{{\cal O}}
\def\CP{{\cal P}}	\def\CQ{{\cal Q}}	\def\CR{{\cal R}}
\def\CS{{\cal S}}	\def\CT{{\cal T}}	\def\CU{{\cal U}}
\def\CV{{\cal V}}	\def\CW{{\cal W}}	\def\CX{{\cal X}}
\def\CY{{\cal Y}}	\def\CZ{{\cal Z}}

\def\rvac{\hbox{$\vert 0\rangle$}}
\def\lvac{\hbox{$\langle 0 \vert $}}
\def\comm#1#2{ \BBL\ #1\ ,\ #2 \BBR }
\def\2pi{\hbox{$2\pi i$}}
\def\e#1{{\rm e}^{^{\textstyle #1}}}
\def\grad#1{\,\nabla\!_{{#1}}\,}
\def\dsl{\raise.15ex\hbox{/}\kern-.57em\partial}
\def\Dsl{\,\raise.15ex\hbox{/}\mkern-.13.5mu D}
%
%
%
\font\numbers=cmss12
\font\upright=cmu10 scaled\magstep1
\def\stroke{\vrule height8pt width0.4pt depth-0.1pt}
\def\topfleck{\vrule height8pt width0.5pt depth-5.9pt}
\def\botfleck{\vrule height2pt width0.5pt depth0.1pt}
\def\Zmath{\vcenter{\hbox{\numbers\rlap{\rlap{Z}\kern
0.8pt\topfleck}\kern 2.2pt
                   \rlap Z\kern 6pt\botfleck\kern 1pt}}}
\def\Qmath{\vcenter{\hbox{\upright\rlap{\rlap{Q}\kern
                   3.8pt\stroke}\phantom{Q}}}}
\def\Nmath{\vcenter{\hbox{\upright\rlap{I}\kern 1.7pt N}}}
\def\Cmath{\vcenter{\hbox{\upright\rlap{\rlap{C}\kern
                   3.8pt\stroke}\phantom{C}}}}
\def\Rmath{\vcenter{\hbox{\upright\rlap{I}\kern 1.7pt R}}}
\def\Z{\ifmmode\Zmath\else$\Zmath$\fi}
\def\Q{\ifmmode\Qmath\else$\Qmath$\fi}
\def\N{\ifmmode\Nmath\else$\Nmath$\fi}
\def\C{\ifmmode\Cmath\else$\Cmath$\fi}
\def\R{\ifmmode\Rmath\else$\Rmath$\fi}

\def\barray{\begin{eqnarray}}
\def\earray{\end{eqnarray}}
\def\beq{\begin{equation}}
\def\eeq{\end{equation}}

\def\no{\noindent}
\def\n{\noindent}

\title{A model of a $2d$ non-Fermi liquid with 
$SO(5)$ symmetry, AF order, and  a d-wave 
SC  gap}
\author{Eliot Kapit and Andr\'e  LeClair}
\affiliation{Newman Laboratory, Cornell University, Ithaca, NY} 
\date{May 2008}

\bigskip\bigskip\bigskip\bigskip

\begin{abstract}

Demanding a consistent quantum field theory description of  spin $\inv{2}$  particles near 
a circular Fermi surface in $2d$ leads to a unique fermionic theory  
 with relevant quartic interactions which has an emergent 
Lorentz symmetry and  automatically has 
an $Sp(4) = SO(5)$ internal symmetry.   
   The interacting theory has a low-energy
interacting fixed point and is thus a non-Landau/Fermi liquid. 
   Anti-ferromagnetic (AF)
and superconducting (SC) order parameters are bilinears in the fields and 
form the $5$-dimensional vector representation of $SO(5)$.
An AF phase occurs at low doping which terminates in a first order transition.    
We incorporate  momentum dependent  scattering of Cooper pairs near the Fermi surface
to 1-loop and derive a new kind of SC gap equation beyond mean field  with a 
d-wave gap solution.    Taking into account the renormalization group (RG)
scaling properties near
the low energy fixed point,  we calculate  the complete phase diagram as a function
of doping,  which shows
some universal geometric features.    The d-wave 
SC dome terminates on the over-doped side at the fixed point of the RG,
which is  a quantum critical point.    
Optimal doping is estimated to occur just below  $3/2\pi^2$.   
The critical
temperature for SC at optimal doping  is set mainly  by  the 
universal nodal Fermi velocity and
lattice spacing,  and  is estimated to average  around $140K$ 
for LSCO.  The pseudogap energy scale is identified with
the RG scale of the coupling.

\end{abstract}


\maketitle

\def\Tr{\rm Tr} 
\def\xvec{{\bf x}}
\def\kvec{{\bf k}}
\def\kvecp{{\bf k'}}
\def\omk{\om{\kvec}} 
\def\dk#1{\frac{d\kvec_{#1}}{(2\pi)^d}}
\def\2pid{(2\pi)^d}
\def\ket#1{|#1 \rangle}
\def\bra#1{\langle #1 |}
\def\vol{V}
\def\adag{a^\dagger}
\def\rme{{\rm e}}
\def\Im{{\rm Im}}
\def\pvec{{\bf p}}
\def\fermiS{\CS_F}
\def\cdag{c^\dagger}
\def\adag{a^\dagger}
\def\bdag{b^\dagger}
\def\vvec{{\bf v}}
\def\muhat{{\hat{\mu}}}
\def\vac{|0\rangle}
\def\pcut{{\Lambda_c}}
\def\chidot{\dot{\chi}}
\def\gradvec{\vec{\nabla}}
\def\psitilde{\tilde{\Psi}}
\def\psibar{\bar{\psi}}
\def\psidag{\psi^\dagger} 
\def\m{m_*}
\def\up{\uparrow}
\def\down{\downarrow}
\def\Qo{Q^{0}}
\def\vbar{\bar{v}}
\def\ubar{\bar{u}}
\def\smallhalf{{\textstyle \inv{2}}}
\def\smallsqrt{{\textstyle \inv{\sqrt{2}}}}
\def\rvec{{\bf r}}
\def\avec{{\bf a}}
\def\pivec{{\vec{\pi}}}
\def\svec{\vec{s}} 
\def\phivec{\vec{\phi}}
\def\daggerc{{\dagger_c}}
\def\Gfour{G^{(4)}}
\def\dim#1{\lbrack\!\lbrack #1 \rbrack\! \rbrack }
\def\qhat{{\hat{q}}}
\def\ghat{{\hat{g}}}
\def\nvec{{\vec{n}}}
\def\bull{$\bullet$}
\def\ghato{{\hat{g}_0}}
\def\r{r}
\def\deltaq{\delta_q}
\def\gcharge{g_q}
\def\gspin{g_s}
\def\deltas{\delta_s}
\def\gQC{g_{AF}} 
\def\ghatqc{\ghat_{AF}}
\def\xqc{x_{AF}}
\def\mhat{\hat{m}}
\def\xup{x_2}
\def\xdown{x_1}
\def\sigmavec{\vec{\sigma}}
\def\xopt{x_{\rm opt}}
\def\Lambdac{{\Lambda_c}}
\def\angstrom{{{\scriptstyle \circ} \atop A}     }
\def\AA{\leavevmode\setbox0=\hbox{h}\dimen0=\ht0 \advance\dimen0 by-1ex\rlap{
\raise.67\dimen0\hbox{\char'27}}A}


\section{Introduction}

In the renormalization group (RG) framework, 
Landau's theory of Fermi liquids is characterized by the irrelevance of the 
interactions of particles near the Fermi surface, 
in other words the low energy fixed point
is simply a free theory of fermions.   
The underlying reasons for the wide success of Landau/Fermi liquid theory
are well-understood\cite{Benfatto,Shankar,Polchinski,Weinberg,Weinbergbook},
and consequently the known models of non-Landau/Fermi liquids are relatively
rare. (Henceforth referred to simply as non-Fermi liquids.)
  An important exception is the Luttinger liquid 
and other related models consisting of quartic interactions of Dirac fields 
in $d=1$ spatial dimension.    Here the non-Fermi liquid behavior
can be attributed to the fact that in $1d$, quartic interactions of Dirac fields
are marginal operators in the RG sense.  In higher dimensions
quartic  interactions  of Dirac fermions are irrelevant and this is one of the 
reasons why  candidate non-Fermi liquid models  were not  found in the previous
works.   Whereas more exotic non-Fermi liquid models have been
proposed which typically involve  gauge fields,  the lack of 
non-Fermi liquid models in $2d$ appears paradoxical when one
considers even the simplest models of itinerant electrons with
quartic interactions,  such as the Hubbard or t-J model,   which are believed
to be at strong coupling.   Since such models have been
proposed as good starting points for thinking about high $T_c$ superconductivity
in the cuprates\cite{Anderson,Anderson2,Anderson3}, 
 it is certainly worthwhile   to continue to try  
and construct relatively
non-exotic models of continuum fermions  with quartic interactions that
have some resemblance to the Hubbard model and have non-Fermi
liquid behavior in the normal state.

Though the search for a novel kind of non-Fermi liquid in $2d$ 
provided one of the main initial motivations  for 
the formulation of the model that will be presented and analyzed in
this work,  the model turns out to have many unexpected  bonus features,
almost all of which are intrinsic to $2d$.   
We list the most prominent:

\vfill\eject

\n\bull ~~    The 4-fermion interaction is unique
for spin $\smallhalf$ electrons and automatically has $SO(5)$ symmetry. 
  In $2d$ the interactions are relevant
and the model has a low energy interacting fixed point with non-classical
exponents which can be computed perturbatively.  

\bigskip
\n\bull ~~ The model generalizes  to $N$ flavors,  where $N=2$ corresponds
to spin $\smallhalf$ electrons,  and has $Sp(2N)$ symmetry.   
Since $Sp(4) = SO(5)$ this provides a underlying framework based
on a microscopic theory  for
exploring the ideas of Zhang based on $SO(5)$\cite{Zhang,Zhang2}.   
In particular one can derive the effective Ginzburg-Landau theory.

\bigskip

\n\bull ~~  Because of the $SO(5)$ symmetry the model naturally  has
both anti-ferromagnetic (AF) and superconducting (SC) order 
parameters that form the 5-dimensional vector representation of
$SO(5)$.  For repulsive interactions the model has AF order 
and no SC order in mean field approximation.  

\bigskip

\n\bull ~~  When one incorporates momentum dependent scattering to 1-loop
to go beyond mean field,  an attractive d-wave channel opens up
and the momentum dependence of the gap can be calculated. 
This d-wave SC phase terminates on the over-doped side at the RG fixed point,
which is a quantum critical point.   Due to mathematical properties of the
d-wave gap equation,  it also terminates on the under-doped side yielding a
``dome''.   Due to the properties of the RG flow,  this attractive d-wave instability
exists for arbitrarily strong repulsive interactions at short distances.    

\bigskip

\n\bull ~~ Although the model may be at arbitrarily strong coupling at 
short distances,  the low energy fixed point is at a relatively
small coupling $\approx 1/8$,  and this renders 
the model perturbatively calculable.  We are thus able to
calculate the main features of the complete phase diagram
as a function of a doping variable, 
including the phase boundary  of the d-wave superconducting  dome
and estimate the optimal doping fraction, which is near
$3/2\pi^2 \approx .15$.    This phase diagram
depends on a single parameter $0<\gamma<1$  which 
encodes the ratio of the strength of the coupling at short  versus long distances.  
In the figure below,  we summarize the results of our calculations
for $\gamma =1$,  which corresponds to infinite coupling at short distances.  
The overall scale of temperature is set by the universal nodal Fermi
velocity and the lattice spacing.

\begin{figure}[htb] 
\begin{center}
\hspace{-15mm}
\psfrag{X}{$h = {\rm doping}$ }
\psfrag{A}{$0.5$ }
\psfrag{B}{$.06$ }
\psfrag{C}{$h_{AF}$}
\psfrag{D}{$h_*$}
\psfrag{n}{$h_1$}
\psfrag{F}{$SC$} 
\psfrag{G}{${\rm pseudogap}$}
\psfrag{H}{$T_{pg}$ }
\psfrag{I}{$AF$}
\psfrag{T}{$\deltas', \deltaq'$}
\psfrag{N}{$T_N$}
\psfrag{S}{$T_c$}
\includegraphics[width=12cm]{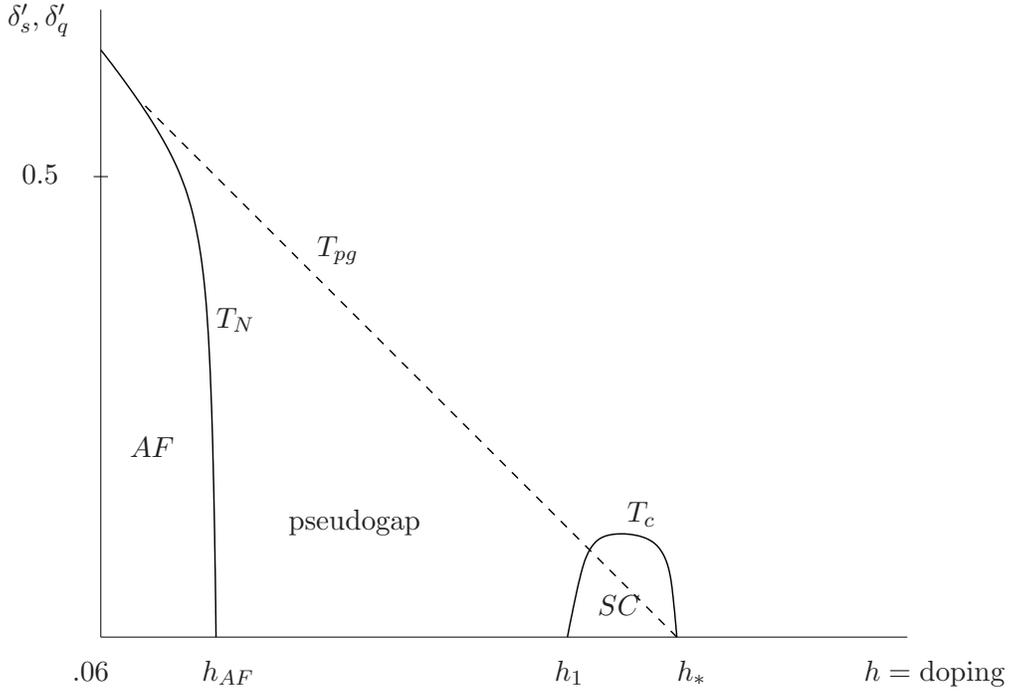} 
\end{center}
\caption{{\it Calculated} phase diagram as a function of hole doping,
which depends on a single parameter $0<\gamma<1$ determined
by the strength of the interaction at short distances.
We set $\gamma =1$ corresponding to infinitely strong
coupling.  The vertical axis represents the low energy scale relative
to the cut-off.  
What is shown are solutions $\deltas'$ and $\deltaq'$ 
of the AF and d-wave gap equations
(\ref{AF.1b},\ref{orb.7}) in units of the cut-off $\Lambda_c$.
The critical temperatures are proportional to these gaps
with constants of order unity which we estimate,
eqns. (\ref{Neeltemp},\ref{TcSC}). 
The overall temperature scale is determined by
the universal nodal Fermi velocity and lattice spacing
(\ref{Tcev}).   
The running coupling $g$ in the gap equations
is in terms of $x$ in eq. (\ref{Phase.2a}),
where $x$ is the inverse dimensionless coupling. The hole-hoping 
$h(x)$ is the one loop corrected expression  (\ref{dope.2}).  
The straight line $T_{pg}$ is the renormalization group scale 
corresponding to the energy scale of the coupling,  
eq. (\ref{Phase.1b}), and represents the boundary to the
pseudogap region. The AF transition point at $h_{AF} = 3/ 4\pi^2$ 
is first order.   The SC transition at $h_* = 3/2\pi^2$ is second-order
and corresponds to the fixed point of the renormalization group,
i.e. a quantum critical point.  The transition point $h_1 \approx 0.13$
is not universal, but relies on mathematical properties of the 
gap equation.} 
\vspace{-2mm}
\label{FigureIntro} 
\end{figure}

\bigskip

\bigskip

Many of the above  properties were highlighted  on a list of the most important
features of high $T_c$ superconductivity  compiled early on in the 
subject\cite{Anderson,Anderson3}.     Because of the importance
of the $CuO_2$  planes,    high $T_c$ is believed to be essentially
a $2d$ phenomenon.    This, along with the detailed properties of
the solution of our model, in particular the  phase diagram,
led us to propose it as a model of high $T_c$ 
superconductivity\cite{KapitLeClair}. 
    If our theory  turns out to be the  correct description,  
it reveals  that the phenomenon of high $T_c$ superconductivity
is remarkably universal,   with a single energy scale,  and its main features follow from
the existence of the low energy fixed point in $2d$. 
It is truly a beautiful phenomenon that has managed to 
realize some subtle theoretical loopholes in  the usual requirements
of unitarity,   the spin-statistics theorem,   and the Mermin-Wagner
theorem,  which 
are  only possible
in $2d$.      
Our theory  represents a significant departure from 
the models considered thus far in connection with high $T_c$,
which are reviewed in \cite{Kivelson,Leggett,WenLee,Norman},
along with  reviews of experimental results.      On the other hand,
we believe it  represents a particular scaling limit of the
Hubbard model at and just below half-filling, and is in this
sense conservative in comparison with other more exotic ideas, 
and is thus in line with the early ideas  concerning
the r\^ole  of AF order and the Heisenberg and Hubbard 
models\cite{Anderson,Anderson2,Scalapino}. 
However our model isn't simply  a direct scaling limit of the
Hubbard model  with no attention paid to the Fermi surface, 
since the latter only has an $SO(4) = SU(2) \otimes SU(2)$ symmetry,
whereas our theory has the $SO(5)$ symmetry.  
In our theory  the ``fermion sign'' problem is solved by doing analytic, perturbative
calculations in a fermionic theory from the beginning,  and relatively
simple 1-loop calculations already reveal the main features.

Irregardless of whether our model has been exactly realized in the laboratory,  
it  can serve as a useful tool for exploring many of 
the paradigms in the area of strongly correlated electrons 
and also for developing new methods.  For instance,  we develop
new gap equations  that take  into account higher order scattering
of Cooper pairs near the Fermi surface.   Our analysis shows clearly
how in $2d$ one can obtain a momentum dependent gap 
with a d-wave structure
from a rotationally invariant continuum field theory, i.e. without
an explicit  lattice that breaks the rotational symmetry.   This is interesting
especially since the precise origin of the d-wave symmetry of the SC gap
has been unclear.   We also show how to introduce
doping in terms of the coupling and RG scale, and 
a small non-zero temperature as a relativistic mass coupling.  

For the remainder of this introduction we outline the 
organization of the paper and summarize our main results.  
In section II we motivate the model by showing how it can approximately
describe particles and holes near a circular Fermi surface.    The manner in which
we expand around the Fermi surface is in the same spirit as in
\cite{Benfatto,Shankar,Polchinski,Weinberg,Weinbergbook} but differs in some
important ways.  For a single spin-less fermion one thereby
obtains a free hamiltonian of particles and holes with a massless,
i.e. relativistic dispersion relation.     In section III 
we insist on a local quantum field description of the effective
theory near the Fermi surface with a consistent quantization. 
Since the particles are massless, the only known candidate
field theories are either Dirac or ``symplectic'' fermions, which differ
primarily by being first versus  second order in space-time derivatives 
respectively.    For the remainder
of the paper we focus on  symplectic fermions  since
unlike the Dirac fermions,  the interactions are relevant.   The model was first proposed in this context  
by one of us\cite{LeClair1}, where the groundwork  was done on the low energy 
non-Fermi liquid fixed point and in part the  AF properties; at the time  the SC properties
were unknown.     
As explained in the present paper,  the central
idea of this previous work,  that the AF order parameter is bilinear in
symplectic fermion fields and that the low energy RG fixed point
describes a quantum critical point, appears to be correct; 
however as we will see,  the quantum critical point terminates the SC rather
than AF phase.  
Quantum critical points in the  context of high $T_c$  were emphasized earlier
by Vojta and Sachdev\cite{Vojta}.   
 The issue of 
the unitarity of our  theory was mostly resolved in
\cite{Neubert} by noting that the hamiltonian is 
pseudo-hermitian and this is sufficient for a unitary
time evolution.   In this paper, the expansion around the
Fermi surface provides a new view  on the pseudo-hermiticity
and it is explained how it is related to the kinematics of
particles versus  holes.    
The critical exponents were computed to 2-loops in \cite{Neubert},
which corrected some errors in\cite{LeClair1}. 
In this paper 
 we analyze many more properties, 
in particular  the  AF and d-wave SC ordering properties
for the first time.

Since the consistency of the quantization of a fermionic theory
with a lagrangian that is second order in time derivatives 
is at the heart of the unitarity issue,  in section IV we work
out  in detail the $d=0$ dimensional quantum mechanical case
where all the subtle consistency issues are present.  In this section
we also construct the conserved charges for the $Sp(2N)$ symmetry.  
In section V the field theory version in $d$ spatial dimensions
is defined and spin and charge are identified for the case of
$N=2$.  In this section we also define the $SO(5)$ order parameters
for AF and SC order. 

In section VI we sketch an argument that the resistivity is linear in temperature
in the limit of no interactions.  In the next section 
  we  consider small thermal perturbations
near $T=0$.  By comparing with the specific heat of a degenerate
electron gas, we argue that a small non-zero temperature
can be incorporated as a coupling in the lagrangian corresponding
to relativistic mass $m=\alpha T$ and we estimate the  constant $\alpha$.  

In section VIII the mean field analysis is carried out with 
potential competition
between AF and SC order.  As we explain,  these two phases
actually do not compete in our model in this approximation.
   As a check of the formalism,
we reproduce some of the basic features of the BCS theory
for an s-wave gap 
in the case of an attractive coupling in section IX.   
For repulsive interactions we find only AF order is possible in mean field approximation 
and this is studied in section X.   
There we first argue that this phase must be anti-ferromagnetic
by comparing our model with the low energy non-linear sigma model
description of the Heisenberg anti-ferromagnet.   
This gives another motivation for  our model at half-filling
away from the circular Fermi surface,  and explains how the same
model can interpolate between a SC phase and an AF one.   
We argue that the AF phase terminates in a first-order phase transition.  
 The AF gap  is then the solution of 
a transcendental equation that is analyzed in various limits.

In section XI,  orbital symmetries of momentum dependent gaps
are studied in a model independent way and we explain how
a d-wave gap can arise.   This analysis is based on a 
 gap equation which is derived in Appendix A.  
In section XII  we compute the 1-loop contributions to the
scattering of Cooper pairs and show that at low energies the
d-wave channel is attractive if the number of components $N<3$.
Since the theory is free for $N=1$, this means that only
the physically relevant $N=2$ case has d-wave SC.  This also means
that the d-wave pairing cannot be studied with large $N$ methods.

Section XIII is devoted to describing our RG prescription which
is specific to $2d$.    This is necessary for a proper understanding
of the phase diagram.   In section XIV we present 
global features of the phase diagram,  which is characterized by
some universal geometric relations, 
and bears a striking resemblance to the cuprates. 
The SC phase terminates at a second order phase transition 
precisely at the low energy RG fixed point,  and is thus
a quantum critical point.   
We also estimate optimal hole doping.  
In section XV we present detailed numerical solutions to the
AF and SC d-wave gap equations at non-zero temperature.
For reasonable values of the lattice spacing and universal nodal 
Fermi velocity, 
we estimate $T_c \approx 140K$ on average  for SC in LSCO.      
In section XVI  we  describe  the interpretation of
the pseudogap  within our model.

Although we do not give a complete and rigorous derivation of our model
from lattice fermion models,  in order to motivate and point out relations, 
 we have collected 
some known results about the latter in Appendix B.

\section{Expansion around the Fermi surface}

\subsection{Kinematics}

Let us first ignore spin and consider a single species of fermion
described by the free hamiltonian in momentum space
\beq
\label{E.1}
H =  \int (d^d \kvec)  \( \vep(\kvec ) - \mu \) 
\cdag_\kvec c_\kvec 
\eeq
where $\mu$ is the chemical potential,  we have defined
$(d^d \kvec) \equiv d^d \kvec / (2\pi)^d $,  and 
\beq
\label{E.2}
\{ \cdag_\kvec , c_{\kvec'} \}  = (2\pi)^d \delta^{(d)} (\kvec - \kvec')
\eeq
At finite density and zero temperature,  all states with $\vep \leq \vep_F$
are filled,  where the Fermi energy $\vep_F$ depends on the 
density,  and at zero temperature $\mu = \vep_F$.   
The Fermi surface $\fermiS$ is the manifold of points 
$\kvec_F$ satisfying $\vep (\kvec_F ) = \vep_F$.   

\begin{figure}[htb] 
\begin{center}
\hspace{-15mm}
\psfrag{A}{$k_x$ }
\psfrag{B}{$k_y$}
\psfrag{S}{$\fermiS$}
\psfrag{p}{$\pvec$}
\psfrag{k}{$\kvec$} 
\psfrag{f}{$\kvec_F$}
\includegraphics[width=10cm]{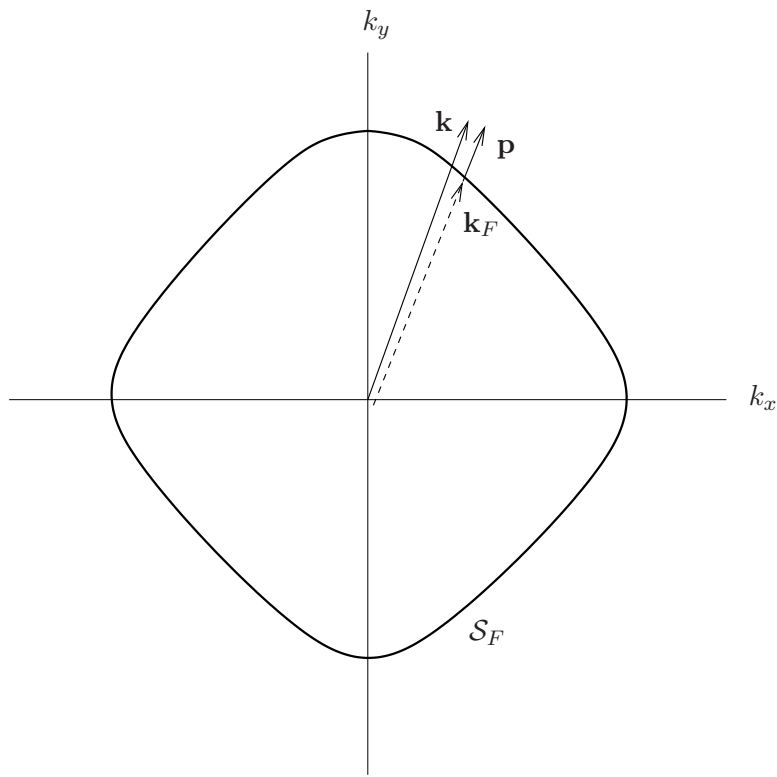} 
\end{center}
\caption{Expansion around the Fermi Surface} 
\vspace{-2mm}
\label{Figure1} 
\end{figure}

We wish to consider a  band of energies near $\vep_F$ as shown in
Figure \ref{Figure1}  for $d=2$.   Let $\kvec$ be any wave-vector in such a
band,  and let $r(\kvec)$ denote a ray from the origin to
infinity along the direction of $\kvec$.     We further
assume that the Fermi surface is sufficiently smooth, such  that $r(\kvec )$ 
intersects $\fermiS$ only once.   The latter implies that
$\kvec$ can be {\it uniquely}  expressed as
\beq
\label{E.3}
\kvec =  \kvec_F (\kvec) + \pvec (\kvec) 
\eeq
where $\kvec_F (\kvec )$ is the vector from the origin to 
the intersection of $r(\kvec)$ with $\fermiS$.   Whereas the 
two vectors $\kvec$ and $\kvec_F (\kvec)$ are by construction
parallel,  the vector $\pvec (\kvec )$ is either parallel or
anti-parallel to $\kvec$.   Let us fix $\pvec$ to be a small
vector parallel to $\kvec$, i.e. pointing radially outward, 
 as shown in Figure \ref{Figure1}.  Since now $\kvec_F (\kvec)$ is uniquely
determined by $\pvec$,  we may write $\kvec_F (\pvec )$.    Furthermore,
since the 
particles below the Fermi surface correspond to $-\pvec$, 
the energies near the Fermi surface are approximately given by
\beq
\label{E.4}
\vep (\kvec ) =  \vep_F \pm  \pvec \cdot \vvec_F (\kvec )
\eeq
where $\pm$ corresponds to above or below the Fermi surface, and 
\beq
\label{E.5}  
\vvec_F (\kvec)  = \vec{\nabla} \vep (\kvec ) \vert_{\kvec_F}  
\eeq
is the Fermi velocity normal to $\fermiS$.   
\bigskip

Let us now assume that the Fermi surface is rotationally invariant,
i.e. $\vep (\kvec)$ depends only on $|\kvec|$.  In $2d$ the
Fermi surface $\fermiS$ is thus a circle.    This leads to the
simplification that $k_F = | \kvec_F (\kvec) |   $ and
$v_F = |\vvec_F (\kvec )| $ are independent of $\kvec$.  
For any $\kvec$ in the band,  
\beq
\label{E.5b}
\kvec = (k_F \pm p ) \hat{p}  
\eeq
where $p= |\pvec|$, and $\pvec = p \hat{p}$.   
 Furthermore,
since $\vvec_F$ is normal to $\fermiS$,  the energies are 
linear in $|\pvec|$:
\beq
\label{E.6}
\vep (\kvec ) = \vep_F \pm v_F \, | \pvec | 
\eeq
For non-relativistic particles with $\vep (\kvec) = \kvec^2 / 2\m$, 
$v_F = k_F/\m $.

Since the map from $\kvec$ to $\pvec$ is one-to-one,  we can
define the following operators:
\beq
\label{E.7}
a_\pvec =  c_{\kvec_F + \pvec }  ,  ~~~~~
b_\pvec = \cdag_{\kvec_F - \pvec} 
\eeq
where it is implicit that $\kvec_F$ depends on $\pvec$.
The above is a canonical transformation, since
\beq
\label{E.7b}
\{ \adag_\pvec , a_{\pvec'} \}  = \{ \bdag_\pvec , b_{\pvec'} \}
= (2\pi^d) \delta^{(d)}(\pvec - \pvec' ) 
\eeq    
After normal ordering,  the hamiltonian for the particles in
the band is defined to be
\beq
\label{E.8}
H =  \int_{|\pvec| < \pcut}  (d^d \pvec)  \[ 
(v_F |\pvec | - \muhat )  \adag_\pvec a_\pvec + 
(v_F |\pvec| + \muhat ) \bdag_\pvec b_\pvec  \]  
\eeq
where
$\muhat = \mu - \vep_F$ is zero at zero temperature. 
For the remainder of this paper we mostly set $\muhat =0$.    
Since we are only interested in a band of energies near
the Fermi surface,  we have introduced a cut-off $\pcut$. 
The vacuum $\vac$  is defined to satisfy
$a_\pvec \vac  = b_\pvec \vac = 0$.     This 
corresponds to $c_{\kvec_F + \pvec} \vac = \cdag_{\kvec_F - \pvec } 
\vac = 0$,  which correctly implies that all states with 
$\vep < \vep_F$ are filled.  The $a_\pvec$ and $b_\pvec$ thus
correspond to particles and holes respectively.

There is an approximation made in obtaining the above hamiltonian
having to do with the density of states, and this is crucial
to understanding how our expansion differs from previous works.    
In  the rotationally invariant case, for particles above the
Fermi surface: 
\beq
\label{E.9}
\int d^d \kvec = \int d\Omega \int dp \,  (k_F +  p)^{d-1}  
\eeq
where $d\Omega$ are angular integrals.   Note that 
due to eq. (\ref{E.5b}),   the angular integrals for
$\kvec$ and $\pvec$ are identical. 
At least two approximations to the above are meaningful.  
The first favors low energies where one  approximates  $k_F + p\approx k_F$.
This is the approximation that is commonly made in
the literature\cite{Benfatto,Shankar}.    
One the other hand, 
 expanding
out the $(k_F + p)^{d-1}$, at high energies  the leading term  is  $p^{d-1}$,
and 
is the most sensitive to the short distance physics 
and spatial dimensionality. 
A possible short-coming of the first, low-energy approximation
is that the high energy physics is discarded from the beginning.
It cannot be recovered by the RG flow to low energies since
the latter is irreversible.   
In the physical problem we are considering,  the short-distance
physics of the strong Coulomb repulsion is known to be important
for understanding the AF phase,  so it makes sense to 
adopt an approximation that favors high energies from the beginning
and to then incorporate their effects by an RG flow to lower energies. 
We thus keep the   
 most important term at short distances and 
 set 
$d^d \kvec = d^d \pvec$,  i.e. $k_F + p \approx p$.      
This is in line with the usual  RG idea that it
is important to fix  the high-energy physics as
accurately as possible, and then flow down to lower energies. 
Finally our choice is necessary for 
the  $2d$ effective field theory description in the next section.
However one should not conclude  that every theory with a circular Fermi
surface can be described by a relativistic field theory. 
One signature of a relativistic description is 
a density of states that is linear in energy:
$\int d^2 \pvec =  2\pi v_F  \int  d\vep \, \vep$.     
  
The above expansion around the Fermi surface is thus not identical
to the expansion in \cite{Benfatto,Shankar,Polchinski,Weinberg,Weinbergbook},
where  the integration over $\pvec$ is taken to
be normal to $\fermiS$ times the angular integrations, 
and it is assumed that $p\ll k_F$.  This leads to 
the choice  $\int d^d \kvec  =  \int d\Omega  \int dp   \, k_F^{d-1}$, 
and the constant $k_F$ is absorbed into the definitions of the
operators. 
Thus in the  approach  followed in 
\cite{Benfatto,Shankar,Polchinski,Weinberg,Weinbergbook}, 
although the angular integrals obviously depend on $d$,  
the scaling analysis of the $p$ dependence leads to marginal
4-fermion interactions for any $d$,  and the resulting theory is
effectively $1$-dimensional, or a collection of such theories,
one for each angular direction.  Notably,  it was not possible 
to obtain a non-Fermi liquid based on 4-fermion interactions 
in this approach\cite{Shankar}.    

In contrast,  in the approach developed in this paper there 
is  a strong dependence on $d$, as in other critical phenomena,
and this will turn out to be very important.  In particular,  it 
leads to a non-Fermi liquid in $2d$.
There are other important
justifications for this choice.    In particular,  near half-filling 
where the interacting lattice model can be mapped to the Heisenberg anti-ferromagnet,  
there is  known to be a relativistic description of the low energy,
long wavelength limit in terms of the $O(3)$ non-linear sigma model.
It will be shown in section X that our choice of field theory 
near the circular Fermi surface can be extrapolated to half-filling
in that an independent derivation of it can be provided exactly
at half-filling.

For general processes, physical momentum  conservation of
the $\kvec$'s is not equivalent 
to $\pvec$ conservation.   However,  consider a zero-momentum
process proportional to $\delta(\sum_i \kvec_i)$.  If the 
$\kvec$'s are all  exactly on the Fermi surface,  then
this implies $\sum_i \kvec_F (\kvec_i ) =0$.  For even numbers
of particles, since all vectors on the Fermi surface have
the same length,  this is satisfied by pairs of particles with
opposite $\kvec_F$.  Allowing now small deviations  $\pvec_i$ from 
the Fermi surface, one has   
\beq
\label{Cons}
\delta (\sum_i \kvec_i )  = \delta(\sum_{\rm particles } \pvec_i 
- \sum_{\rm holes }  \pvec_i )
\eeq
Because of the particle/hole transformation for the $b$'s in
eq. (\ref{E.7}),  this is equivalent to overall $\pvec$ conservation.
Note that by construction,  it is not possible for the momentum
of a particle and a hole to add up to zero,  so in the above
$\delta$-function,  holes are paired with other holes,  and
particles with other particles.  
Therefore, spatial  translational invariance of our local field theory
will ensure physical momentum conservation of the $\kvec$'s 
for this class of processes.   

\begin{figure}[htb] 
\begin{center}
\hspace{-15mm}
\psfrag{A}{$\fermiS$ }
\includegraphics[width=12cm]{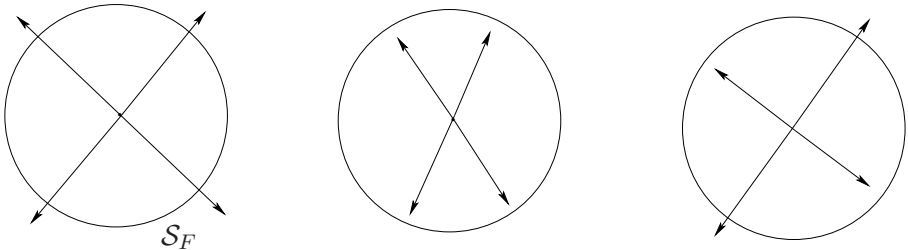} 
\end{center}
\caption{Allowed processes} 
\vspace{-2mm}
\label{Figure2} 
\end{figure}

\begin{figure}[htb] 
\begin{center}
\hspace{-15mm}
\includegraphics[width=12cm]{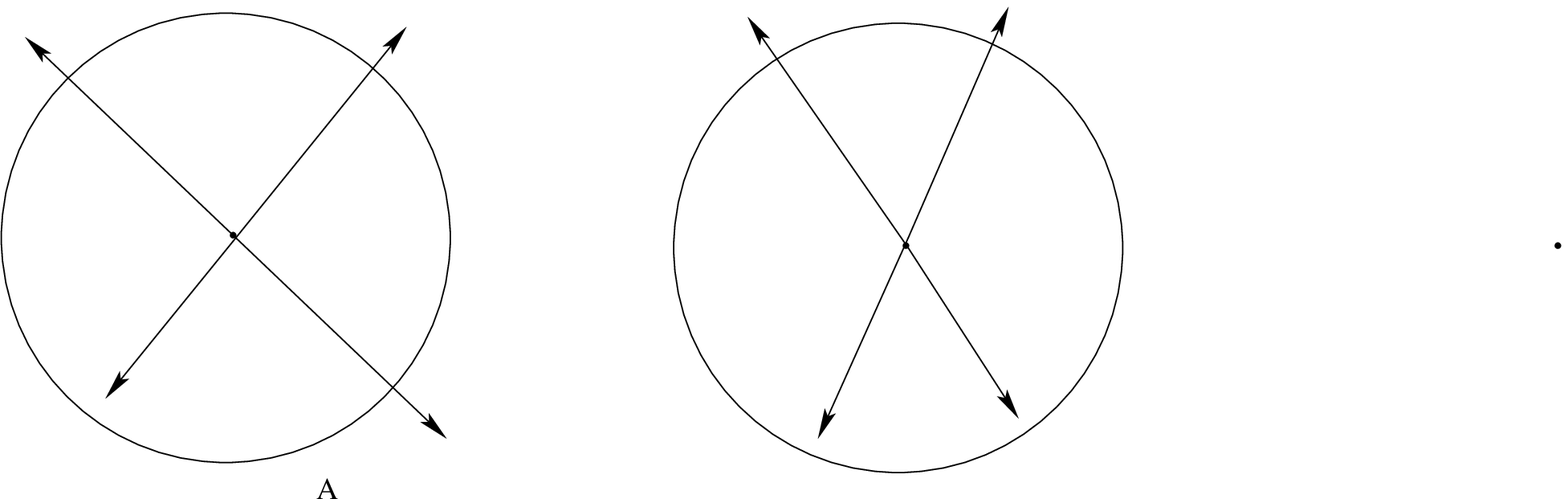} 
\end{center}
\caption{Un-allowed processes} 
\vspace{-2mm}
\label{Figure3} 
\end{figure}

The important allowed processes are shown in Figure \ref{Figure2}.  Examples 
 of  un-allowed process are  shown in Figure \ref{Figure3}.  
The distinction between allowed and un-allowed processes
can be made explicit by introducing an operator $C$ 
that distinguishes particles and holes:
\beq
\label{pso}
C \, a_\pvec  \, C =  a_\pvec, ~~~~~~~
C \, b_\pvec  \, C = -b_\pvec 
\eeq
where $C$ is a unitary operator satisfying $C = C^\dagger$
so that $C^2 =1$.   An eigenstate with pairs of particles and/or pairs of  holes is 
then required to have $C=1$.   We will return to this in
connection with the pseudo-hermiticity of symplectic fermions
in the sequel.

\subsection{Lattice fermions}

\begin{figure}[htb] 
\begin{center}
\hspace{-15mm}
\psfrag{A}{$\fermiS$ }
\includegraphics[width=6cm]{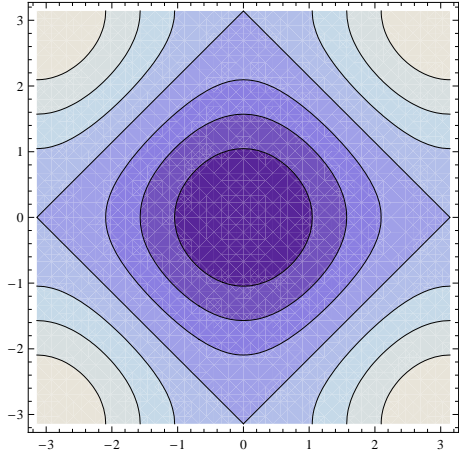} 
\end{center}
\caption{Fermi surface contours for free lattice fermions in $2d$ with lattice spacing $a=1$.} 
\vspace{-2mm}
\label{Figure4} 
\end{figure}

In the sequel, our field theory model will be related to 
lattice models of intinerant electrons  
such as the Hubbard model, although we do not claim a precise equivalence.
   The known $2d$ square  lattice model 
results we will need to make the comparison are all contained in Appendix B.
In this section we consider only the free, hopping term.  
In momentum space the 1-particle energy is (\ref{latticeep}) 
\beq
\label{lattice.1}
\vep_\kvec = -2t ( \cos k_x a +  \cos k_y a ) 
\eeq
where $a$ is the lattice spacing.   Equal energy contours
in the first Brillouin zone are shown in Figure \ref{Figure4}.
The Fermi surface at half-filling is the square diamond with
corners on the $x,y$ axes.  
Note that one does not have to be very far below half-filling
for the contours to be approximately circular.  
The free local field theory model in the next section can thus be viewed
as an approximate effective theory for free particles on the lattice 
below half-filling.    

An important point is that our model is not simply a direct 
continuum limit  of the lattice model since,  without additional care, 
the latter does not
take into account the Fermi surface at finite density.  
For instance, 
whereas the Hubbard model has at most an $SO(4)=SU(2) \otimes SU(2)$ symmetry\cite{YangZhang},
our continuum model has the larger $SO(5)$ symmetry.     
Furthermore, as will be explained in section X,  the success of our model
can be attributed to the fact that an alternative justification of it
can be given right at half-filling so that it can actually interpolate 
between half-filling and below.

\section{Requirements on the free local field theory}

The main requirements we impose for  a local field theory description of the
last section are:

\noindent (i)  The theory has a lagrangian description with 
a consistent quantization.   

\noindent (ii)  In momentum space the hamiltonian reduces to 
equation (\ref{E.8}) for particles and holes of energy 
$v_F |\pvec| $.   The latter is a relativistic dispersion relation
for massless particles.

In order to motivate our arguments, let us start from  
non-relativistic particles with $\vep_\kvec = \kvec^2 / 2\m$.
The second-quantized description consists of a single field 
$\Psi (\xvec , t)$ with lagrangian
\beq
\label{R.1}
L =  \int d^d \xvec \( i \Psi^\dagger \d_t  \Psi -  \inv{2\m} 
\gradvec \Psi^\dagger \cdot \gradvec \Psi \)
\eeq
The field has the momentum space expansion
\beq
\label{R.2}
\Psi (\xvec, t) =  \int (d^d \kvec)   
\,   c_\kvec  \,  e^{ - i \vep_\kvec  + i \kvec \cdot \xvec }  
\eeq
Expanding around the Fermi surface one finds
\beq
\label{R.3} 
\Psi (\xvec , t) =  e^{-i \vep_F t }  
\int  (d^d \pvec)     \, 
\tilde{\Psi}_\pvec  (\xvec , t )  \,   e^{i \kvec_F (\pvec ) \cdot \xvec } 
\eeq
where 
\beq
\label{R.4} 
\tilde{\Psi}_\pvec (\xvec , t) =  
a_\pvec \,  e^{-i v_F |\pvec| t + i \pvec\cdot \xvec }  
+  \bdag_\pvec \, e^{i v_F |\pvec| t - i \pvec\cdot \xvec} 
\eeq

We wish to find an effective theory for $\tilde{\Psi}$,  which 
satisfies the relativistic wave equation:
\beq
\label{R.5} 
\( \d_t^2 - v_F^2  \gradvec^2 \)  \tilde{\Psi}_\pvec (\xvec , t) = 0
\eeq
Thus, due to the kinematics of the expansion around the Fermi surface
we identify an emergent Lorentz symmetry.  In $2d$ this Lorentz symmetry
is $SO(3)$.  
The case of $d=1$ is special in that the Fermi surface consists of
only two disconnected points $\kvec_F  = \pm k_F$, and the 
decomposition (\ref{R.3}) naturally separates into left and
right movers.   For higher dimensions there is no such
separation since all points on the Fermi surface are related by
spatial  rotations and continuously connected.  
There are many additional reasons why $d=1$ is the exceptional case,
and these will be pointed out where appropriate in the sequel since
some of the literature attempts to draw analogies between $1d$ and
$2d$.

There are only two known  candidate field theories which differ
in whether the lagrangian is first or second order in derivatives.  
First consider the case of first-order.   One  then needs to factor 
the operator $\d_t^2 - \gradvec^2$ into two first-order multiples.  
The only way to accomplish this is to promote $\tilde{\Psi}$ 
to a multi-component field and introduce a matrix representation 
$\gamma^\mu$, $\mu= 0, 1, .., d$, 
of the Clifford algebra:
\beq
\label{R.6} 
\{ \gamma^\mu , \gamma^\nu \}  = 2 \eta^{\mu \nu}
\eeq
where $\eta^{\mu\nu} = {\rm diag} (1,-1,., -1)$.   
One then has
\beq
\label{R.7}
\sum_{\mu , \nu} \gamma^\mu  \gamma^\nu  \d_\mu \d_\nu  
=  \sum_\mu  \d^\mu \d_\mu  =  \d_t^2 - \gradvec^2  
\eeq
(We have adopted the relativistic notation $x^\mu = (x^0, x^1 , .., x^d)
= (t, \xvec )$ and $\d_\mu = \frac{ \d }{\d x^\mu } $, 
and henceforth, repeated indices are implicitly summed over.)  
The lagrangian is then the standard first-order Dirac lagrangian:
\beq
\label{R.8} 
L =  \int d^d \xvec  ~  i \,  \bar{\psi}  \gamma^\mu \d_\mu  \, \psi 
\eeq
where $\psibar = \psidag \gamma^0$.   
The smallest representation of the Clifford algebra is
two dimensional:  $\gamma^0 = \sigma_z, \gamma^1 = i\sigma_x$,
and $\gamma^2 = i \sigma_y$ where $\vec{\sigma}$ are the standard Pauli
matrices.  
In this simplest case, although
 there is a doubling of components,  they are constrained
by the Dirac equation of motion and the spectrum still consists of
one species of particles and holes with hamiltonian
eq. (\ref{E.8}).  

Interactions are relevant to the low energy physics if 
the operator characterizing them has scaling dimension
less than  $d+1$.  
The classical scaling dimension of the Dirac  field $\psi$ is
$d/2$ in $d$ spatial dimensions,  thus a quartic interaction 
has dimension $2\cdot d$ and is thus only relevant  for $d<1$.   
It is not even perturbatively renormalizable for $d>1$.
Thus, the Dirac theory should lead to ordinary Landau-Fermi liquid behavior in $2d$.  
It is noteworthy that once again $d=1$ is special, and this
helps to explain how for instance the $1d$ Hubbard model 
can be mapped onto interacting Dirac fermions,
and the low energy fixed point found using special bosonization
techniques and spin-charge separation\cite{Affleck}.  Simply based on the fact 
that the interactions in the Hubbard model are strong in $2d$,
one can rule out a description in terms of Dirac fields 
with quartic interactions since the latter are irrelevant.  
Furthermore,  it is already understood that one normally needs additional special properties
in order to obtain the first-order Dirac theory.  For example,  it
is known to arise when one expands around special nodes (Dirac points)  on the Fermi surface
for a hexagonal lattice\cite{Semenoff}, as in graphene,  and the 
multiple components of the Dirac field correspond to different sub-lattices.
There is no reason to expect 
this here for a square lattice,  as in the cuprates.

The other candidate field theory is second-order in derivatives,
 with kinetic term
\beq
\label{symp.1}
S = \int dt \, d^d \xvec  ~ \d_\mu \chi^-  \d^\mu \chi^+
\eeq
For $\chi^\pm$ fermionic (Grassman) fields,  this is a very unconventional
theory,  since  it potentially has problems with the spin-statistics theorem
and unitarity;  in  high energy elementary particle theory it usually corresponds to ghost fields. 
   These issues will be discussed in detail and resolved completely
in the next two sections.   Here,  let us give the main arguments for 
why this should be the right starting point:

\n (i)  As shown in the next two sections,   the free theory in momentum space
corresponds precisely to the hamiltonian (\ref{E.8})  for particles and holes
near a circular Fermi surface.   This is of course a perfectly hermitian theory
with no negative norm states.

\n (ii)  The fundamental 
field $\chi$  has scaling dimension $(d-1)/2$ and thus  quartic interactions
have dimension $2(d-1)$ which is actually relevant for $d<3$. 
Thus it can have non-Fermi liquid behavior.   

\n (iii) Although we are led to consider this model for the nearly circular
Fermi surface below half-filling,   a simple argument leads to the same 
model at half-filling.   It is well-known that a low-energy description of
excitations above the staggered AF state is described by the $O(3)$ non-linear
sigma model for a field $\phivec$ constrained to have constant length 
with the action 
\beq
\label{symp.2}
S =  \int dt \, d^d \xvec  ~  \d_\mu \phivec \cdot \d^\mu \phivec   
\eeq
(See Appendix B.)  
In our model the anti-ferromagnetic order parameter $\phivec$ is bilinear
in the fields $\phivec = \chi^- \sigmavec \chi^+ /\sqrt{2} $.   The non-linear
constraint on the $\phivec$ fields follow from imposing  a similar constraint
on the $\chi$ fields:   $\chi^- \chi^+ = {\rm constant}$.  This was pointed out
in\cite{LeClair1}.  Inserting this
into the above action one finds that one obtains the second order 
action (\ref{symp.1}) for the $\chi$ fields up to irrelevant operators
(eq. (\ref{half.5}) below).   Thus the symplectic fermion model
with interactions can in principle describe AF order,  and in the sequel we
will show that this is indeed the case. 
  This is explained in greater detail in section X.

Since $v_F$ only serves to convert dimensions of time and space,
it plays the r\^ole  of the speed of light;  we can set it to unity
since it can always be restored by dimensional analysis. 
The important point here is that $v_F$ is fixed and universal 
in our model,  in particular it does not depend on the coupling.
In the sequel the doping will be related to the coupling,  so
$v_F$ does not depend on doping either.   For high $T_c$ materials,
we believe the ``speed of light''  has actually already been measured\cite{FermiV}.  
The Fermi surface at half-filling is closest to the nearly
circular surface below it in the nodal  $\kvec = (0,0)$ to $(\pi , \pi)$ direction,
so our $v_F$ should correspond to this nodal Fermi velocity.  
Remarkably, the 
latter was measured to be universal at low energies,  i.e. independent of doping in
\cite{FermiV}.   
Whereas this universality of $v_F$ has not been explained theoretically up to now, 
it is a necessary aspect of our theory.  
Taking the slope $\d E / \d k $ of the curves
in\cite{FermiV} we estimate 
$v_F \approx 1.4 \,ev\AA =210 km/s$ for LSCO. 
As we will show in section XV,  this gives very reasonable estimates of $T_c$. 
Another signature that the system may be in a relativistic regime 
is a density of states that is linear in energy,
as explained in the last section.

\section{Symplectic Fermion Quantum Mechanics}

As stated above,  symplectic fermions are primarily characterized 
by a lagrangian that is second-order in space {\it and}  time  derivatives.   Since
this is unfamiliar to most readers and 
there are some delicate issues in the quantization of such 
theories,  let us first start with the simplest case of $d=0$ quantum
mechanics.

\subsection{Canonical quantization}

In order to draw comparisons,  let us first consider a first
order lagrangian as in the Dirac theory:
\beq
\label{S.1}
L =  \sum_{\alpha=1}^N  \( i   \cdag_\alpha  \d_t  c_\alpha  -  
\omega  \cdag_\alpha c_\alpha \)  
\eeq
It is well-understood that this lagrangian has two consistent
quantizations,  i.e. one can impose either canonical commutation
relations $[  c_\alpha , \cdag_\beta ] =  \delta_{\alpha, \beta}$
or canonical anti-commutation relations $\{ c_\alpha , \cdag_\beta \} 
= \delta_{\alpha, \beta} $.    In both cases the hamiltonian 
is $H =  \sum_\alpha  \omega  \cdag_\alpha c_\alpha$.   
It's clear that both options are possible since eq. (\ref{R.1}) 
is a proper second quantized description of either bosons or fermions.
For future reference we note that the model has a manifest 
$SU(N)$ symmetry.    

The second order bosonic  version of the above is just the ordinary harmonic
oscillator with $L = ( (\d_t q)^2 - \omega^2 )/2 $.    
Since the first-order lagrangian can be consistently quantized 
as a fermion or boson,  one expects that the second-order case
should also be quantizable as a fermion,  and as we now describe,
this is indeed the case.    In order to have a fermionic version,
we need at least 2 degrees of freedom since  fermionic variables  square to zero.   
   Let us
therefore  consider the lagrangian
\beq
\label{S.1b}  
L  =  \chidot^-  \chidot^+  -  \omega^2  \chi^- \chi^+ 
\eeq
where $\chi$ are Grassman variables:
\beq
\label{S.1c}
\{ \chi^i , \chi^j \}  = 0,   
\eeq
which implies $(\chi^-)^2  = (\chi^+ )^2 = 0$,  and we have defined 
$\chidot =  \d_t \chi $.    
The canonical momenta  are $p^- =  \d L / \d \chidot^- =  \chidot^+$
and $p^+ =  \d L / \d \chidot^+ =  - \chidot^-$,  which leads to the
canonical anti-commutation relations
\beq
\label{S.2}
\{ \chi^- ,  \chidot^+ \}  =  - \{ \chi^+ , \chidot^- \}  = i 
\eeq
The  canonical  hamiltonian is simply
\beq
\label{S.3}
H = \chidot^- \chidot^+  +  \omega^2  \chi^- \chi^+ 
\eeq

The equation of motion is $(\d_t^2 + \omega^2 ) \chi = 0$.
Because this is second-order,  the mode expansion involves both
positive and negative frequencies:  
\barray 
\nonumber 
\chi^- (t)  &=&  \inv{\sqrt{2\omega}} \(  \adag \, e^{-i \omega t} 
+ b \, e^{i \omega t}  \)  
\\ 
\label{S.4}
\chi^+ (t)  &=&  \inv{\sqrt{2\omega}} \( - \bdag \, e^{-i \omega t} 
+ a \, e^{i \omega t}  \)  
\earray
The canonical anti-commutation relations (\ref{S.2})  then
require
\beq
\label{S.5}
\{ a , \adag \}  =  \{ b, \bdag \}  = 1
\eeq
with all other anti-commutators equal to zero.   
The hamiltonian is 
\beq
\label{S.6}
H = \omega ( \adag a +  \bdag b -1 )  
\eeq

\subsection{Pseudo-hermiticity}

The only subtle aspect of the above quantization is the extra minus
sign in the expansion of $\chi^+$ in eq. (\ref{S.4}), which was 
necessary in order to have the canonical relations (\ref{S.5}).  This minus
sign implies that $\chi^+$ is not the hermitian conjugate of
$\chi^-$.     One can  understand this feature more clearly,
and also  keep track of it,  with the operator $C$ that distinguishes
particles and holes in eq. (\ref{pso}): 
\beq
\label{S.7}
\chi^+ = C (\chi^- )^\dagger C. 
\eeq
In terms of the original $\chi$ variables,  the hamiltonian
is pseudo-hermitian,  $H^\dagger = C H C$.   However 
after using the equations of motion and expressing it in terms
of $a,b$'s,  since it is quadratic in $b$'s,  the hamiltonian
(\ref{S.6})
is actually hermitian.   This issue will be revisited when 
interactions are introduced in the next section.

\subsection{Symmetries}

We now study the symmetries of the $N$-copy theory.   
Introduce variables $\chi^i_\alpha$,  $i=-,+$, $\alpha= 1,2,..,N$
and define the lagrangian
\beq
\label{S.8}
L = \inv{2} \sum_{i,j,\alpha}    \ep_{ij} \(  \chidot^i_\alpha \chidot^j_\alpha 
- \omega^2  \chi^i_\alpha \chi^j_\alpha \) 
\eeq
where $\ep_{ij}$ is the $2\times 2$ anti-symmetric matrix 
$\ep_{-+} = - \ep_{+-} = 1$. 
The hamiltonian is 
\beq
\label{Hsp}
 H = \inv{2}  \sum_{i,j,\alpha}    
\ep_{ij} \(  \chidot^i_\alpha \chidot^j_\alpha 
+  \omega^2   \chi^i_\alpha \chi^j_\alpha \) 
\eeq
  Arrange $\chi^i_\alpha$ into a 
$2N$-component vector and consider the transformation 
$\chi \to M \chi$,  where $M$ is a $2N$ dimensional matrix.  
Then the lagrangian is invariant if $M^t \ep_N M = \ep_N$ where
$\ep_N = \ep \otimes 1_N$, and $M^t$ is the transpose.
  This implies that $M$ is an element of
the group $Sp(2N)$ of dimension $N(2N+1)$.
Interestingly,   of the classical Lie groups,  $Sp(2N)$ is the only
one that doesn't play any known r\^ole in elementary particle physics\cite{Georgi}.  
 Note that the bosonic version of the theory
with $2N$ real components has the symmetry $O(2N)$ with
dimension $N(2N-1) + 1$, thus the fermionic version always
has a larger symmetry.  The N-component symplectic fermion 
also has a larger symmetry than the N-component first order
fermionic action which has a  $U(N)$ symmetry or an
$O(2N)$ symmetry if the complex fermions are rewritten 
in terms of $2N$ real fields.   

The conserved charges that generate the $N(2N+1)$ dimensional Lie
algebra of $Sp(2N)$ are easily constructed.   Define
\barray
\nonumber 
Q^{0}_{\alpha \beta}  &=& -i \( \chi^-_\alpha \chidot^+_\beta 
+  \chi^+_\beta \chidot^-_\alpha \) 
\\ 
\label{S.9}
Q^{-}_{\alpha \beta}  &=& -i \( \chi^-_\alpha \chidot^-_\beta 
+  \chi^-_\beta \chidot^-_\alpha \) \\
\nonumber
Q^{+}_{\alpha \beta}  &=& i \( \chi^+_\alpha \chidot^+_\beta 
+  \chi^+_\beta \chidot^+_\alpha \) 
\earray
One can easily verify that the hamiltonian $H$ commutes with all
the charges $Q$ using
\barray
\nonumber
\{ \chi^-_\alpha , \chidot^+_\beta \} &=& 
-\{ \chi^+_\alpha, \chidot^-_\beta \} 
= i \delta_{\alpha\beta} 
\\
\label{S.10}
\{ \chi^i_\alpha, \chi^j_\beta \} &=&  \{ \chidot^i_\alpha, \chidot^j_\beta \} 
= 0 
\earray
In ``momentum'' space,  the charges are 
\barray
\nonumber
Q^{0}_{\alpha\beta}  &=&  \adag_\alpha a_\beta - \bdag_\beta b_\alpha 
\\ 
\label{S.11} 
Q^{-}_{\alpha\beta}  &=&  \adag_\alpha b_\beta + \adag_\beta b_\alpha
\\
\nonumber
Q^{+}_{\alpha\beta}  &=&  \bdag_\alpha a_\beta + \bdag_\beta a_\alpha
\earray
and they satisfy the hermiticity properties
\beq
\label{S.12}
( Q^{0}_{\alpha \beta} )^\dagger = Q^{0}_{\beta \alpha} 
, ~~~~~
(Q^{-}_{\alpha \beta} )^\dagger =  Q^{+}_{\alpha \beta} 
\eeq

\section{Field theory version with spin and interactions}  

\subsection{Lagrangian and Hamiltonian}

The field theory version in $d$ spatial  dimensions follows
straightforwardly from the above $d=0$ case with the addition 
of spatial or  momentum integrals.     Introducing fields
$\chi^i_\alpha (\xvec , t)$,   the action 
is 
\beq
\label{S.13}
S = \inv{2} \int dt \, d^d \xvec \,    \sum_{i,j,\alpha} \ep_{ij} 
\(  \d^\mu  \chi^i_\alpha \d_\mu \chi^j_\alpha 
- m^2  \chi^i_\alpha  \chi^j_\alpha \) 
\eeq  
and 
the equations of motion are
\beq
\(  \d^\mu \d_\mu  + m^2 \) \chi = 0.
\eeq

The momentum space expansion is 
\barray
\nonumber
\chi^- (\xvec ,t )  &=& \int \frac{ (d^d \pvec)}{ \sqrt{2\omega_\pvec}}
\(  \adag_\pvec  \, e^{-i p \cdot x }   +  b_\pvec \,  e^{i p \cdot x}  \) 
\\
\label{S.15}
\chi^+ (\xvec ,t )  &=& \int \frac{ (d^d \pvec)}{ \sqrt{2\omega_\pvec}}
\( - \bdag_\pvec  \, e^{-i p \cdot x }   +  a_\pvec \,  e^{i p \cdot x}  \) 
\earray
where $\omega_\pvec = \sqrt{\pvec^2 + m^2 }$ and 
$p\cdot x \equiv \omega_\pvec t - \pvec \cdot \xvec $.   
(We do not display the $\alpha$ indices since they just correspond to
$N$ identical copies.) 
The  canonical anti-commutation relations
are 
\beq
\label{S.17}
\{  \chi^- (\xvec , t ) , \chidot^+ (\xvec', t) \} 
= - \{ \chi^+ (\xvec , t), \chidot^- (\xvec', t ) \} = i \delta^{(d)} 
(\xvec - \xvec' ) 
\eeq
which in momentum space leads to
\beq
\label{S.16}
\{ a_\pvec , \adag_{\pvec'} \} =  \{ b_\pvec , \bdag_{\pvec'} \} 
= (2\pi)^d  \delta^{(d)} ( \pvec - \pvec' )
\eeq
(All other anti-commutators are zero.)  

The hamiltonian is 
\beq
\label{S.18}
H =  \int  (d^d \pvec)   
\sum_{\alpha = 1}^N   \omega_\pvec 
\(  \adag_{\pvec, \alpha} a_{\pvec , \alpha}  
+ \bdag_{\pvec, \alpha} b_{\pvec , \alpha}   \) 
\eeq
   In the limit $m\to 0$, we obtain the 
effective hamiltonian near the Fermi surface in eq. (\ref{E.8}), 
as desired.  
The mass $m$ in this section is  an infra-red regulator
and is unrelated to the non-relativistic mass  $\m$ above.
In section  VII we will show that it can be viewed as
proportional to the temperature.  

The field theory has the same $Sp(2N)$ symmetry as the
quantum mechanical version.  The expressions for the 
conserved charges are identical  to the eqs. (\ref{S.9},\ref{S.11}) 
with additional integrals over $\xvec$ or $\pvec$.

There is a unique 4-fermion interaction that preserves the 
$Sp(2N)$ symmetry:
\beq
\label{Sint}
S_{\rm int} = -  \pi^2 g   \int dt \, d^d \xvec  \,
\( \chi^- \ep_N \chi^+  \)^2  =  - 4 \pi^2 g \int dt \, d^d \xvec ~
\( \sum_\alpha \chi^-_\alpha \chi^+_\alpha \)^2  
\eeq   
(We have included an overall $\pi^2$ so that the RG equations below 
for $g$ have no $\pi$'s;   our convention is the same as in 
\cite{Neubert}.) 
For $N=2$,  even without imposing
the $Sp(4)$ symmetry,  there is a unique interaction due to
fermionic statistics since there are only $4$ independent fields,
which implies that higher order terms beyond quartic interactions are
zero by Fermi statistics.   This interaction  is automatically $SO(5)$ invariant.  
Positive $g$ corresponds to repulsive interactions.  
Since the field $\chi$ has classical scaling dimension $(d-1)/2$,
the interaction is a dimension $2(d-1)$ operator which is 
relevant for $d<3$ as previously mentioned.

\subsection{Pseudo-hermiticity}

The symplectic fermion action (\ref{S.13}) has a Lorentz
invariance if $\chi$ is understood to be a Lorentz scalar.  
Since  $\chi$ is fermionic,  to a particle physicist this
model would appear  to violate the spin-statistics theorem.  
There are two separate aspects of this issue.  First of all,
in the condensed-matter context,  rotational spin is an internal
flavor symmetry (spin $\inv{2}$ fermions corresponds to $N=2$)
which  is not viewed as embedded in the Lorentz group.   
This implies that spin $\smallhalf$ particles are not forced
to be described by the first-order Dirac theory.  
There is no violation of the spin-statistics connection in
our theory since we are quantizing spin $\smallhalf$ particles
with fermionic fields. 
The potential problem rather has to do with unitarity, 
as explained in the quantum mechanical case studied in the last
section,  and 
is manifested in the pseudo-hermiticity property 
(\ref{S.7}).   This  explains  how the proof of the spin-statistics
connection is circumvented:  the proof assumes that the hamiltonian
is built out of fields and their hermitian conjugates and thus
doesn't allow for different fields being related by 
pseudo-hermitian conjugation\cite{Weinbergbook}. 
Furthermore, in the end
the free hamiltonian in momentum space is a perfectly hermitian theory
with no negative norm states.

Whereas the free theory is hermitian in momentum space,  
for  the interacting theory, it follows from 
(\ref{S.7}) that  the hamiltonian is pseudo-hermitian:
\beq
\label{pseudo.1}
H^\dagger = C H C
\eeq
where the unitary operator $C=C^\dagger$ and $C^2 =1$.  
This sort of generalization of hermiticity was understood
to give a consistent quantum mechanics long ago by Pauli\cite{Pauli},
and more recently in connection with $\CP \CT$ symmetric
quantum mechanics\cite{Mostafazadeh,Bender}.   Let us summarize
the main properties enjoyed by pseudo-hermitian hamiltonians:

\bigskip

\n (i)  Define a C-hermitian conjugation as follows:
\beq
\label{pseudo.2}
A^\daggerc = C A^\dagger C
\eeq
Then the usual rules are satisfied:
\beq
\label{pseudo.3}
(AB)^\daggerc = B^\daggerc A^\daggerc , ~~~~~
(a A + b B)^\daggerc = a^* A^\daggerc + b^* B^\daggerc
\eeq
  
\bigskip

\n (ii)  Define a $C$-conjugate inner product:
\beq
\label{pseudo.4}
\langle \psi' | \psi \rangle_c \equiv \langle \psi' | C |\psi\rangle
\eeq
Then time-evolution is unitary:
\beq
\label{pseudo.5}
\langle \psi' (t) | \psi(t) \rangle_c  =  \langle \psi' | e^{iH^\dagger t}
C e^{-i H t} | \psi\rangle =  \langle \psi' (0) |\psi(0)\rangle_c 
\eeq

\bigskip

\n (iii) The eigenvalues of $H$ are real: 
\beq
\label{pseudo.6}
(E - E^*) \langle \psi_E | \psi_E \rangle_c =  \langle \psi_E |
(CH - H^\dagger C) |\psi_E \rangle = 0
\eeq

\bigskip

\n (iv)  Diagonal matrix elements of $C$-pseudo-hermitian operators $A = A^\daggerc$
are real.  This will be important in the sequel since it guarantees the reality of 
vacuum expectation values of pseudo-hermitian order parameters. 
For our model
\beq
\label{dagc}
H^\daggerc = H, ~~~~~~~  (\chi^-)^\daggerc = \chi^+ , ~~~~~
(\chi^- \chi^+)^\daggerc = \chi^- \chi^+ 
\eeq

\bigskip

In the present work there is a new aspect of the pseudo-hermiticity
that relates to the kinematics of the expansion around the
Fermi surface.   As explained in section II,  for the 4-particle
processes near the Fermi surface,  conservation of  $\pvec$ relative
to the Fermi surface 
is equivalent to  conservation of  the physical $\kvec$ momentum if 
particles are paired with particles and holes with holes.   
Since $C = \pm 1$ for particles versus  holes,  on physical grounds
we should  restrict to  eigenstates with even numbers of holes 
and even numbers of particles 
with $C=1$.   Let $|\psi_E \rangle$ denote an eigenstate of $H$
which is also an eigenstate of $C$.   Then 
$H^\dagger |\psi_E \rangle =  C^2 H |\psi_E \rangle = H|\psi_E \rangle$.
Thus for the eigenstates of interest,  $H = H^\dagger$.

\subsection{Charge and Spin}

In order to describe spin $\inv{2}$ electrons,  as usual we
treat the spin as a flavor and thus consider the $N=2$ theory.
The symmetry of the free theory is $Sp(4)=SO(5)$.   
A subgroup of this large (10-dimensional) symmetry can be
identified with rotational spin and charge.  

It was pointed
out in \cite{Neubert} that there are potentially two ways to
identify electronic spin, and the focus in that work was 
 the $SU(2)$ subalgebra that exists for all $N$ and 
acts on the $\pm$ indices of $\chi^\pm_\alpha$.      
It turns out that the other  identification is
more natural in the present context since, as explained in
section II,  the two 
components corresponding to the $\pm$ indices are already
necessary for the expansion of a  single spinless fermion
near the Fermi surface.  
There is actually  an analog of the  identification of
spin and charge for arbitrary $N$.   
Let  $M=e^{-m}$ be an element of the group $Sp(2N)$ in the defining
$2N$ dimensional representation.  Using the relation
\beq
\label{CS.1}
M^t \ep_N M = \ep_N,
\eeq
 the elements of the Lie
algebra satisfy 
\beq
\label{CS.2}
m^t \ep_N  = - \ep_N m,
\eeq
  and 
  a basis  is
the following:  
$m\in \{ 1\otimes a,  \sigma_x \otimes a_x , \sigma_y \otimes a_y , 
\sigma_z \otimes a_z \}  $, where
$\sigma_i$ are the Pauli matrices,  $a$ is an $N\times N$ 
dimensional anti-symmetric matrix,   and $a_i$ are $N \times  N$ symmetric
matrices\cite{Georgi}.    
 Clearly $Sp(2N)$ has an $SU(2)^{\otimes N} $ sub-algebra
generated by $\sigma_i \otimes I^{(\alpha)}$ where 
$I^{(\alpha)} = {\rm diag} (0,0,..,1,0,..,0)$ with the $1$ in
the $\alpha$-th entry.  For $N=2$ this corresponds to 
an $SO(4) = SU(2) \otimes SU(2)$ symmetry.   However since this
sub-algebra does not mix the flavors,  it  is  not the right
sub-group for identification with spin as a flavor.    
The correct identification
is the $SU(N)$ sub-algebra generated by $1\otimes a$ and $\sigma_z \otimes
a_z$ where now $a_z$ is also traceless.    There is also 
a $U(1)$ which commutes with the $SU(N)$ corresponding to 
$\sigma_z \otimes 1_N$.      The $SU(N)$ is generated by the 
charges $Q^{(0)}_{\alpha\beta}$,  which form a closed algebra:
\beq
\label{SC.1}
\[  \Qo_{\alpha\beta} , \Qo_{\alpha' \beta' } \] 
= i \( \delta_{\alpha \beta'} \Qo_{\alpha'  \beta} - 
\delta_{\beta \alpha'} \Qo_{\alpha \beta'}  \)
\eeq
The $U(1)$ we identify with electric charge 
is generated by $Q_e = \sum_\alpha \Qo_{\alpha\alpha}$.   

Let us now return  to the physically interesting case of 
$N=2$ and label the two components as $\alpha = \up , \down$,
corresponding to up and down spins.   
 The $SU(2)$ spin symmetry
is generated by 
\barray
\nonumber
Q_z &=& \smallhalf \( \Qo_{\up\up} - \Qo_{\down\down} \) 
=  - {\textstyle \frac{i}{2} } 
\int d^d \xvec \, 
\( \chi^-_\up \chidot^+_\up + \chi^+_\up \chidot^-_\up 
- \chi^-_\down \chidot^+_\down - \chi^+_\down \chidot^-_\down \) 
\\ 
\label{SC.2}
&=& \int (d^d \pvec)\( \smallhalf ( \adag_{\pvec\up} a_{\pvec\up}
 - \adag_{\pvec\down} a_{\pvec\down} ) 
- \smallhalf  ( \bdag_{\pvec\up} b_{\pvec\up} - \bdag_{\pvec\down} 
b_{\pvec\down} )\) 
\\
\nonumber 
Q_+ &=& \smallsqrt  \Qo_{\up\down} 
= -{\textstyle \frac{i}{\sqrt{2}}} 
\int d^d \xvec  \( \chi^-_\up \chidot^+_\down  
+ \chi^+_\down \chidot^-_\up \) 
= \smallsqrt 
\int (d^d \pvec ) 
( \adag_{\pvec\up} a_{\pvec\down} - \bdag_{\pvec\down} b_{\pvec\up} )
\\
\nonumber
Q_- &=& \smallsqrt  \Qo_{\down\up} 
= -{\textstyle \frac{i}{\sqrt{2}} } 
\int d^d \xvec \( \chi^-_\down \chidot^+_\up + 
\chi^+_\up \chidot^-_\down \) 
= \smallsqrt  \int (d^d \pvec) 
( \adag_{\pvec\down} a_{\pvec\up}  - \bdag_{\pvec\up} b_{\pvec\down} ) 
\earray
satisfying the $SU(2)$ Lie algebra: 
\beq
\label{SC.3}
[ Q_z , Q_\pm ] = \pm Q_\pm , ~~~~~[Q_+ , Q_- ] = Q_z 
\eeq
(As before $\chidot = \d_t \chi$.) 
The $U(1)$ charge  is generated by
\barray
\nonumber
Q_e  = \Qo_{\up\up} + \Qo_{\down\down} &=& 
- i  
\int d^d \xvec \( \chi^-_\up \chidot^+_\up + \chi^+_\up \chidot^-_\up 
+ \chi^-_\down \chidot^+_\down + \chi^+_\down \chidot^-_\down \) 
\\ \label{SC.4}
 &=& 
\int (d^d \pvec ) ~ \( 
(\adag_{\pvec\up} a_{\pvec\up} + \adag_{\pvec\down} a_{\pvec\down} ) 
- ( \bdag_{\pvec\up} b_{\pvec\up} + 
\bdag_{\pvec\down} b_{\pvec\down} ) \) 
\earray
The conserved electric current corresponding to the above charge
is 
\beq
\label{ecurrent}
J^{e}_\mu = -i \sum_{\alpha = \up , \down} 
\( \chi^-_\alpha  \d_\mu \chi^+_\alpha   + \chi^+ _\alpha \d_\mu \chi^-_\alpha  \) 
\eeq

The fields $\chi^\pm$ have electric charge $Q_e = \pm 1$.  
Commutations of the fields with the $SU(2)$ generators 
shows that $(\chi^-_\up , \chi^-_\down )$ form a doublet
whereas $(\chi^+_\up , \chi^+_\down )$ is the conjugate:
\beq
\label{conju}
\[ Q_z , \chi^\pm_\up \] = \mp  \smallhalf \chi^\pm_\up , ~~~~~
\[ Q_z , \chi^\pm_\down \]  = \pm  \smallhalf \chi^\pm_\down
\eeq 
Note that the above identification is consistent with 
$a$ being particles and $b$ holes:   $Q_z | a_\up \rangle = 
\inv{2} | a_\up \rangle$  whereas $Q_z |b_\up \rangle = -\inv{2} 
| b_\up \rangle$.   Also,  the $a$-particles have $Q_e=1$ whereas
the $b$ have $Q_e=-1$.

The additional $6$ conserved charges with electric charge
$\pm 2$  that complete the $SO(5)$
Lie algebra are 
\barray
Q^{\pm}_{\up\up} &=& -2i \int d^d \xvec \, \chi^\pm_\up \chidot^\pm_\up, ~~~~~
Q^\pm_{\down\down} = -2i \int d^d \xvec \, \chi^\pm_\down \chidot^\pm_\down 
\\ 
Q^\pm_{\up\down} &=& -i \int d^d \xvec \( 
\chi^\pm_\up \chidot^\pm_\down + \chi^\pm_\down \chidot^\pm_\up \) 
\earray
These  symmetries flip the charge and spin of the fields, for instance:
\beq
\label{flip}
\[ Q^+_{\up\up} , \chi^-_\up \] = 2 \chi^+_\up , ~~~
\[Q^+_{\up\down} , \chi^-_\up \] = \chi^+_\down , ~~~
\[Q^+_{\up\down} , \chi^-_\down \] = \chi^+_\up 
\eeq

There is no real separation of spin and charge degrees of freedom
in our model at this stage,   unlike Dirac fermions in $1d$, where 
this separation relies on
bosonization.    However as we explain in section VIII,  
the Goldstone bosons are  electrically neutral but carry spin
quantum numbers.   

\subsection{$SO(5)$ order parameters}

The 4 fields $\chi^\pm_\up , \chi^\pm_\down$ are in the
4-dimensional spinor representation of $SO(5)$.    
(See e.g. \cite{Georgi}).    The important operators
are bilinears,  which decompose as ${\bf 4} \otimes {\bf 4} =
{\bf  1}  \oplus  {\bf 5}  \oplus {\bf  10}$,
where ${\bf 1}$ is the singlet $\chi^- \chi^+$,  the ${\bf 5}$ is the
vector representation, and ${\bf 10}$ the adjoint.   The ${\bf 10}$ corresponds
to the currents constructed above.   
The ${\bf 5}$ will serve as  the order parameters of our model 
in the sequel. 
A triplet of fields $\vec{\phi} = (\phi_x, \phi_y, \phi_z)$ 
transforming under the 3-dimensional vector representation of
the spin $SU(2)$ is the following:
\beq
\label{triplet}
\phi^+ = \chi^-_\up \chi^+_\down ,  ~~~~
\phi^- = \chi^-_\down \chi^+_\up , ~~~~
\phi_z = \smallsqrt \( \chi^-_\up \chi^+_\up - \chi^-_\down \chi^+_\down \)
\eeq
where $\phi^\pm =  (\phi_x \pm i \phi_y)/\sqrt{2}$.   
Note that these fields are electrically neutral.  
Let us also define two $SU(2)$ singlets that carry electric charge
$Q_e = \pm 2$:
\beq
\label{chargeor}
\phi^+_e  = \chi^+_\up \chi^+_\down ,  ~~~~~~
\phi^-_e  = \chi^-_\down \chi^-_\up 
\eeq
Using the commutation relations of the $SO(5)$ charges with 
the $\chi$-fields,  one finds that  the 5 order parameters 
\beq
\label{allfive}
\vec{\Phi} = (\phi_x , \phi_y , \phi_z , \phi^+_e , \phi^-_e ) 
\eeq
transform under the 5-dimensional vector representation of $SO(5)$.   
For instance 
$\[ Q^+_{\up \up } , \phi^-_e \] = 2 \phi^- $
and $\[ Q^+_{\up\up} , \phi^+ \] = 2 \phi^+_e $. 
The ordering of fermionic operators was chosen such that 
$(\vec{\Phi})^\daggerc = \vec{\Phi}$, more precisely 
$\phivec^\daggerc = \phivec$ and $(\phi_e^+)^\daggerc = 
\phi_e^-$,  which guarantees that
vacuum expectation values of $\vec{\phi}$ are real 
and those of $\phi_e^\pm$  are complex conjugates.     
An $SO(5)$ invariant is the following
\beq
\label{invar}
\vec{\Phi} \cdot \vec{\Phi} \equiv 
\phi^+ \phi^- + \phi^- \phi^+ + \phi^2_z  - \phi^+_e \phi^-_e 
- \phi^-_e \phi^+_e  
\eeq
Finally the interaction lagrangian density 
 eq. (\ref{Sint}) can be expressed as 
\beq
\label{Svec}
\CL_{\rm int} = \frac{8 \pi^2 g}{5}  \,
   \vec{\Phi}\cdot \vec{\Phi}  = -8 \pi^2 g \, 
\chi^-_\up \chi^+_\up  \chi^-_\down \chi^+_\down 
\eeq

\subsection{Low energy fixed point}

The Feynman rules for the theory are the same as 
for a Lorentz invariant scalar with $\phi^4$ interaction, which in
practice is quite simple\cite{Weinbergbook,Peskin},  
{\it except for some all important minus signs}.  
In the sequel we mainly work in euclidean space $t \to - i t$  
since this simplifies solving the gap equations and is also appropriate
for finite temperature.      Where appropriate
we will return to real time (Minkowski space) for certain physical quantities.   
In particular the propagators in euclidean space are
\beq
\label{propogate}
\langle \chi^-_\alpha (x) \chi^+_\beta (0) \rangle 
= - \langle \chi^+_\alpha (x) \chi^-_\beta (0) \rangle
= - \delta_{\alpha\beta}  \int \frac{d^D p}{(2\pi)^D} \, e^{-i p \cdot x} ~
\inv{p^2 + m^2 } 
\eeq
where $D=d+1$, and they respect causality.  
In the usual euclidean conventions the 4-vertex for the interaction in (\ref{Sint})  
is $-4\pi^2 g$. (See Figure \ref{Figure8}.)

Our RG prescription will be specialized precisely to $2d$ 
and is described in detail in section XIII 
after some  1-loop diagrams are explicitly calculated. 
In  order to obtain a clear physical picture, it
will turn out to be important to carry out the RG directly
in $2d$.  
However the RG can also be studied perturbatively
in an epsilon expansion around $d=3$.  For future reference 
we summarize the main results obtained in \cite{LeClair1,Neubert}
from  the epsilon expansion.  
We will not use these results very much in the sequel;
they are included as a guide to the sign and strength of the anomalous
corrections to scaling.   
Since the perturbative expansion differs from that of bosonic
scalars only by fermionic minus signs,  the standard
methods for the bosonic  $O(M)$ vector models apply. 
Specifically, these are 
models of an $M$-vector of scalar fields $\phivec$ with interaction
$(\phivec \cdot \phivec )^2$. In $D=d+1 =3$, the low energy
theory is the Wilson-Fisher fixed point\cite{WilsonFisher} 
describing classical  $3d$ magnets.     
The anomalous dimensions of the most important operators
were computed to 2-loops in \cite{Neubert}.  As it turns out,
most of the results can be obtained from known results for
the $O(M)$ models analytically continued to $M=-2N$.    
Besides the $\chi$ field itself  and mass term  $\chi^- \chi^+$ (which will be the thermal perturbation),
an additional important class of operators  are the bilinears
that correspond to order parameters,  in particular the  $\phivec$ 
of the $SO(5)$ order parameters when $N=2$,
of the form $\chi^- \sigmavec  \chi^+$ where $\sigmavec$
 is a Pauli  matrix.  
These have no analog that is known to be physically meaningful for 
the $O(M)$ models.  It should be emphasized that the quantum critical
exponents of our $N=2$ model are completely unrelated to 
the usual Wilson-Fisher 
exponents of the bosonic $O(3)$ model,  since here the $O(3)$ vector is a composite
bilinear operator and it's scaling dimension must be calculated in the
fundamental $\chi$-theory.

Let $\dim{X}$ denote
the scaling dimension of $X$ in inverse length, i.e. energy  units.  
One can also define a correlation length exponent based on the
mass $m$:
\beq
\label{anom.4}
\xi \sim m^{-\nu} 
\eeq
where $\nu = 1/\dim{m}$:
\beq
\label{anom.5}
\nu^{-1} = (d+1 - \dim{\chi^- \chi^+})/2
\eeq
Specializing the results in \cite{Neubert} to 
 $N=2$ and  $2d$ one obtains
\beq
\label{anom.3}
\dim{\chi} \approx  \frac{15}{32} , ~~~~~
\dim{\chi^- \chi^+} \approx \frac{5}{8} , ~~~~~
\dim{\chi^- \sigmavec \chi^+ } \approx  \frac{3}{2}, ~~~~~
\nu \approx \frac{16}{19} 
\eeq
Note that whereas $\chi^- \chi^+$ decreases in dimension from
the classical value $1$,  $\chi^- \sigmavec \chi^+$ increases.

Some remarks concerning the $d=1$ case are  again appropriate. 
It is well-known that the low energy fixed points of 
the bosonic $O(M)$ vector models do not extend down to $d=1$, i.e. $D=2$.
For general $M$ this can be viewed as a manifestation of the
Mermin-Wagner 
result\cite{CMW},  which states that spontaneous
symmetry breaking in a $D$-dimensional euclidean field theory is
not possible in $D=2$.   On the other hand,  the models do
have a conformally invariant fixed point 
for $-2 < M < 2$ in $2D$\cite{Nienhuis}.   Since our $N$-component
symplectic fermion model is formally equivalent in perturbation
theory to $O(-2N)$, this suggests that our models do not extend
to $d=1$ except possibly for $-1 < N < 1$.   This helps to explain
why for instance it has never been considered before in connection
with lattice fermion models in $1d$.

\vfill\eject

\section{Resistivity in the normal state}

\def\qvec{{\bf q}}

In this section we give a rough calculation of the temperature dependence
of the conductivity when the interactions are negligible. 
The rigorous study of this question requires  a full finite
temperature treatment of the Kubo formula, which is known to be
quite subtle, and therefore beyond the scope of the present article.
We hope to return to this in a future work,  however in this
section we present the following  non-rigorous scaling argument.

Start with a version of the zero temperature Kubo formula 
\beq
\label{Kubo.1}
\sigma_{ij} (\qvec,\omega ) = \inv{\omega} \int_0^\infty dt  \, e^{i\omega t}
\int d^2 \xvec  e^{-i\qvec \cdot \xvec} ~
\langle  \[ J_i (\xvec , t)  , J_j (0) \]  \rangle 
\eeq
where $i,j=x,y$ are the spatial components of the currents $J_\mu$
given in (\ref{ecurrent}).

 It is well known that  the  frequency
$\omega =0$, which corresponds to the DC conductivity, is a delicate limit.
Let us first set $\omega =0$ in the integrand.   
Using the propagators in eq. (\ref{propogate}) one finds 
\beq
\label{Kubo.2} 
\int dt  \int d^2 \xvec  ~ \langle  
\chi^-  \d_\mu \chi^+ (x) \,  \chi^- \d_\nu \chi^+ (0)  \rangle 
=  \int_0^\pcut \frac{d^3 p}{(2\pi)^3} ~
\frac{ p_\mu p_\nu }{p^4}
\eeq
where we have set $m=0$.  
The above integral is proportional to $\Lambda_c$.   
  Since the conductivity
$\sigma$ is dimensionless in $2d$,  one must 
have that it is proportional to $\Lambda_c / T$ where
$T$ is the temperature.   For reasons that are unclear,  this apparently amounts to setting the overall
$1/\omega$ in eq. (\ref{Kubo.1}) equal to $T$.  
Taking into account the two spin components, and the 4 terms of
the above form,  one obtains 
\beq
\label{Kubo.3}
\sigma_{xx} =  \frac{4}{3\pi^2}  \frac{\Lambda_c}{T}
\eeq
Therefore the  resistivity ($1/\sigma$) is linear in the temperature $T$.

\section{Thermal perturbations and anomalous specific heat}

The parameters of our model thus far are the Fermi
velocity $v_F$,  which we have set equal to $1$, 
the cut-off $\pcut$,   the coupling $g$, and the
infra-red regulator mass $m$.      The 
Fermi energy $\vep_F$ can be viewed as implicit
in the cut-off if the latter is taken to be the 
 frequency $\omega_D = v_F k_F$.    
The density of the free electron gas is only a function of 
$k_F$:
\beq
\label{thermal.00}
\frac{N}{V} =  2 \int_0^{k_F}  \frac{d^d \kvec}{(2\pi)^d} =
\frac{4 \pi^{d/2}}{\Gamma (d/2)} k_F^d 
\eeq
Thus the density can be varied by varying the cut-off,
and since $g$ is proportional to the cut-off in $2d$, 
equivalently  by varying $g$.  (See the discussion of the RG in
section XIII.)

In this
section we suggest how  to introduce a {\it small} non-zero temperature
as the mass $m$.  This is similar to how temperature appears in
the Landau theory for continuous phase transitions in $O(M)$ magnets\cite{LL},
where there also temperature corresponds to a coupling in an effective action.  
However there are some important differences,  since in the latter 
the coupling is $T-T_c$, whereas here it will be proportional to $T^2$.   
Since we have built our model by expanding around the zero
temperature Fermi surface,  it is not obviously consistent  to incorporate
a finite temperature by starting over and  formulating a finite-temperature
version of the $\chi$ fields with Matsubara frequency summations,
ignoring that they are effective fields.  
 Rather, since a small non-zero temperature amounts to 
a small distortion of the Fermi surface,  it could  correspond
to an additional coupling in the lagrangian.  If this is correct,
then it should be possible to obtain known results
in the limit where the interaction is turned off.   Suggestive
of this possibility is the fact that in natural units the
 leading contribution to
the specific heat of
a degenerate electron gas can be expressed entirely in terms of
$k_F$ and $T$,  i.e. the dependence on the electron mass 
is only through the Fermi velocity $v_F = k_F/ m_*$.  
We emphasize that this way of introducing temperature 
is expected to be valid for temperatures near zero,
and cannot replace a full-fledged finite temperature formalism
at arbitrary temperatures.   It will however be useful for
exploring the temperature dependence of low temperature gaps 
in the sequel.

In euclidean space, introducing a finite temperature $T$ is
known to correspond to compactifying the euclidean time
to a circle of circumference $1/T$.  Let us consider adding
a thermal perturbation to the euclidean action of the form
\beq
\label{thermal.0}
\delta S =  \int dt \, d^d \xvec ~  g_T  \CO_T (x)
\eeq
where $g_T$ is a coupling and $\CO_T$ is the 
``thermal operator''.    
Since the euclidean functional integral is over $e^{-S}$,
the correction to the free energy $F= -T \log Z$ 
 to lowest order in $g_T$ is 
$\delta F =  T g_T \int dt \, d^d \xvec \langle \CO_T \rangle$.  
 Since the euclidean 
space-time volume is $V^{(D)} = V /T$ where $V$ is the spatial volume,
one finds
\beq
\label{thermal.0b}
\frac{\delta F }{V} =  g_T \langle \CO_T \rangle 
\eeq

This leads us to identify the thermal perturbation with a mass term:
\beq
\label{thermal.0c}
g_T =  (\alpha T)^2 , ~~~~~ \CO_T = \chi^- \chi^+ \equiv 
 \sum_\alpha  \chi^-_\alpha  \chi^+_\alpha 
\eeq
The above relation is consistent with dimensional analysis
for $\alpha$ a dimensionless parameter.   
The specific heat $C_V$ at constant volume is 
$C_V = - T \frac{\d^2 F}{\d T^2}$. 
Since the propagator goes as $-1/p^2$ as $m\to 0$ (See
eq. (\ref{propogate}) above)  one has 
\beq
\label{thermal.0d}
\langle \CO_T \rangle = 
\sum_{\alpha = \up , \down}  \langle \chi^-_\alpha \chi^+_\alpha \rangle =  
- 2 \int_0^\pcut \frac{ d^{D} p}{(2\pi)^{D}} ~  \inv{p^2} ,
\eeq
where the factor of $2$ comes from spin up and down.   
If the cut-off $\pcut$ is equated with the  frequency $k_F$,
then this leads to specific heat that is linear in $T$:
\beq
\label{thermal.3}
\frac{C_V}{V} =  \frac{\alpha^2 T }{(d-1) 2^{d-1} \pi^{(d+1)/2} 
\Gamma ( {\textstyle \frac{d+1}{2} } ) } \, 
\frac{ k_F^{d-1}}{v_F}
\eeq
(We have re-introduced the Fermi velocity for the sake of 
comparison.)    
The above dependence on $T, k_F$ and $v_F$ is the correct one, i.e.
it is the same as for a non-relativistic degenerate electron gas
near $T=0$.   Repeating the standard calculation of the specific heat
of the electron gas to order $T/\vep_F$ 
(see for instance \cite{LL})  for arbitrary $d$, expressing
the result in terms of $k_F$ instead of the density,   and requiring the
result to match eq. (\ref{thermal.3}) fixes the constant $\alpha$:
\beq
\label{thermal.4}
\alpha^2  =  \frac{ \pi^2 (d-1) \Gamma ({\textstyle \frac{d+1}{2}})}{3}
\eeq
For $d=2$,  $\alpha =\pi^{5/4}/ \sqrt{6} \approx 1.7$.  
The case $d=1$ is noteworthy since $\alpha =0$, and will be commented on
below.

The $k_F$ dependence of the specific heat in eq. (\ref{thermal.3})
is a direct consequence of the scaling dimension $d-1$ of $\CO_T$,
which is twice the scaling dimension of $\chi$.  
When one includes the quartic interaction,  at the low energy fixed
point this scaling dimension has anomalous corrections 
and this should lead to anomalous $T$ dependence of the specific heat.
The scaling dimension of $\CO_T$ was computed perburbatively in an epsilon-expansion
\cite{LeClair1,Neubert}, and the results  summarized in the last section
for arbitrary $N$.  For $N=2$, $\CO_T$ has scaling dimension
approximately equal to $5/8$ in the epsilon expansion.

We can provide  a naive estimate of the anomalous $T$ dependence of the specific heat. 
   Let $\dim{\CO_T}$ denote
the scaling dimension of $\CO_T$.  Then if we assume  the only effect of the anomalous
corrections is to replace $k_F^{d-1}$ by $k_F^{\dim{\CO_T}}$, then since
$C_V/ V$ has scaling dimension $d$,  this requires 
$C_V  \propto T^{d-\dim{\CO_T}}$.  It is not clear that our previous assumption
is correct however,  since  in general $C_V $ could contain terms
$m^x k_F^y$ with $x+y = d$.   Nevertheless,  using our estimate of $5/8$ for
the dimension of $\CO_T$,  this gives $C_V \propto T^{11/8}$. 
Thus $C_V / T \propto T^{3/8}$.     Though the exponent $3/8$ should perhaps
not be taken as very accurate,  the  general  point is that since
$\dim{\CO_T}$ is shifted downward  to $5/8$ from the classical value $1$,
this shows that $C_V /T $ should vanish as $T\to 0$.   This shift downward
is entirely due to a  fermionic minus sign\cite{LeClair1}.   
We hope to study this more carefully in future work.

Finally it is important to note that the manner in which we
have introduced temperature allows,  at least computationally,
 for phase transitions that 
break the continuous $SU(2)$ and $U(1)$ symmetries in $d=2$. 
The Mermin-Wagner result\cite{CMW} is  the statement  that spontaneous symmetry breaking
is not possible in a $1+1$  dimensional (space plus time)
 quantum mechanical system 
at zero temperature because of infra-red divergences that 
plague the existence of Goldstone bosons.   An example of such a divergence  is
in eq. (\ref{thermal.0d}) for $D=2$.    In the Matsubara
approach to finite temperature,  time is compactified into a circle
of circumference $1/T$ and the discrete Matsubara frequencies are
summed over.  Thus the arguments of the theorem  in principle apply to 
a finite temperature system in $2+1$ dimensions at each 
Matsubara frequency. 
On the other hand,  in our model temperature appears as a coupling
in the theory,  as in classical statistical mechanics in $3d$. 
Furthermore,  as discussed in section V,  our model breaks down in $d=1$, 
since the fixed point is lost,  and a manifestation of this is the
vanishing of $\alpha$ in $d=1$.

In the sequel,  wherever $m$ is non-zero,  it should be thought of as
representing a small non-zero temperature.

\section{Mean field analysis}

Because of the $SO(5)$ symmetry,  the interaction term in
the lagrangian can be expressed in terms of either the
magnetic or electric order parameters:
\beq
\label{mean.1}
\CL_{\rm int.} = - 8 \pi^2 g  ~ 
\chi^-_\up \chi^+_\up  \chi^-_\down \chi^+_\down =  
8 \pi^2 g ~\vec{\phi} \cdot \phivec  /3  =  - 8 \pi^2 g~ \phi^+_e \phi^-_e 
\eeq
This implies that magnetic  and SC order may in principle compete. 
In this section we study this in mean field approximation. 

Introduce auxiliary fields $\vec{s} , q^\pm$  coupled to
the order parameters with the action:
\beq
\label{mean.2}
S_{\rm aux.}  =  \int dt \, d^d \xvec \(
\sqrt{2}  \vec{s}\cdot \vec{\phi} - \inv{8\pi^2 g_s } \svec \cdot \svec 
+ q^+ \phi^-_e  + q^- \phi^+_e - \inv{8\pi^2 g_q } q^+ q^- \)
\eeq
Variations $\delta S_{\rm aux.} = 0$ imply
\beq
\label{mean.3}
q^\pm = 8\pi^2 g_q  \phi^\pm_e  , ~~~~~ \svec = 8\pi^2 g_s  \phivec / \sqrt{2}
\eeq
Plugging this back into the action, one finds that the 
interaction is recovered if 
\beq
\label{mean.4}
g_q  - 3 g_s /2  = - g
\eeq

The effective action for the auxiliary $\svec, q^\pm$ fields follows
from performing the fermionic gaussian integrals over the $\chi$ fields.
Let us pass to euclidean space with the usual prescription
$t\to -i t$,  $iS \to -S$.  We will refer to the $D=d+1$
euclidean coordinates as simply $x$.   Then the effective action 
$S_{\rm eff}$ is defined as 
\beq
\label{mean.5}
e^{-S_{\rm eff} (s , q)} = 
 \int D\chi  \, e^{ - S_{\rm aux.} (\chi, s,q) 
- S_{\rm free} (\chi) } 
\eeq
where $S_{\rm free}$ is the free action for the $\chi$ fields.   
The result is
\beq
\label{mean.6}
S_{\rm eff} =  \int d^D x \( \inv{8\pi^2 g_s } \svec \cdot \svec + 
\inv{8\pi^2 g_q } q^+ q^- \)   - \inv{2} \Tr \log A  
\eeq
where the operator $A$ is a differential operator that depends on 
$\svec , q^\pm$.    For constant $\svec$ and $ q^\pm$, the 
 $\Tr \log A$ can be
computed in $D$-dimensional euclidean  momentum space since the derivatives
are diagonal:  $\langle p |  \d | p \rangle = i p \langle p | p \rangle $.  
For constant fields it is meaningful to define the effective potential
$V_{\rm eff} =  S_{\rm eff} / V^{(D)}$ where 
$V^{(D)}$ is the $D$-dimensional volume.   At finite temperature
for instance $V^{(D)} = V \beta$ where $V$ is the usual $d$ dimensional
volume and $\beta$ the inverse temperature.   
Using $\langle p | p \rangle =  V^{(D)}/(2\pi)^D$,  one obtains
\beq
\label{mean.7}
V_{\rm eff} =  \inv{8\pi^2 g_s }  \svec\cdot \svec + \inv{8\pi^2 g_q } 
q^+ q^-   - \inv{2}  \int \frac{d^D p}{(2\pi)^D} ~ 
\Tr \log A (p) 
\eeq
In the basis $(\chi^-_\up, \chi^+_\up, \chi^-_\down, \chi^+_\down)$
the anti-symmetric matrix $A$ is the following:
\beq
\label{mean.8}
A(p) =  \( \matrix{
0& p^2 + m^2 - s_z & q^+ & - \sqrt{2} s^- \cr
-p^2 -m^2 + s_z & 0 & \sqrt{2} s^+ & - q^- \cr
-q^+ & -\sqrt{2} s^+ &0 & p^2 + m^2 + s_z \cr
\sqrt{2} s^- & q^- & -p^2 -m^2 -s_z & 0 \cr } \)
\eeq
Using $\Tr \log A = \log Det A$, one finds
\beq
\label{veffexp}
V_{\rm eff}  =  \inv{8\pi^2 g_s }  \svec\cdot \svec + \inv{8\pi^2 g_q } 
q^+ q^-  - \int \frac{d^D p}{(2\pi)^D}  \, 
\log \( (p^2 + m^2)^2 + q^+ q^- - \svec\cdot \svec \) 
\eeq

The gap equations follow from setting the variation of 
$V_{\rm eff}$ with respect to $\svec$ and $q^\pm$ separately equal to zero.
The result is the following for $m=0$:
\barray
\label{mean.8b}
\svec  &=& - 8\pi^2 g_s  \int  \frac{d^D p}{(2\pi)^D} ~ 
\frac{\svec}{p^4 + q^+ q^- -s^2 }
\nonumber \\
q^\pm &=&
8\pi^2 g_q  \int  \frac{d^D p}{(2\pi)^D}  \frac{q^\pm}{p^4 + q^+q^- - s^2}
\earray  
 There is no simultaneous solution
with both $q^\pm$ and $\svec $ non-zero unless one fine tunes 
to the $SO(5)$ invariant point $g_s = - g_q$.
  In fact, there is no true
competition between AF and SC order in these equations,
and the distinction between $g_s$ and $g_q$ is somewhat fictitious,
since for one sign of the coupling there are AF solutions and no
SC solutions,  and visa-versa if the sign is flipped. 
We thus consider solutions with either pure SC order
($s=0$) or pure magnetic order ($q=0$).  
To further clarify  the structure  of the gap equation in the sequel,  let us 
separate the spatial  and temporal parts of the $D$-dimensional
momentum vector $p$ as $p = (\omega , \kvec )$. 
(This  notation is different from that of section II
where there $\kvec$ was a physical momentum;  here and henceforth it 
is relative to the Fermi surface.)  
 We also restore the
mass $m$.  
For pure SC order one then obtains the gap equation:
\beq
\label{mean.9}
1 =   8\pi^2 g_q   \int  \frac{d\omega \, d^d \kvec }{(2\pi)^{d+1}}
\, \inv{ (\omega^2 + \kvec^2 + m^2 )^2 + q^2 }
\eeq
where $q^+ = q^- = q$.   
For pure magnetic order one instead has
\beq
\label{mean.10}
1 =  - 8\pi^2 g_s   \int  \frac{d\omega \, d^d \kvec }{(2\pi)^{d+1}}
\, \inv{ (\omega^2 + \kvec^2 + m^2 )^2  -s^2  }
\eeq

It is important to note the asymmetry in the signs of the
above SC versus  magnetic gap equations,  which is ultimately
traced to the signs in the $SO(5)$ invariant (\ref{invar}).  
Thus, although the model has an $SO(5)$ symmetry that rotates
the SC and magnetic order parameters, the gap equations
are not invariant under the exchange of $s$ and $q$.   
This implies that the AF gap is related by symmetry to 
a conventional s-wave gap obtained when one flips the sign
of the coupling,   described in the next section. 

Let us consider now the implications of Goldstone's theorem.   
A  non-zero vacuum expectation value in any one direction
of the 5-vector $\vec{\Phi}$  preserves an $SO(4)$ subgroup of 
$SO(5)$.  Since ${\rm dim}(SO(5))  -  {\rm dim} (SO(4)) = 4$,
there are potentially 4 Goldstone bosons.     One of these 
is associated to the $U(1)$,  and when coupled to the electromagnetic field
is eaten up by the Anderson-Higgs mechanism.    This leaves a spin-1  
triplet of Goldstone bosons  associated with the 3-vector $\phivec$,
and these are the closest thing to spinons in our model.   
The effective theory for these modes follows from the above effective
potential   (\ref{veffexp})   for  $\svec$.     For non-constant fields
it contains the kinetic energy term 
$\d_\mu  \svec \cdot  \d_\mu   \svec$.

\section{Conventional s-wave superconductivity in $3d$ for
attractive interactions}

It is well understood that a gap that spontaneously breaks 
the $U(1)$ symmetry in the presence of  electromagnetic gauge potential automatically
has the characteristic  electromagnetic properties  of
a superconductor,  i.e. Meissner effect,  etc.    
This precursor to the Higgs mechanism
can be understood in the original Ginzburg-Landau theory\cite{Ginzburg}.
A clear explanation of this can be found in Weinberg's book\cite{Weinbergbook}. 

We are primarily interested in positive coupling $g$ since
this corresponds to repulsive interactions. (In the next section 
we describe  how this should correspond  to the positive $U$ Hubbard model.)  However
as a check of our formalism thus far,  let us consider
attractive interactions, i.e. negative $g$, or negative $U$ Hubbard, which should
reveal the usual s-wave SC instability of the BCS theory.  
As our analysis shows,  symplectic fermions give
a proper field theoretic description of conventional
s-wave superconductivity which has not been considered before.

\begin{figure}[htb] 
\begin{center}
\hspace{-15mm}
\psfrag{X}{$g_q$}
\psfrag{Y}{$\frac{\Delta}{\Lambda_c}$}
\psfrag{A}{$1$}
\psfrag{B}{$1$}
\psfrag{C}{$2$}
\includegraphics[width=8cm]{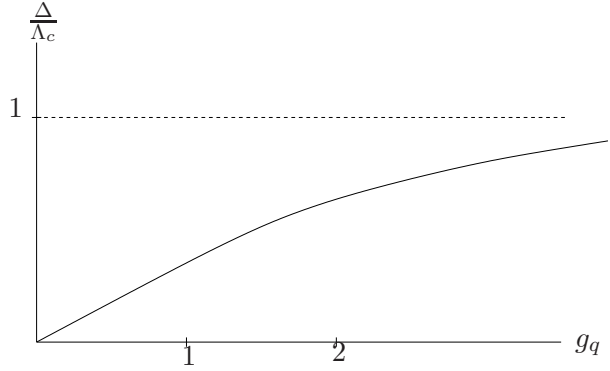} 
\end{center}
\vspace{-2mm}
\caption{s-wave gap as a function of the coupling.} 
\vspace{-2mm}
\label{Figure5} 
\end{figure}

As expected there is no s-wave superconductivity  for
positive $g$ since  the coupling $g_q$ is negative and the 
gap equation  of the last section has no solutions. For pure SC at 
negative $g$, 
    equation (\ref{mean.4}) then implies
$g_q = - g$ is positive.   Specializing to $d=3$ one should
recover some of the basic features of the BCS theory.  
The gap equation reads
\beq
\label{warm}
\frac{4}{g_q} =  \log (1 + \Lambda_c^4/q^2 ) 
\eeq
As the cut-off goes to infinity,  the divergence can be absorbed
into the coupling by defining $1/g_q (\pcut) = \log \pcut$.
This gives the 1-loop correction to the RG beta function:
$\beta (g_q) = - d g_q / \log \pcut = g_q^2$ and is consistent
with the results in \cite{LeClair1,Neubert}.   A more complete RG prescription
will be described in section XIII where 1-loop diagrams are computed.
Note also that positive $g_q$ is
marginally relevant, i.e. $g_q$ increases as the energy is lowered.  
Since $q$ has dimension two, let us define a gap $\Delta = \sqrt{q}$
with units of energy.   
When the cut-off is large compared to $\Delta$ the solution
to the gap equation is approximately
\beq
\label{warm.2}
\Delta = \sqrt{q} = \pcut  e^{-1/g_q}
\eeq
which is characteristic of the BCS theory, i.e. $\Delta$ vanishes 
as $g_q$ goes to zero,  and saturates at the cut-off when 
$g_q$ goes to infinity.  
The zero temperature gap as a function of $g_q$ behaves as in
Figure \ref{Figure5}.

\begin{figure}[htb] 
\begin{center}
\hspace{-15mm}
\psfrag{X}{$m/\Lambda_c $}
\psfrag{Y}{$\frac{\Delta(m)}{\Lambda_c}$}
\psfrag{A}{$1$}
\psfrag{B}{$1$}
\psfrag{C}{$g=1$}
\psfrag{D}{$g=2$}
\psfrag{E}{$g=4$}
\includegraphics[width=8cm]{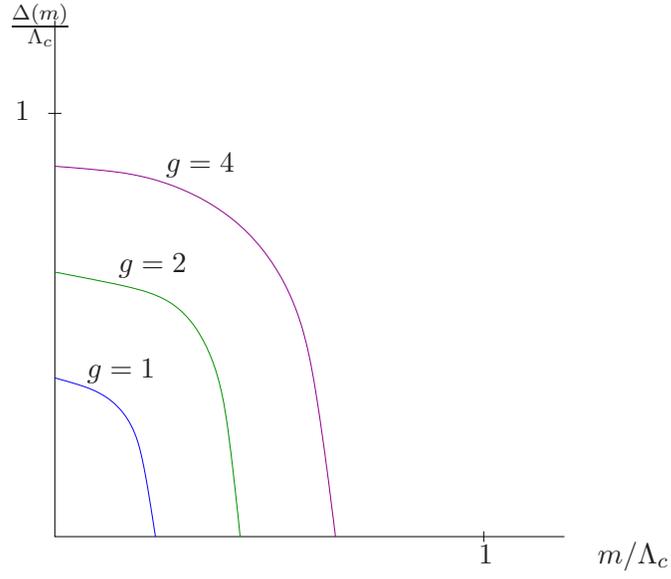} 
\end{center}
\vspace{-2mm}
\caption{s-wave gap as a function of the mass $m$.} 
\vspace{-2mm}
\label{Figure6} 
\end{figure}

Let us now restore the infra-red regulator mass $m$ and
confirm the idea  that it can be viewed as  proportional to
the temperature, as long as the latter is close enough to zero.    
   As described in section VII,  
$m = \alpha T$ where $\alpha = \pi \sqrt{2/3}$ in
$3d$.    Restoring the mass,  the gap equation now reads in $3d$:
\beq
\label{temp.1}
\inv{g} = \inv{4} \log \( \frac{ q^2 + (m^2 + \Lambda_c^2 )^2 }{q^2 + m^4  } \) 
+ \frac{m^2}{2q} \[ \tan^{-1}( m^2 /q) - \tan^{-1}  ((m^2 + \Lambda_c^2)/q)  \]
\eeq
When $m$ is too large, the RHS can go through zero and change sign,
at which point the solution is lost.  Numerical solutions of the
gap $\Delta$ as a function of $m$ is shown for several values of $g$
in Figure \ref{Figure6}.

Let $m_c (g)$ denote the value of $m$ where the gap disappears.
Then,  one expects the critical temperature to be given by
a formula of the form
\beq
\label{TcBCS}
T_c  = \frac{m_c}{\alpha} =  \frac{c_g}{\alpha} \Delta_0
\eeq
where $\Delta_0$ is the gap at $m=0$ and $c_g$ is a constant
of order unity.  
This is similar to the relation in the BCS theory where 
$T_c \approx \Delta_0 / 1.76$.    The constant $c_g = m_c/\Delta_0$ can 
be estimated numerically and shows a weak dependence on $g$,
as Table I shows.

\begin{center}
\begin{tabular}{|c||c|c||c|}
\hline\hline
$g$    &  $\Delta_0 /\Lambda_c$     &  $m_c / \Lambda_c$  &  $c_g = m_c / \Delta_0$ \\ 
\hline\hline 
$0.5$   & $0.135$                     &  $0.083$             &   $0.61$                 \\ 
\hline
$1.0$   & $0.370$                     &  $0.235$             &   $0.64$                 \\
\hline
$2.0$   & $0.629$                     &  $0.440$             &   $0.70$                 \\
\hline
$4.0$   & $0.873$                     &  $0.660$             &   $0.76$                 \\  
\hline
$10.0$   & $1.19$                     &  $0.99$             &   $0.83$                 \\
\hline
$20.0$   & $1.46$                     &  $1.27$             &   $0.87$                 \\
\hline
$100.0$   & $2.22$                     &  $2.09$             &   $0.94$                 \\
\hline\hline 
\end{tabular}
\end{center}

\section{Anti-ferromagnetic phase (Mott-Hubbard insulator)}

\subsection{Symplectic fermions from the Hubbard model at half-filling}

As stated in the introduction, one cannot derive our model by taking
a direct scaling limit of a lattice fermion model like the Hubbard model;
this should be clear from the fact that they have different symmetries,
i.e. $SO(4)$ versus  $SO(5)$.   The main reason for this is that 
one needs to take special care of the Fermi surface in taking such 
a continuum limit.   Nevertheless,  we can further motivate our model
by comparison with lattice models as follows. 
Since our model has no explicit lattice,  the meaning of
the $\vec{s} \cdot \vec{s}  \neq 0$ phase can only be understood
by comparing with low energy, continuum descriptions of 
magnetism.     It is well-known\cite{Haldane,Affleck} that excitations above
the anti-ferromagnetic N\'eel state of the Heisenberg model
 are described by the non-linear
$O(3)$ sigma-model.  The basic facts about lattice models we
need for this  discussion are collected in Appendix B. 
Since the $\nvec$ field
corresponds to our $O(3)$ vector of order parameters $\phivec$ 
in eq. (\ref{triplet}), let us use that notation. 
  
At half-filling and strong coupling,  the Hubbard model
can be mapped to the Heisenberg model and the order parameter
$\phivec$ is constrained to have fixed length
$\phivec\cdot \phivec = {\rm constant}$.  
In our model, 
the magnetic order parameter can be expressed as 
\beq
\label{half.1}
\phivec = \inv{\sqrt{2}} \chi^- \sigmavec \chi^+ 
\eeq
where $\sigmavec$ are the Pauli matrices.   
A fixed length constraint on $\phivec$ is equivalent to
a similar constraint on the $\chi$'s.  Using 
\beq
\label{half.2}
\sigmavec_{ij} \cdot \sigmavec_{kl} = 2 \delta_{il} \delta_{jk} - \delta_{ij}
\delta_{kl}
\eeq
one can show
\beq
\label{half.3}
\phivec\cdot \phivec  = -\frac{3}{2} (\chi^- \chi^+ )^2 
\eeq
where $\chi^- \chi^+ = \sum_{\alpha = 
\up , \down} \chi^-_\alpha \chi^+_\alpha$. 
Therefore if one imposes
\beq
\label{half.4}
\chi^-_\up \chi^+_\up + \chi^-_\down \chi^+_\down =  i h \pcut 
\eeq
for some constant $h$,  then $\phivec\cdot \phivec  = 3 h^2 \Lambda_c^2 /2$.  
(The $i$ is consistent with the propagators in Minkowski space,  
eq. (\ref{pseudogap.1}).)

At half-filling there is a simple argument that leads to symplectic
fermions.  Imposing the above half-filling constraints in the
lagrangian:
\beq
\label{half.5}
\d \phivec \cdot \d \phivec =  4 i h \pcut \( \d \chi^- \d \chi^+ \)
~~+ ~~~~~{\rm {irrelevant ~ operators}}
\eeq
where the additional irrelevant operators can be expressed in
terms of  dimension  $4$  
current-current interactions $J_\mu J^\mu$.   
Thus, due to the second-order nature of the non-linear sigma model,
one obtains a fermionic theory that is second order in space and time derivatives.  
Furthermore,  
let us recall now that 
 we showed in sections II and III how symplectic fermions
also arise when one expands around a circular Fermi surface. 
Since for lattice fermions 
the nearly circular Fermi surface occurs inside the 
half-filling diamond,  this shows how symplectic fermions can
actually extend below half filling.  Above half-filling  (electron rather than hole doping) 
the 
same formalism applies where now one expands around the circular
Fermi surface centered on the node $\kvec = (\pi , \pi)$.     

Relaxing the constraint (\ref{half.4}) moves us away from half
filling.   A meaningful measure of the degree of filling that
we will utilize in the sequel is the following.   Specializing
the formula 
(\ref{thermal.0d}) to $2d$ in euclidean space and  with cut-offs, 
one finds to zero-th order:
\beq
\label{half.6}
h \equiv  - \inv{\Lambda_c}   
\langle \chi^- \chi^+ \rangle =  -2 \int_\Lambda^{\Lambda_c} \frac{d^3 p}{(2\pi)^3} 
~ \inv{p^2}  =  
\inv{\pi^2} \( 1 - \frac{\Lambda}{\Lambda_c} \) 
\eeq
Thus, increasing $\Lambda/ \Lambda_c $ corresponds to lowering 
the density below half-filling, so that $h$ is a measure of hole
doping. 
We will return to this point when we analyze the phase diagram
as a function of doping in section XIV, and will also include
1-loop corrections.

The coupling $g$ should be  proportional to the Hubbard coupling $U$
(See Appendix B).     It can be estimated below half-filling
near the circular Fermi surface.    From the lattice kinetic energy
eq. (\ref{lattice.1}),  $\vep (\kvec) \approx t a^2 \kvec^2$ for small $k$,
which gives a Fermi velocity $v_F = 2 t a^2 k_F$.    Since $g$ has
inverse length units,  $U= v_F g$ has units of energy.  This gives
\beq
\label{HubbardU} 
g = \frac{U}{2 t (k_F a) } \inv{a}
\eeq
where $a$ is the lattice spacing.

\subsection{Solutions of the gap equation}

We now study the solutions of the pure AF order gap equation
(\ref{mean.10}) at $m=0$, which should correspond to zero temperature. 
Because of the two additional minus signs in the AF gap equation
in comparison with the attractive SC one studied in the last section,
the interpretation of its solutions is somewhat subtle, and perhaps
the most delicate point in this whole paper.    First of all,  
the gap equation definitely has solutions for $g_s$ positive due to
the two compensating minus signs.  However because of the pole
at $p^4 = s^2$ we require that $s> \Lambda_c^2$ where $\Lambda_c$ is
the upper momentum cut-off.     
A proper understanding of the solutions requires the introduction
of both UV and IR cut-offs and to implement the RG directly in
$2d$.    The precise description of our RG prescription is
postponed until we calculate some explicit Feynman diagrams 
and is described in section XIII.

When $\gcharge =0$ the equation (\ref{mean.4}) gives the relation
between $g$ and $\gspin$ and they are both positive.    
However the resulting relation is purely classical
and specific to the auxiliary field construction.   Instead
we fix the relation between $\gspin$ and $g$ by requiring
consistency with the perturbative RG.    To lowest
order this requires $\gspin = 2 g$, in order for eq. 
(\ref{AF.2}) to be consistent with eq. (\ref{RG.2d}).
   The gap equation now
reads 
\beq
\label{AF.1}
\inv{g} = -8  \int_0^\pcut  dp  \,  \frac{p^2}{p^4 - s^2}  
\eeq
Let us define
\beq
\label{AF.1a}
s = \deltas^2 \, \Lambda_c^2 
\eeq
The result of doing the integral is
\beq
\label{AF.1b}
\frac{\Lambda_c}{g} = \frac{4}{\deltas} \( \inv{2}  
\log \( \frac{\deltas + 1}{\deltas -1} \) - \tan^{-1} 1/\deltas \)
\eeq

Since the $\log$-term is positive in the above equation,
there are solutions for $\deltas > 1$.   In fact,  when 
the coupling  is lowered, the $\log$ term dominates,  and 
the solution approaches $\deltas = 1^+$ 
and remains there for arbitrarily small $g$.    
This behavior is not physically sensible since
the gap should vanish as the coupling $g$ vanishes.  
To resolve this puzzle,  first note that when $s=0$ the
gap equation has IR divergences that need to be regulated.
Let us therefore introduce a low-energy cut-off $\Lambda$.
Setting $s=0$ in eq. (\ref{AF.1}), the result 
can be written as
\beq
\label{AF.2}
\inv{g} + \frac{8}{\Lambda} = \frac{8}{\pcut}  
\eeq
The left hand side is precisely an expansion of the running
coupling $g(\Lambda)$, since by eq. (\ref{RG.2}), 
$1/g(\Lambda) \approx (1+8g/\Lambda)/g$.  
Denoting the solution as $\gQC$, 
\beq
\label{AF.3}
\gQC (\Lambda) =  \frac{\pcut}{8}  
\eeq
Our  interpretation of this value of the coupling $\gQC$
is that $s$ should vanish for $g< \gQC$ since
$s=0$ is a consistent solution at $g=\gQC$.

The value of the coupling $\gQC$ is closely related, but not the same as 
 the low-energy fixed point value of $g$.  (See 
in section XIII.)   Namely, at 1-loop the fixed point
is at $g_* = \ghat_* \Lambda$ where $\ghat_* = 1/8$,  thus the low energy
fixed point occurs at $g_* = \frac{\Lambda}{\Lambda_c}  g_{AF}$.  
As we will see, the low energy quantum critical point at $g_*$  is 
a second-order continuous phase transition  
that terminates the super-conducting phase on the 
over-doped side.  
This leads us to propose  that  the termination point  $\gQC$ 
of the AF phase 
is a first-order transition,  i.e. the gap drops discontinuously to zero.

Let us return now to solutions of the gap equation with $s\neq 0$.  
The solution  in the two asymptotic limits
$g/\Lambda_c  \to 0, \infty$ have simple expressions.  
It is useful to use the identity
$\inv{2} \log (\deltas +1)/ (\deltas -1) = \tanh^{-1} 1/\deltas$.
When $g \to \infty$, $\deltas$ is large.
Using $\tanh^{-1} 1/\deltas - \tan^{-1} 1/\delta \approx
 2/ (3 \deltas^3)$, 
one obtains 
\beq
\label{AF.6}
\deltas \approx  \( \frac{8 g}{3 \Lambda_c} \)^{1/4} ,  
 ~~~~~~~ (g  / \Lambda_c  \to \infty)
\eeq

\begin{figure}[htb] 
\begin{center}
\hspace{-15mm}
\psfrag{A}{$\frac{g}{\Lambda_c}$}
\psfrag{B}{$\deltas$}
\psfrag{C}{$\frac{\gQC}{\Lambda_c}   \approx \inv{8}$}
\psfrag{D}{$\sim  \frac{3}{8}$}
\psfrag{E}{$\sim 1.28 \, (g/ \Lambda_c)^{1/4}$}
\psfrag{F}{$1$}
\includegraphics[width=12cm]{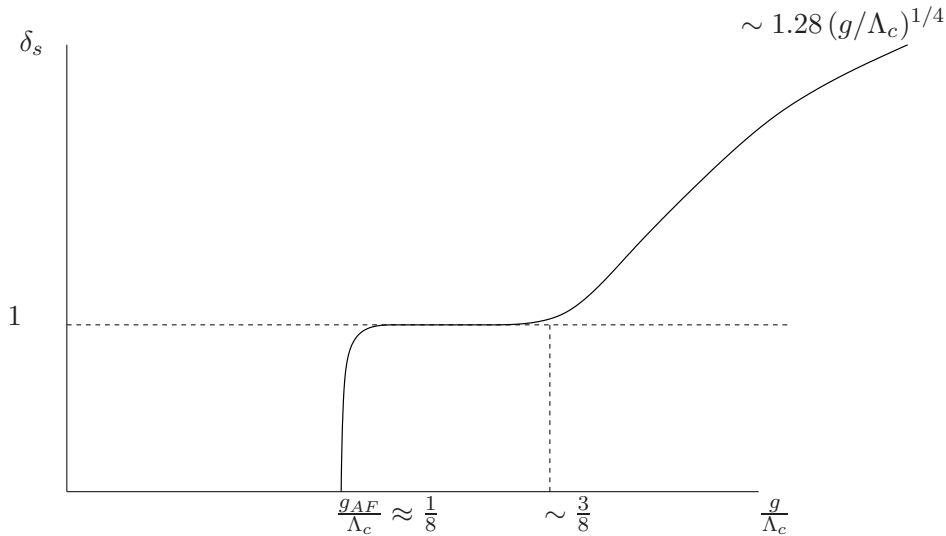} 
\end{center}
\vspace{-2mm}
\caption{Solutions to the anti-ferromagnetic gap equation.} 
\vspace{-2mm}
\label{Figure7} 
\end{figure}

As $g$ decreases,  $\deltas$ saturates to $1$ since
the argument of the $\tanh^{-1}$ must be less than 1 otherwise
the solution is complex. 
Where the solution starts to flatten out can be approximated 
by extrapolating eq. (\ref{AF.6}) down to $\deltas=1$, i.e.
around $g/ \Lambda_c =3/8$.     
Using  $\tanh^{-1} 1/\deltas   
\approx \inv{2} \log (2/(\deltas -1))$ when $\deltas \approx 1$, 
one obtains
\beq
\label{AF.7}
\deltas \approx  1 + 2 e^{-\Lambda_c /2g } , ~~~~~~
( g / \Lambda_c < 3/8 )
\eeq
This behavior is shown in Figure \ref{Figure7} where we have 
incorporated the drop to zero at $\gQC$.

\section{d-wave gap equation in  $2d$}

In the last section we derived gap equations in the
approximation that the gap has no momentum dependence.  
This is essentially equivalent to assuming the lowest
order scattering of pairs is momentum independent.  
In Appendix A we show how to incorporate momentum dependent
scattering and derive the following form of gap equation:
\beq
\label{orb.1}
q(\kvec) =  -  \int \frac{ d\omega \, d^d \kvec'}{(2\pi)^{d+1}} 
~  G(\kvec , \kvec' ) ~  \frac{ q(\kvec' )}{(\omega^2 + \kvec'^2)^2 + 
q(\kvec' )^2 }
\eeq
where the kernel $G$ is given by a Green function related to the
scattering of pairs with momenta $\pm \kvec$ and $\pm \kvec'$. We have
set the mass $m=0$;  it can be restored by $\omega^2 \to \omega^2 + m^2$. 
(As explained in Appendix A, 
$q(k)$ is not simply  the Fourier transform of $q(x)$.) 
 
For our model the kernel will be computed to 1-loop in the next section.
In this section we analyze the orbital properties of the gap
in $2d$ in a model independent way based only on the structure
of the gap equation and its symmetries.   Similar arguments
apply to a BCS type of gap equation.   

Let us assume that $G$ is symmetric,  $G(\kvec , \kvec') =
G(\kvec' , \kvec)$.   If $G$ is also rotationally invariant,
then its angular dependence arises only through the dependence
on $\kvec\cdot \kvec' = k k' \cos (\theta - \theta')$.   
The kernel and gap can thus be expanded as follows:
\barray
\nonumber
G(\kvec, \kvec' ) &=& \sum_{\ell =0}^\infty  G_\ell (k , k') \cos \ell
(\theta - \theta')
\\ 
\label{orb.2}
q(\kvec) &=& \sum_{\ell =0}^\infty q_\ell (k) \cos \ell \theta
\earray

Substituting the above expansions into the gap equation one sees
that due to the non-linearity different $\ell$ can mix.   
For simplicity consider a single channel, i.e. assume $G$ has only
one term in its expansion at fixed $\ell$.   
The angular integral $\int d\theta'$ can be performed and turns out to be
 independent
of $\ell$:
\beq
\label{orb.3}
\int_0^{2\pi} d \theta ~ \frac{ \cos^2 \ell \theta }{1+ a \cos^2 \ell \theta}
= \frac{2\pi}{a} \( 1 - (1+a)^{-1/2} \)
\eeq
The result is
\beq
\label{orb.4}
q_\ell (k) =  - \inv{(2\pi)^2}  \int_{-\infty}^\infty d\omega
\int_0^\infty 
   dk' \, k'   \, 
G_\ell (k, k')  \inv{ q_\ell (k') } 
\(  1-  \frac{ \omega^2 + k'^2 }{\sqrt{ (\omega^2 + k'^2)^2 + q_\ell^2 (k') }}
\) 
\eeq

\def\deltac{\deltaq}


It's important to note that although $G(\kvec , \kvec')$ varies in
sign due to the oscillating cosine,  the sign of $G_\ell (k,k')$ 
is meaningful and determines whether the $\ell$ channel is 
attractive or repulsive.  Negative $G_\ell$  corresponds
to an attractive channel.      Furthermore,  any $\ell = 0,1,2,..$
is in principle allowed. 

The channel $\ell =2$  can arise rather naturally from a term
in the kernel of the form 
$-2 g_2 (\kvec \cdot \kvec')^2 = -g_2 k^2 k'^2 (1+ \cos 2\theta)$
for $g_2$ a constant, which gives rise to both $\ell = 0,2$ with
the same sign.   As we will show in the next section,  for our 
model $\ell =2$ is the first attractive channel.   
In particular $G_2$  has the form 
\beq
\label{orb.5}
G_2 (k, k') =  - 8 \pi^2  g_2 k^2 k'^2 
\eeq
with $g_2$ a positive constant which we will calculate in the next section. 
This leads to 
a solution of the pure d-wave gap equation of the form
\beq
\label{orb.6}
q(\kvec)  = \deltaq^2  \,  k^2 \cos 2\theta   = \deltaq^2
 \,  (k_x^2 - k_y^2) 
\eeq
where $\deltaq$ is a constant satisfying the integral equation:
\beq
\label{orb.7}
\deltaq^4 =  2 g_2 \int_0^\pcut   d\omega \,  dk^2 \,  
\(  1 -  \frac{ \omega^2 + k^2 }
{\sqrt{ (\omega^2 + k^2 )^2 + \deltaq^4 k^4}}  
\)
\eeq
The dependence on $\kvec$ for $\ell=2$ is of the same form 
as a particular linear combination of $\ell =2$ spherical harmonics in $3d$,
thus we refer to it as $d_{x^2 - y^2}$, or simply d-wave, 
as in the literature;   $\ell = 0,1$ can be referred to as $s$ and
$p$ wave.

The above gap equation has some interesting properties,  in particular,
$\deltaq = 0$ when $g_2$ is too small.  
Since this kind of gap equation must be regularized in the UV,
we are led to define
\beq
\label{SCgap.0}
g_2 =  \frac{\ghat_2}{\Lambda_c^3}
\eeq
where $\ghat_2 $ is dimensionless. 
 To estimate the lowest
value of $\ghat_2$ with non-zero gap,  the integrand in the above equation
can be expanded in powers of $\deltaq$. 
 Keeping terms of order $\deltaq^8$,
one finds that for $\deltaq$ small it behaves as
\beq
\label{SCgap.1}
\deltaq \approx \[ \frac{a}{a'} \( 1- \frac{1}{a\ghat_2} \) \]^{1/4} 
\eeq
where
\beq
\label{SCgap.2}
a = (8+\pi - 8\log 2)/12 , ~~~~~a' = 3(104+ 5\pi - 128\log 2)/384
\eeq
Thus
\beq
\label{SCgap.3}
\deltaq = 0 ~~~~~~~{\rm for}~~~ \ghat_2 <  \inv{a}  \approx 2.15
\eeq


For $\ghat_2$ large,  one finds
\beq
\label{SCgap.4}
\deltaq \approx (2 \ghat_2 )^{1/4} 
\eeq
However as we will explain in section XIV,   there will
also be an upper threshold that comes about when the RG is
properly implemented, and this leads to a SC dome.

\section{1-loop scattering and the gap equation kernel}

\subsection{Feynman diagrams} 

\begin{figure}[htb] 
\begin{center}
\hspace{-15mm}
\psfrag{A}{$ p_1 , \alpha $ }
\psfrag{B}{$ p_2 , \beta $}
\psfrag{C}{$p_4 , \beta$}
\psfrag{D}{$p_3 , \alpha $}
\psfrag{E}{$= - 4 \pi^2 g $} 
\includegraphics[width=6cm]{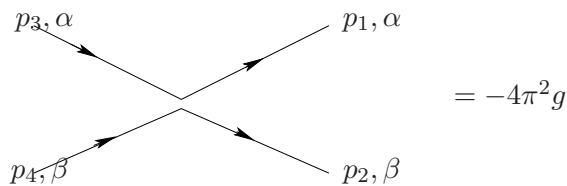} 
\end{center}
\caption{Interaction vertex} 
\vspace{-2mm}
\label{Figure8} 
\end{figure}

In this section we derive the kernel for our model and show
that it has an attractive d-wave channel.  Let $\Gfour (p_1, p_2,
p_3 , p_4 )$ denote the 4-particle vertex function 
with the overall $(2\pi)^D \delta^{(D)} (p)$ removed. 
(Apart from this overall factor,  $\Gfour$ differs from
$\Gamma^{(4)}$ in Appendix A  by an overall sign since the Feynman rule
vertex is defined to be negative in euclidean space.) 
The arrows in the figure indicate how the flavor indices
$\alpha, \beta = 1,..,N$ are contracted.  They also indicate
the flow of charge since the interaction is proportional to 
$( \sum_\alpha \chi^-_\alpha \chi^+_\alpha )^2$.    This structure
is shown in Figure \ref{Figure8},  which indicates the interaction 
 of the euclidean space
Feynman rules.  

\begin{figure}[htb] 
\begin{center}
\hspace{-15mm}
\psfrag{A}{$+$ }
\psfrag{B}{$+$}
\psfrag{C}{$2-N$}
\psfrag{D}{$1$}
\psfrag{E}{$1$} 
\includegraphics[width=14cm]{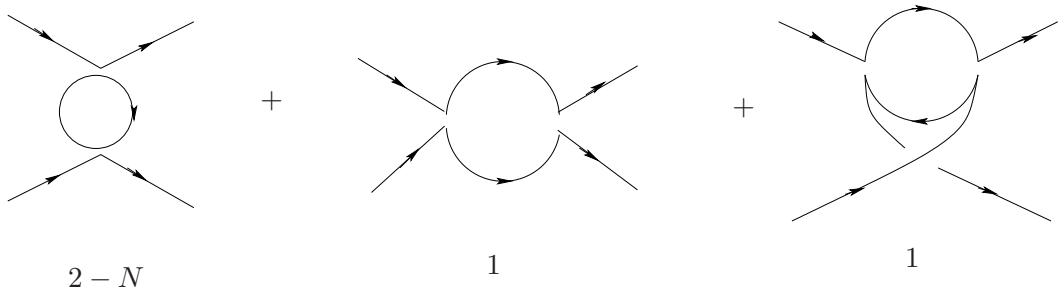} 
\end{center}
\caption{1-loop Feynman diagrams with group theory factors} 
\vspace{-2mm}
\label{Figure9} 
\end{figure}

The 1-loop  contributions to $\Gfour$ are displayed in Figure \ref{Figure9}
and consist of 3 separate channels which differ in how the
$\alpha,\beta $ flavor indices are contracted and also in
their momentum dependence.    It is interesting to carry out
this part of the calculation for arbitrary $N$.   
The $Sp(2N)$ group theory factors are $2-N, 1, 1$ respectively
for the three diagrams, where the $-N$ dependence comes about
from the closed loop and a fermionic minus sign.

To this order one thus has
\beq
\label{loop.1}
\Gfour (p_1 , p_2 , p_3 , p_4 ) =  4 \pi^2 g  
- 2 (4\pi^2 g)^2  \[ 
(2-N) f(p_{13}^2  ) + f (p_{12}^2 )  + f (p_{14}^2 ) 
\]
\eeq
where $p_{ij} = p_i + p_j$ and $f$ is the function:
\barray
\label{loop.2}
f (p^2) &=&  \int  \frac{d^D \ell}{(2\pi)^D}  \, 
\inv{[\ell^2 + m^2] [ (\ell+p)^2 + m^2 ]}  
\\ \nonumber
&=&  
\int  \frac{d^D \ell}{(2\pi)^D}
\int_0^1 dx  
\inv{ \[ \ell^2 + x(1-x) p^2 + m^2 \]^2 }
\earray

Since we are interested in low energies,  it is meaningful
to expand the integrand in powers of $p^2 / \ell^2$.
At zero temperature, we can also set $m=0$.    One finds
\beq
\label{loop.3}
f (p^2) = \int  \frac{d^D \ell}{(2\pi)^D}
\,  \inv{\ell^4}  
\( 1  - \inv{3}  \frac{p^2}{\ell^2}  +  \inv{10} \frac{p^4}{\ell^4} + ...\)
\eeq
As explained in Appendix A,  the gap equation kernel is obtained
by specializing to pairs of opposite  momenta.
We thus fix    
\beq
\label{loop.4}
p_1 = - p_2  = p,    ~~~~~
p_3 = -p_4 = p' 
\eeq
There are also 2 inequivalent ways 
in which to contract the vertex with external 
legs which leads us to define
\beq
\label{loop.5}
G(p , p' ) =  \Gfour (p, -p, p' , -p' ) + (p' \to -p') 
\eeq

The momentum independent part is the following:
\beq
\label{loop.5b}
G(p,p')\vert_{p,p'=0}  =  8 \pi^2 g  - 64 \pi^4 g^2 (4-N) 
\[  \frac{d^D \ell}{(2\pi)^D} \inv{\ell^4} \] 
\eeq
The  second term in the above equation can be absorbed
into a redefinition of the coupling $g$.    It in fact is
just the 1-loop contribution to the RG beta-function and
we will return to it in section XIII where 
our RG prescriptions will be fixed.  

The order $p^2$ term is
\beq
\label{p2}
G(p, p')\vert_{p^2}  =  \frac{64 \pi^4 (3-N)}{3} \( p^2 + p'^2 \) 
\[  \frac{ d^D \ell}{(2\pi)^D} \inv{\ell^6} \]
\eeq
According to the derivation in Appendix A, for static gaps,   the kernel $G(\kvec , \kvec')$
in the gap equation is simply $G(p,p')$ with the time components
of $p$ disregarded,   i.e. $p=(0,\kvec)$.   This implies that
though $G(\kvec , \kvec')$ is closely related to the S-matrix, it is
not identical since the external momenta are not on-shell.  
The above term is then
a repulsive correction to the s-wave contribution,  which is already
repulsive at tree level.     

Finally we come to the $p^4$ term:
\beq
\label{p4}
G(\kvec ,\kvec') \vert_{k^4} =  - \frac{32 \pi^4 (3-N)}{5} \( (\kvec^2 + \kvec'^2)^2 + 4 (\kvec\cdot 
\kvec')^2 \) 
\[  \frac{d^D \ell}{(2\pi)^D} \inv{\ell^8} \]  
\eeq
Using
\beq
\label{coss}
(\kvec \cdot \kvec')^2 = k^2 k'^2 \cos^2 (\theta - \theta') =  
k^2 k'^2 (\cos 2(\theta - \theta') +1 )/2
\eeq 
one sees that this term gives 
 both s and d-wave contributions.  
  Whether the channel is attractive
depends on the sign of   the factor $3-N$.  
For $N=2$,  the  above 1-loop s-wave correction is attractive but since
the leading tree-level term is repulsive,  it is unlikely that
it could lead to s-wave pairing.   
The leading contribution to the d-wave term is attractive 
for $N<3$.   This is rather interesting since this d-wave instability
is invisible to large $N$ methods.   In the sequel we will make the approximation
of the last section and not consider possible mixing of the s and d-wave gaps.

The d-wave term 
$G_2 (k, k')$ is of the form in eq. (\ref{orb.5}) 
with 
\beq
\label{loop.8}
g_2  =   \frac{ (3-N)  8\pi^2  g^2}{5}   \int  \frac{ d^3\ell}{ (2\pi)^3 } 
\, \inv{\ell^8} 
\eeq
The gap thus has the characteristic d-wave form in eq. (\ref{orb.6})
where the constant $\deltaq$ is a solution to eq. (\ref{orb.7}).

\section{Renormalization group specifically in $2d$}

In the present context it is most appropriate to 
adopt the Wilsonian effective action view of the RG.  
In this approach there are two energy scales to consider,
a fixed cut-off scale $\pcut$ and a running scale $\Lambda < \pcut$.
Integrating out high energy modes leads to couplings that
depend on the running scale $\Lambda$.  Wherever possible, 
one lets the cut-off $\pcut$ go to infinity.   

For illustration, consider first the contribution of loop modes with
$\Lambda < \ell < \pcut $ to the  momentum-independent part of $G$ in $3d$:
\beq
\label{RG.1}
G(p,p') =  8 \pi^2 \( g - (4-N) g^2  \int_\Lambda^\pcut  \frac{d\ell}{\ell}
\) 
\eeq
The effect of these high energy modes is to modify the coupling
according to $g(\Lambda) = g + g^2 (4-N) \log \Lambda / \pcut $.  
The beta function is then $- dg / d \log \Lambda = - (4-N) g^2 $,
in agreement with results in\cite{LeClair1,Neubert}.   
An equivalent manner to describe this RG prescription is to
let $\int d\ell \to \int_\Lambda^\pcut d\ell$ and then
set the cut-off $\pcut \to \infty$ and keep the dependence
on the renormalization energy scale $\Lambda$.   

Let us now specialize this to $2d$.    The momentum independent 
part of G is now the  following
\beq
\label{RG.2}
G = 8 \pi^2 \( g - 4(4-N)g^2 / \Lambda \) \equiv 8\pi^2 \, g(\Lambda) 
\eeq
where we have  sent   the cut-off $\pcut$  to $\infty$.  
This gives $-dg / d\log \Lambda = -4(4-N)g^2 / \Lambda$,  
which  is consistent with the fact that $g$ has units of energy.
The latter is an important feature of our model,  and 
in order to deal with it,  we define a dimensionless
coupling $\ghat$:
\beq
\label{RGmore}
g(\Lambda)   = \Lambda \,  \ghat (\Lambda)
\eeq
The beta function for $\ghat$ is now 
\beq
\label{RG.2b}
-\Lambda  \frac{d\ghat }{d\Lambda} =  \ghat  - 4(4-N) \ghat^2 
\eeq
where the linear term just reflects the classical dimension 1 of $g$. 
There is a fixed point at 
\beq
\label{RGmore.2}
\ghat_* =   \inv{4(4-N)}
\eeq
where the above is only approximate due to corrections beyond 1-loop.

\begin{figure}[htb] 
\begin{center}
\hspace{-15mm}
\psfrag{X}{$\Lambda$}
\psfrag{C}{$\ghat (\Lambda)$}
\psfrag{A}{$\ghat_* =  \inv{8}$}
\psfrag{B}{$\pcut$} 
\includegraphics[width=12cm]{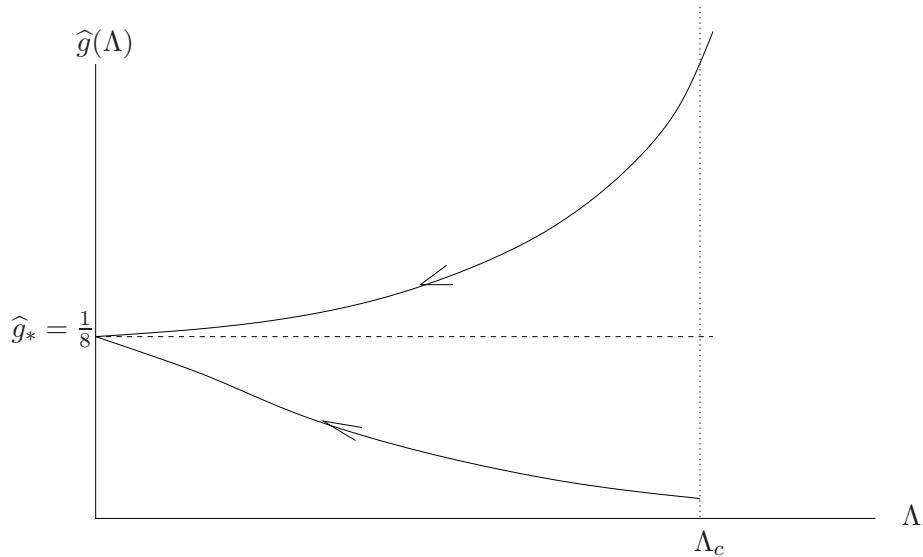} 
\end{center}
\vspace{-2mm}
\caption{RG flow to $\inv{8}$ at low energies
from either strong or weak coupling at short distances.} 
\vspace{-2mm}
\label{Figure10} 
\end{figure}

The solution to the  RG flow equation is then
\beq
\label{RG.2d}
\ghat (\Lambda) =  \frac{ \pcut \ghato}{\Lambda + 4(4-N) (\pcut - 
\Lambda) \ghato } 
\eeq
The fixed point value $\ghat_*$ is reached 
 irregardless
of whether the initial coupling $\ghato$ at the cut-off $\pcut$ 
is large or small,  as long as it is positive.     This 
behavior is sketched in Figure \ref{Figure10}.  

The coupling $g_2$ that enters the d-wave gap equation 
 can now be expressed as 
\beq
\label{RG.7}
g_2 =  \frac{4}{25}  \frac{\ghat^2}{\Lambda^3}
\eeq
where we have safely taken $\pcut$ to $\infty$ in eq. 
(\ref{loop.8}) 
and  set $N=2$.

\section{Scaling and global features of  the phase diagram}

\def\ratio{\gamma}

\subsection{RG scaling}

For the purposes of comparison with real compounds, 
our model has only two independent parameters.
First  is the strength of the coupling $\ghat_0$ 
at the short distance scale $\pcut$.   Second is
the overall scale $\Lambda_c$.   
The complete phase diagram falls into place  when one takes
into account the RG scaling.      Throughout this section
it is implicit that we are only working to 1-loop
and all equalities should be taken as approximations.  

The structure of the phase diagram is most  transparent
and the universal features more clearly revealed 
if one works with the dimensionless inverse couplings: 
\beq
\label{Phase.1}
x = \inv{\ghat} , ~~~~~ x_0 = \inv{\ghat_0}  ,  ~~~~~x_* = \inv{\ghat_*} 
\eeq
Throughout this section we fix $N=2$ 
where  $x_* =  8$ is the fixed point value.   
The scale $\Lambda$ of $g = \Lambda \ghat$ 
can now be expressed in a simple way in terms of $x$ using
eq. (\ref{RG.2d}):
\beq
\label{Phase.1b}
\frac{\Lambda}{\Lambda_c} = \frac{ x_* -x }{x_* - x_0} 
\eeq
The variable $x$ should not be confused with the conventional
doping variable in the literature,  although it is closely related;
the hole doping is the variable $h$ below.  
The above linear relation is specific to $2d$ and is crucial
to understanding the overall structure of the phase diagram.  
Since the  line $\Lambda / \Lambda_c$  represents the scale of the
coupling,  below this line energies are comparable to the coupling
and the non-Fermi liquid properties begin to reveal themselves.
We have labeled this region the pseudogap as in the literature
and will return to it in section XVI.   
Note that when $x = x_0$,  $\Lambda / \Lambda_c =1$.

As we defined them, 
both the AF and SC gaps  $\deltas$ and $\deltaq$ 
 are in units of the cut-off $\Lambda_c$.    Since we have performed
 an RG transformation of the coupling from $\Lambda_c $  to $\Lambda$, 
 we must also scale  the gaps as follows: 
 \beq
\label{Phase.1c}
\delta_{q,s}' = \frac{\Lambda}{\Lambda_c}  \delta_{q,s}  =  
\( \frac{ x_*-x}{x_* -x_0}\)   \delta_{q,s} 
\eeq
It will also be convenient to define the parameter $\ratio$ which
measures the distance to the fixed point at short distances:
\beq
\label{Phase.2}
\ratio  \equiv  \frac{|x_* - x_0|}{x_*}  
\eeq
where $0 < \ratio < 1$.   
The AF gap equation (\ref{AF.1b}) should  now be solved for
$\deltas' (x)$ where $g$ is replaced by
\beq
\label{Phase.2a}
\frac{g}{\Lambda_c} = \inv{x}  \(  \frac{x_* - x}{x_* - x_0} \) 
\eeq

There are two cases to consider depending on whether the coupling
$\ghat$ is strong or weak at short distances,  i.e. whether
$\ghat_0 $ is above or below the fixed point value $\ghat_*$,
corresponding to the two curves in Figure \ref{Figure10},  which we will
refer to as Type A and B.

\bigskip

\noindent
{\bf Type A:  strong coupling at short distances }  

\bigskip

Here we assume  $\ghat_0 > \ghat_* =  1/8$,  which
implies  $x_0 = (1-\ratio) x_* $.   The scale as a function of $x$ now has
negative slope 
\beq
\label{Phase.3}
\frac{\Lambda}{\Lambda_c} = -\inv{\ratio} \( \frac{x}{x_*} - 1 \) 
\eeq
and since the above ratio is by definition less than 1,  
we have $x_0 < x < x_*$.  
From eq. (\ref{AF.3}) the
first-order transition that terminates the AF phase 
corresponds to  $\ghatqc = \ghat_* \Lambda_c / \Lambda$ 
which translates to
\beq
\label{Phase.4}
\xqc =  \frac{x_*}{1+\ratio}
\eeq
The value $\xqc$ is always to the right of $x_0$ since $\gamma^2 >0$.   

Let us turn now to the SC phase.   The d-wave gap equation 
(\ref{orb.7}) depends on $g_2$,  which by eqns. (\ref{SCgap.0},
\ref{RG.7}) can be expressed as
\beq
\label{Phase.5}
\ghat_2 =   \frac{4\ratio^3}{25 x^2 (1-x/x_*)^3 } 
\eeq
The d-wave gap equation only has solutions if $\ghat_2 $ is positive,
which requires  $x< x_*$.    Furthermore,   $\ghat_2$ must
be above the threshold $\ghat_2 > 1/a$ by eq. (\ref{SCgap.3}).
Thus the d-wave SC gap is non-zero in the range
\beq
\label{Phase.6}
\deltaq \neq 0  ~~~~ {\rm for} ~~~~ \xdown   \, < x < \, x_* 
\eeq
where the lower threshold is a solution to
\beq
\label{Phase.7} 
4a \ratio^3 = 25 \xdown ^2 ( 1 - \xdown  / x_* )^3
\eeq
(The constant $a \approx .466$ is defined in eq. (\ref{SCgap.2}).) 
It should be emphasized that,  unlike the values of $x_0, x_{AF}$ 
and $x_*$ which are determined by the low energy fixed point,
the value $x_1$ is not  universal, i.e. not predicted by the RG itself.  
Since the solution to the above equation is close to $x_*$, 
one can estimate
\beq
\label{Phase.8}
\xdown   \approx  x_* \( 1  - \ratio \( \frac{4 a}{25 x_*^2} \)^{1/3} \)  
\approx (1-0.1 \ratio) x_*
\eeq
In principle it is  mathematically possible to have a non-zero solution 
to the d-wave gap for $x$ sufficiently small,  but this
will tend to be small and inside the AF phase.  

\begin{figure}[htb] 
\begin{center}
\hspace{-15mm}
\psfrag{A}{$1$ }
\psfrag{B}{$x_0$}
\psfrag{C}{$\xqc $}
\psfrag{D}{$x_1$}
\psfrag{E}{$x_*$} 
\psfrag{F}{$\deltaq'$}
\psfrag{G}{$\Lambda / \Lambda_c$}
\psfrag{H}{$\deltas'$}
\psfrag{I}{$AF$}
\psfrag{J}{${\rm pseudogap}$}
\psfrag{K}{$SC$}
\psfrag{x}{$x$}
\psfrag{Z}{${\rm non-Fermi ~ liquid}$}
\includegraphics[width=16cm]{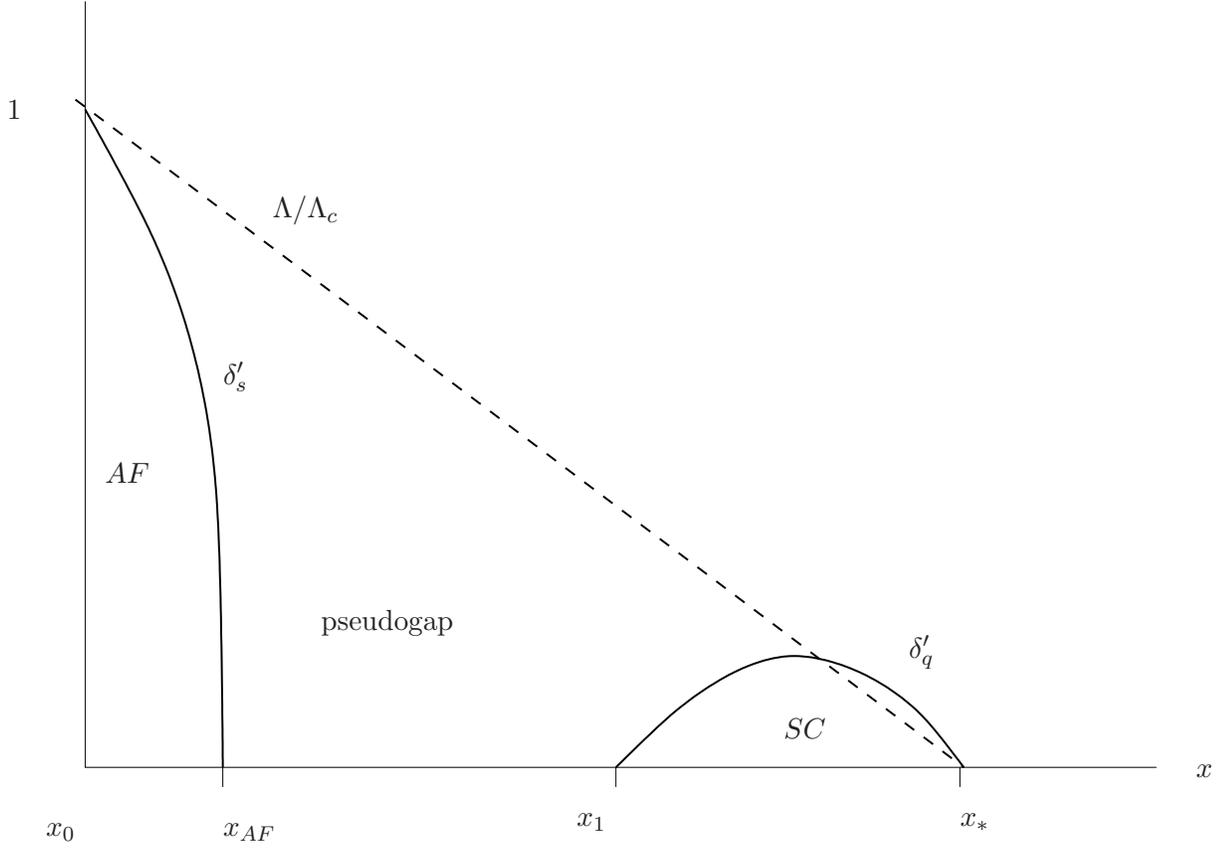} 
\end{center}
\caption{Global phase diagram for Type A.  The slope
of the dashed line is $-\inv{\gamma x_*}$.  The various
transition points are related geometrically by
 $x_0 = (1-\ratio) x_*$, $x_{AF} = x_*/(1+\gamma)$,  and $x_1 \approx (1-.1\ratio)x_*$,
where $x_* = 8$.   The dashed line represents eq. (\ref{Phase.1b}).} 
\vspace{-2mm}
\label{Figure11} 
\end{figure}

The transition  point $\xqc$  is outside the SC dome 
in the approximation we have made,  since $\xqc > \xdown $ would
require $\ratio > 1$.   
These features,  and the geometrical relationships between the various
transition points,  are shown in Figure \ref{Figure11}, which is not 
drawn to scale since the latter depends on $\gamma$.

\begin{figure}[htb] 
\begin{center}
\hspace{-15mm}
\psfrag{X}{$x$ }
\psfrag{A}{$0.5$ }
\psfrag{B}{$3.0$ }
\psfrag{C}{$4.0$}
\psfrag{D}{$8.0$}
\psfrag{E}{$7.1$}
\psfrag{F}{$\deltaq'$} 
\psfrag{G}{$\deltas'$}
\psfrag{H}{$\Lambda /\Lambda_c$ }
\includegraphics[width=12cm]{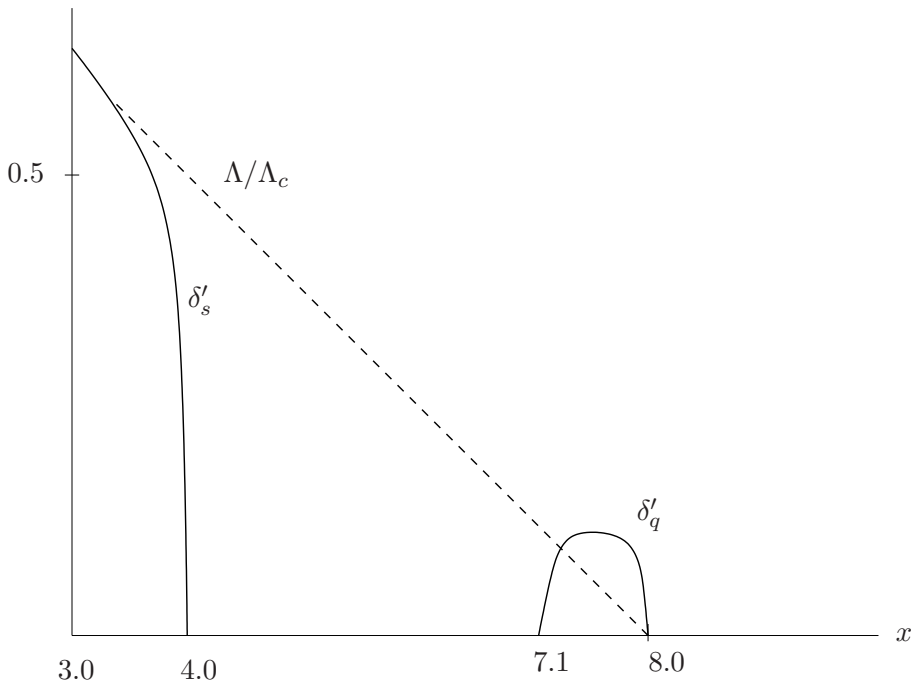} 
\end{center}
\caption{Numerical solutions of the AF and d-wave SC gap
equations for $\gamma = 1$ (Type A).} 
\vspace{-2mm}
\label{Figure12} 
\end{figure}

In Figure \ref{Figure12} we display  numerical solutions to the AF and d-wave SC gap equations
 for $\gamma=1$.
Rough comparison with experimental results  suggests 
the SC dome as we have calculated it  is too narrow, 
 which would imply that we overestimated $x_1$.
However as stated above, the value of $x_1$ is less universal than 
$x_0, x_{AF}$ and $x_*$, and could easily change by improving our 
approximations,  for example taking into account s and d wave mixing, 
or incorporating other effects we have neglected,   such as interplane coupling
and disorder.    
The important point is that there is both an onset and termination of 
the SC phase, and  the termination point on the overdoped side 
is what is universal since it corresponds precisely to a quantum critical point. 
Furthermore,  since $x_{AF} < x_1$,  AF and d-wave SC do not appear to compete
so that the SC is robust.

\bigskip
\noindent
{\bf Type B:  weak coupling at short distances}

In this case  $\ghat_0 < \ghat_*$ at short distances,
and $\ratio = (x_0 - x_*)/x_*$ where in principle
$0<\ratio < \infty$.   Now the slope of $\Lambda$ versus  $x$ 
is positive:
\beq
\label{Phase.8b}
\frac{\Lambda}{\Lambda_c} = \inv{\ratio} \( \frac{x}{x_*} -1 \) 
\eeq
and $x_* < x< x_0$.   
The AF transition point  is at 
\beq
\label{Phase.9}
\xqc =  \frac{x_*}{1-\ratio}
\eeq
Thus $0 < \ratio < 1$ otherwise $\xqc$ goes from $\infty$ to $-\infty$.  

\begin{figure}[htb] 
\begin{center}
\hspace{-15mm}
\psfrag{A}{$1$ }
\psfrag{B}{$x_*$}
\psfrag{C}{$x_2$}
\psfrag{D}{$x_0$}
\psfrag{F}{$\deltas'$} 
\psfrag{E}{$x_{AF}$}
\psfrag{G}{$\Lambda / \Lambda_c$}
\psfrag{H}{$\deltaq'$}
\psfrag{I}{$AF$}
\psfrag{J}{${\rm pseudogap}$}
\psfrag{K}{$SC$}
\psfrag{x}{$x$}
\includegraphics[width=16cm]{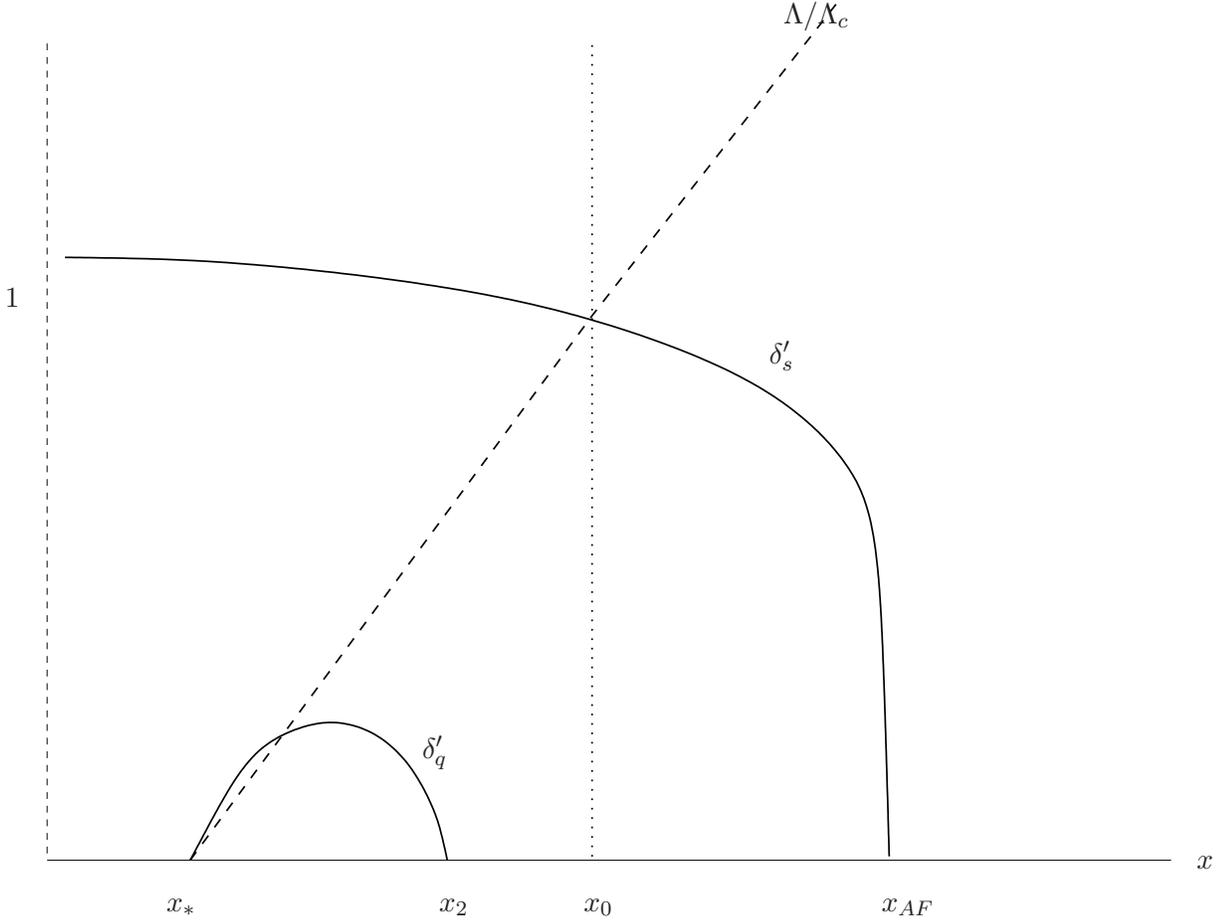} 
\end{center}
\caption{Global phase diagram for Type B. The values shown are related
as follows:   $x_0 = (1+\ratio) x_*$, $x_{AF} = x_*/(1-\gamma)$, 
and $x_2 \approx (1+.1 \ratio) x_*$, 
where $x_* = 8$.  The dashed line has slope $1/(\gamma x_*)$.} 
\vspace{-2mm}
\label{Figure13} 
\end{figure}

There are non-zero solutions to the d-wave gap equation in
the range:
\beq
\label{Phase.10}
\deltaq \neq 0 ~~~{\rm for} ~~~  x_* < x < \xup  
\eeq
where $\xup $ is a solution to eq. (\ref{Phase.7}) with
$\xdown  \to \xup $ and $\ratio \to - \ratio$. 
An approximation of the form (\ref{Phase.8}) also  holds 
 with $\ratio \to - \ratio$.   
With the range of parameters of our model,  $\xqc$ lies
outside the SC dome,  so that the SC dome is completely inside
the AF phase.   These features are shown in Figure \ref{Figure13}.

\subsection{Doping:  1-loop corrections and optimal doping}

The hole-doping variable $h$ defined in section X 
takes the  simple form in terms of the $x$ variables: 
\beq
\label{Phase.2b}
h(x)  = \inv{\pi^2} \(  \frac{x-x_0}{x_* - x_0} \) 
\eeq 
Again, this linear dependence on $x$ is characteristic to $2d$.  
As argued in section X, since increasing $\Lambda / \Lambda_c$ decreases the density
to below half filling,  $h$ is a measure of doping,
where $x=x_0$ should correspond  to  half-filling.  Thus apart from
an overall scale of $1/\pi^2$ and a shift of the origin,  
our previous phase diagrams in terms of $x$ are effectively
in terms of the doping $h$.

Since both the AF and SC termination points $x_{AF}$ and 
$x_*$ are closely related to the RG fixed point, 
the ratio of doping at these two values should be universal.
Let $h_{AF} = h(x_{AF})$ and $h_* = h(x_*)$.   Expressed in
terms of $\ratio$ one has
\beq
\label{moredope.1}
h_{AF}  = \inv{\pi^2} \frac{\gamma}{1+\gamma} , ~~~~~
h_* =  \inv{\pi^2} 
\eeq
which gives the ratio $h_{AF} / h_* =  \gamma / (1+\gamma)$.

\begin{figure}[htb] 
\begin{center}
\hspace{-15mm}
\psfrag{A}{$+$ }
\psfrag{B}{$=\inv{p^2}  + \frac{8 \pi^2 g}{p^4}  \int_\Lambda^{\Lambda_c} 
 \frac{d^3 \ell}{(2\pi)^3} \inv{\ell^2}$}
\includegraphics[width=8cm]{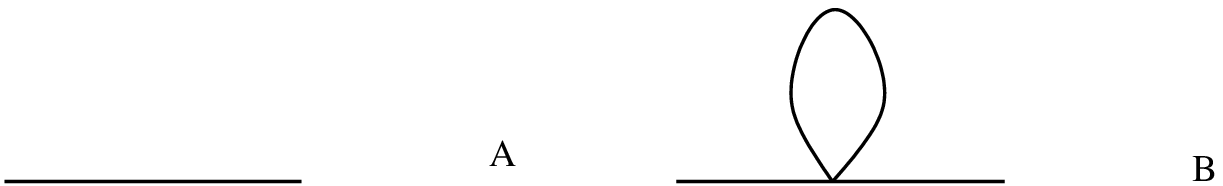} 
\end{center}
\caption{1-loop correction to the propagator} 
\vspace{-2mm}
\label{Figure14} 
\end{figure}

The above formula for $h$ is from eq. (\ref{half.6}) where only
the zero-th order contribution was calculated.     We can
easily include the first order correction in $g$,  which is
the 1-loop self-energy diagram shown in Figure \ref{Figure14}.   
This leads to the following correction:
\beq
\label{dope.1}
h = \frac{2}{\Lambda_c}  \int_\Lambda^{\Lambda_c}  \frac{d^3 p}{(2\pi)^3} \( 
\inv{p^2}   + \frac{4 g (\Lambda_c - \Lambda) }{p^4} \) 
\eeq
There is an important fermionic minus sign in the above expression coming from
the fermion loop.    
Expressing the result of doing the above integral in terms of $x$
using eq. (\ref{Phase.1b}) one obtains
\beq
\label{dope.2}
h(x) = \inv{\pi^2} \( \frac{x-x_0}{x_* - x_0 } \) \[ 1 + \frac{4}{x}
 \( \frac{x-x_0}{x_* - x_0 }\) \] 
\eeq
The corrections to the formula (\ref{moredope.1}) are then 
\beq
\label{moredope.2}
h_{AF} = \inv{\pi^2} \frac{\gamma (2+\gamma)}{2 (1+\gamma)} , ~~~~~
h_* =  \frac{3}{2\pi^2}
\eeq

The optimal doping fraction can be estimated as follows.  
Recall that for Type A,  the SC phase occurs for $x_1 < x < x_*$.
Evaluating $h$ at the lower limit  $x_1$ one finds 
a weak dependence on $\ratio$:
\beq
\label{dope.2b}
h (x_1 )  \approx  \inv{\pi^2}   
\(   \frac{1.3 - 0.09 \gamma}{1- 0.1 \gamma} \) 
\eeq
One thus concludes that optimal doping 
occurs in the tight range:
\beq
\label{dope.3}
 \frac{1.3}{\pi^2}  = .13  <   h_{\rm optimal} <  .15 = \frac{3}{2\pi^2} 
\eeq
This leads to 
\beq
\label{moredope.3}
\frac{h_{AF}}{h_{\rm optimal}} \approx  \frac{\gamma(2+\gamma)}{3(1+\gamma)}
\eeq
For $\gamma = 1/2$ this gives $h_{AF}\approx .04$.

\section{Details on the AF and  d-wave SC gaps and critical temperatures}

In this section we provide detailed numerical solutions to
the gap equations for a variety of $\gamma$ and also 
reintroduce the mass to incorporate a temperature.

\subsection{AF gap}

\begin{figure}[htb] 
\begin{center}
\hspace{-15mm}
\psfrag{A}{$1$ }
\psfrag{B}{$x_0$}
\psfrag{C}{$x_{AF}$}
\psfrag{D}{$\deltas'$}
\psfrag{E}{$AF$}
\includegraphics[width=4cm]{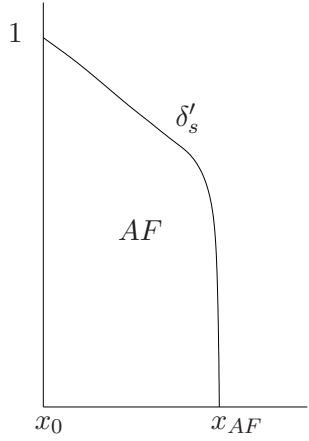} 
\end{center}
\caption{Numerical solutions to the AF gap for $\gamma = 1/2$.
$x_0 = 4$ and $x_{AF} = 5.3$.} 
\vspace{-2mm}
\label{Figure15} 
\end{figure}

Numerical solutions to the AF gap equation (\ref{AF.1b}) expressed 
in terms of $x$ for $\gamma = 1/2$ are shown  in Figure \ref{Figure15}.

To study  finite temperature, we introduce a mass $m=\alpha T$
where $\alpha = \pi^{5/4}/ \sqrt{6} \approx 1.7$ was introduced 
in section VII.  The gap equation becomes
\beq
\label{tempAF.1}
\frac{\Lambda_c}{g}  = \frac{4}{\deltas^2} \( 
\sqrt{\deltas^2 -\mhat^2} \tanh^{-1} \inv{\sqrt{\deltas^2 - \mhat^2 }} 
- \sqrt{\deltas^2 +\mhat^2} \tan^{-1} \inv{\sqrt{\deltas^2 + \mhat^2 }} 
\) 
\eeq
where $\mhat = m/\Lambda_c$.  
Since we have RG scaled the gaps,  in order to make comparisons on the same
scale we need to also  define a scaled mass: 
\beq
\label{tempAF.1b}
m' =  \frac{\Lambda}{\Lambda_c}  m  
\eeq

Numerical study of eq. (\ref{tempAF.1}) shows that the solution
$\deltas'$ vanishes when $m'$ is too large,  consistent with
identification of $m$ with temperature,  as in conventional
superconductivity discussed in section IX.  
Let $m_N'$ denote the value of $m'$ where the solution $\deltas'$
vanishes.   This leads us to define a N\'eel temperature $T_N$  where
the gap vanishes.  As in the BCS theory,  $T_N$ is expected to
be proportional to the zero temperature gap.    Let us define then
\beq
\label{Neeltemp}
T_N =  \frac{m'_N}{\alpha}  =  
c_{AF} (\gamma , x) \, \frac{\deltas' \Lambda_c}{\alpha} 
\eeq
where  
\beq
\label{Neeltemp.b}
c_{AF} (\gamma, x) \equiv  \frac{ m_N' (\gamma, x)}{\Lambda_c \, \deltas'}
\eeq
and  in these equations $\deltas'$ is the zero temperature gap.  
Inspection of the finite temperature gap equation (\ref{tempAF.1}) 
shows that the solution should  disappear when the argument of the 
square-root is negative,  i.e. when $\mhat \approx \deltas$, 
and $c_{AF}\approx 1$.  This is a delicate limit,   
however  we have verified that this is
approximately correct  numerically.

\subsection{SC gap}

\begin{figure}[htb] 
\begin{center}
\hspace{-15mm}
\psfrag{A}{$\deltaq'$ }
\psfrag{B}{$0.1$}
\psfrag{C}{$7.0$}
\psfrag{D}{$x_* = 8$}
\psfrag{E}{$\ratio = 1$}
\psfrag{F}{$\ratio = 0.25$}
\includegraphics[width=15cm]{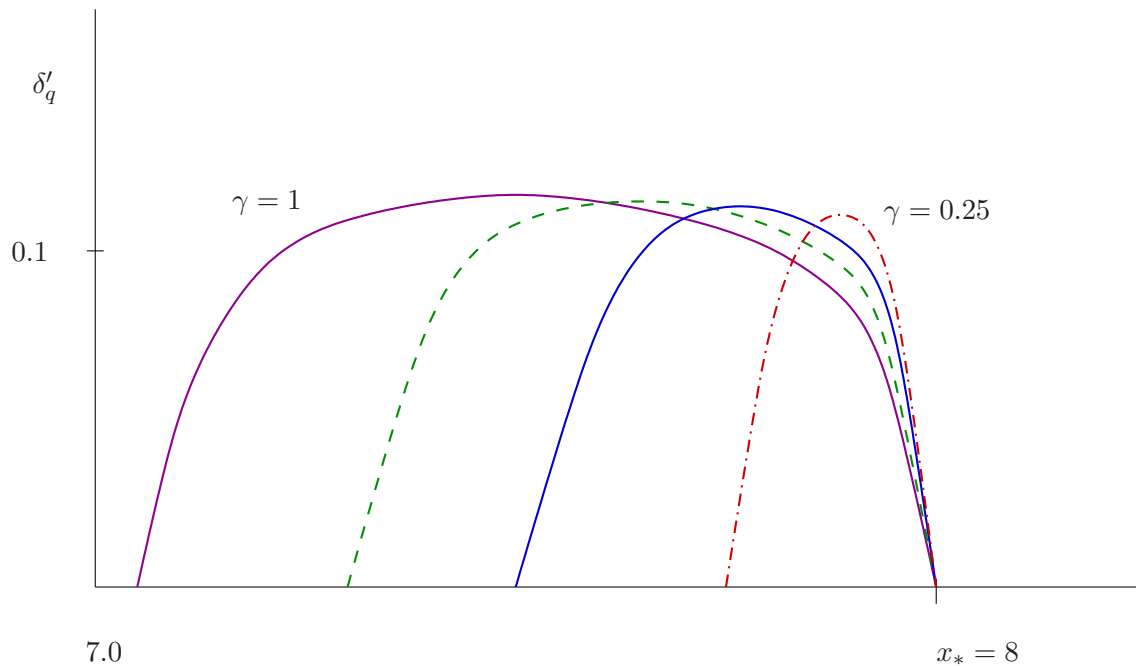} 
\end{center}
\caption{Numerical solutions of the d-wave gap equation for
$\gamma =  \inv{4}, \inv{2}, \frac{3}{4}, 1$.} 
\vspace{-2mm}
\label{Figure16} 
\end{figure}

Numerical solutions to the d-wave gap equation (\ref{orb.7}) expressed in 
terms of $x$ are shown in Figure \ref{Figure16}
 for $\gamma = \inv{4}, \inv{2}, \frac{3}{4}, 1$.   
One sees that the peak value of the gap $\deltaq' \approx 0.11$  is not very sensitive
to $\gamma$.  

\begin{figure}[htb] 
\begin{center}
\hspace{-15mm}
\psfrag{X}{$m'/\Lambda_c$ }
\psfrag{Y}{$\deltaq' (m')$ }
\psfrag{A}{$0.1$ }
\psfrag{B}{$0.1$}
\psfrag{C}{$\ratio=1$}
\psfrag{D}{$\ratio=1/4$}
\psfrag{E}{$m_c' / \Lambda_c $}
\includegraphics[width=10cm]{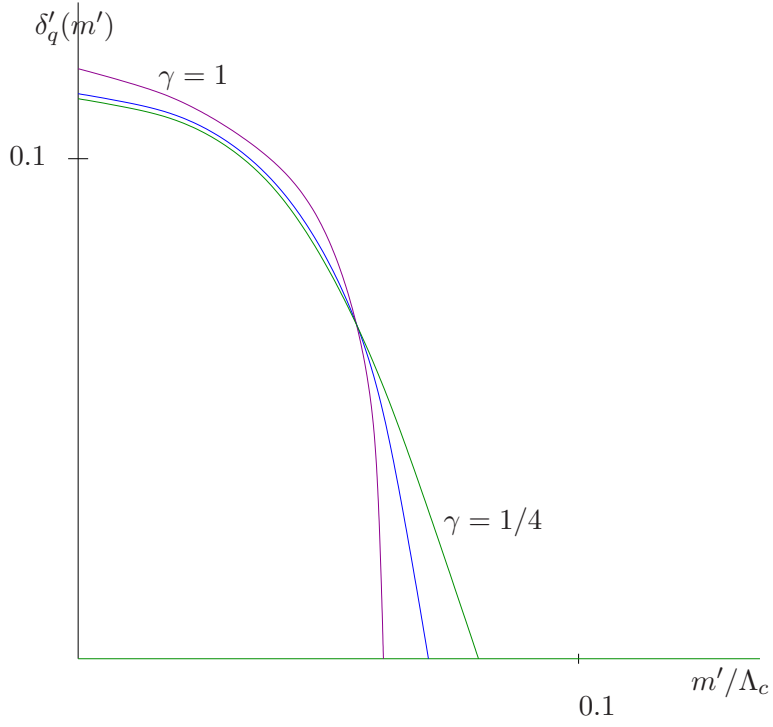} 
\end{center}
\caption{d-wave gap as a function of $m$ for $\gamma = 1/4, 1/2, 1$. } 
\vspace{-2mm}
\label{Figure17} 
\end{figure}

Let us now introduce a temperature $T$ by letting $\omega^2 \to \omega^2 + m^2$
in the gap equation (\ref{orb.7}) where as before $m = \alpha T$.
The behavior of the gap as a function of $T$ is shown in
Figure \ref{Figure17} for $\gamma = \inv{4} , \inv{2} , 1$.   
Let $m_c '$ denote the value of $m$ where the gap vanishes.   
One sees from the figure that here there is a somewhat stronger dependence 
of $m_c$ on the inverse coupling $x$ than for the AF gap.

The critical temperature $T_c$ can now be defined as
\beq
\label{TcSC}
k_B T_c = \frac{m_c'}{\alpha} =   \frac{c_{SC} (\gamma , x)}{\alpha}
 \,  v_F  \hbar  \deltaq' \Lambda_c
\eeq
where
\beq
\label{TcSC.b}
c_{SC} (\gamma , x) =  \frac{m_c'}{\Lambda_c \deltaq' } 
\eeq
is a dimensionless constant.  
In this final formula, we have restored fundamental constants
and the Fermi velocity $v_F$.  
Above, 
$\deltaq'$ is the zero temperature gap,  which depends on $\gamma, x$
as does $m_c'$.

Let $\xopt$ denote the value of $x$ with the largest gap.   The constant
$c_{SC}$ for various $\gamma$ at the corresponding $\xopt$ are shown
in Table 2.  One sees that $c_{SC}$ is of order 1 and has only a weak dependence
on the couplings.

\begin{center}
\begin{tabular}{|c||c|c|c||c|}
\hline\hline
$\ratio$    &  $\xopt$     &  $\deltaq' (\xopt)$   &    $m_c'/ \Lambda_c $    &    $c_{SC}  = m_c'/ \deltaq'  \Lambda_c$ \\ 
\hline\hline 
$1$           & $7.5$        &  $0.118$      &    $0.06$               &    $0.51$                                          \\ 
\hline
$3/4$         & $7.6$        &  $0.116$      &    $0.06$               &    $0.52$                                          \\ 
\hline
$1/2$         & $7.8$        &  $0.114$      &    $0.07$               &    $0.61$                                          \\ 
\hline
$1/4$         & $7.9$        &  $0.113$      &    $0.08$               &    $0.71$                                          \\ 
\hline
$1/8$         & $7.94$       &  $0.113$      &    $0.06$               &    $0.53$                                          \\ 
\hline
$1/16$        & $7.97$        & $0.113$      &    $0.06$               &    $0.53$                                          \\ 
\hline\hline 
\end{tabular}
\end{center}

We can estimate now  the critical temperature
at optimal doping.    
The above table shows that $c_{SC}$ depends relatively  weakly on
$\gamma$.  Furthermore, $\deltaq'$ at optimal doping 
is also relatively constant $\approx 0.11$.   Thus the scale of the
 critical temperature
(\ref{TcSC}) is set  primarily by  $v_F$ and $\Lambda_c$, i.e.
it is only weakly dependent on  the coupling $g$ or the Hubbard
couplings by eq. (\ref{HubbardU});  this is the beauty of 
having a low energy fixed point.   
The scale  $\Lambda_c = 1/a$ is an inverse length $a$,
which  should be on the order of the   lattice spacing.
   
To give a rough estimate of $T_c$,  let us take
the lattice spacing to be  that of the 
$CuO_2$   square lattice $a=3.8 \AA$.
For $v_F$ we use the universal nodal Fermi velocity from\cite{FermiV},
which we estimated to be  
  $v_F \approx 210 km/s$ in section III for LSCO.    
The most unknown quantity is $\alpha$ that  sets the relation between temperature
and mass;  
let  use the estimate of  
$\alpha = \pi^{5/4}/ \sqrt{6}$ from  
section VII.   With the  average values  $\deltaq' =0.11$,  $c_{SC} = 0.6$,
from the above table,   
this gives $T_c \approx 140K$,  which is quite reasonable considering
all of the order $1$ constants we have approximated.  
Note that this result relies on the relatively small value
of $\deltaq'$ we found from numerical solutions of the d-wave
gap equation;  a value $\deltaq'$ of order $1$ would give
$T_c$ higher  by an order of magnitude.  

A useful form of the above equation for $T_c$ is 
\beq
\label{Tcev} 
T_c = c_{SC} \, \frac{v_F}{a} \cdot 650 K
\eeq
where $a$ is in $\AA$ and $v_F$ in $ev-\AA$,  and  we have set $\alpha$ to its estimated value $\alpha  = 1.7$.  
For the range of $c_{SC}$ shown in the above table,
$120K  < T_c < 160K$.  
The maximum $T_c$ 
occurs around $\gamma = 1/4$.  Since this $T_c$ is on the high
side,  this is perhaps because we underestimated $\alpha$
in our simplified inclusion of temperature.  Another possibility
is that  $\Lambda_c$ should instead be set by the average
separation of holes.  At doping $h=.15$ this increases  $a$ by
a factor of about $2.6$ leading to $46K < T_c < 62K$.    

The above formula gives some hints
on how to increase $T_c$:  shorten the lattice spacing,  increase
the Fermi velocity by somehow modifying the effective electron mass
$m_*$,  or tuning the material to $\gamma = 1/4$, which would require
screening the Coulomb potential at short distances.  
In particular $T_c$ should increase with pressure if the main effect
of higher pressure is to reduce the lattice spacing.

\section{The pseudogap region}

It is often suggested in the literature that the pseudogap 
is the key to understanding high $T_c$ since the AF and SC
order condense out of it,  and there has been much speculation
about its nature. (For a review of the experimental data 
see\cite{Hufner}.)  
In this section we briefly discuss the insights on the
pseudogap furnished by our model thus far.  

As stressed throughout this paper,  the essential ingredient
is the non-Fermi liquid at higher temperatures before condensation.  
The pseudogap is 
the region where 
the energy scales are such that this non-Fermi liquid behavior is
visible, i.e. 
there are strong correlations of the electrons  described by our
fields $\chi$,  but the energy is not low enough for them to condense
into an AF or SC ordered state.
The relevant energy scale is $\Lambda$ of the coupling 
$g = \Lambda \ghat$,  thus it is natural to define a pseudogap
temperature $T_{pg}$:
\beq
\label{Tgp}
T_{pg} =  \Lambda =  x g(\Lambda) =  \inv{\gamma} 
\( 1 - \frac{x}{x_*} \)  \Lambda_c  
\eeq
where we have used
eq. (\ref{Phase.2a}).  Thus, as defined,  $T_{pg}$ is proportional the 
coupling $g$ and below this scale the order $g$ corrections are relatively
large.   $T_{pg}$ corresponds to the dashed line in the
figures of section XIV,          
and this line  is seen in 
measurements of various properties  on the underdoped side.
Our  model predicts that the $T_{pg}$ line runs between the top
of the AF gap to precisely the quantum critical point which
terminates the SC phase on the overdoped side,  and the
experiments support this\cite{Hufner}.

Since the  dashed  line in the figures
simply represents  the relevant  energy scale where the interactions become strong,
it does 
not represent a phase transition. 
There are thus 
no spontaneously broken $SU(2)$ or $U(1)$ symmetries just below $T_{pg}$.
The field with non-zero expectation value is the 
$SU(2) \otimes U(1)$ invariant  $\langle \chi^- \chi^+ \rangle$
which by eq. (\ref{half.6}) is related to $\Lambda / \Lambda_c$.  
The spectrum  still  consists  of 
particles and holes with charge $\pm e$, just  strongly coupled. 
Any experiment performed at energies comparable to the scale of the
coupling $g$ is sensitive  to the quantum corrections in our theory,
thus there are a wide variety of probes of $T_{pg}$.

A  possible concrete manifestation of the above pseudogap scale  is a dynamically
generated mass $m$ at zero temperature,  since such a mass is necessarily
proportional to $\Lambda$.     At finite temperature this mass will typically
lead to thermal suppression factors  $e^{-m/T}$ at low temperatures.    
This idea will be investigated in forthcoming work.

\section{Concluding remarks}

To summarize,  we have constructed  a $2d$  relativistic model with 4-fermion interactions 
and analyzed in detail many of its properties which revealed  
a  striking resemblance with the known
features of high $T_c$ superconductivity in the cuprates.
Apart from the overall scale $\Lambda_c$,  the phase diagram can be calculated
as a function of a single parameter $\gamma$, and gives a very good first draft  
of what is observed in experiments.      
Since these features of the model were outlined in the introduction,  
we conclude by summarizing the main theoretical aspects  that  are new 
and responsible for the model's properties.  

\bigskip

\n\bull ~~   In the approach we have taken for expanding around a circular  Fermi surface,  
 there is essentially a unique non-Fermi 
liquid in  $2d$    for spin $\smallhalf$ electrons.
The requirements of a local, rotationally invariant quantum field theory in a sense 
make this theory inevitable.  The rotational invariance is not in conflict
with the existence of a lattice if the wavelengths at low energies are long
compared with the lattice spacing, in fact it is already known that 
the non-linear sigma model description of the Heisenberg magnet is
rotationally invariant. 

\bigskip
\n\bull ~~ 
Our model has  a low energy fixed point 
for the same reason that the $O(M)$ vector models have the Wilson-Fisher fixed point,
since the perturbative expansion differs from the latter only by some fermionic
minus signs.    
The model is perturbatively tractable since the low energy value of the coupling
is relatively small $\approx 1/8$,  even for arbitrarily strong interactions 
at short distances.

\bigskip

\n\bull    ~~   The kind of theory considered here,  namely a Lorentz scalar fermion
with interactions,    has only recently been considered in this 
 context\cite{LeClair1,
Neubert}    because it was  previously  thought to be
inconsistent with the spin-statistics theorem and unitarity.    In fact,  the free version
arises  as  Faddeev-Popov ghosts in gauge theory.     As explained   in \cite{Neubert},
since the theory
turns out to be pseudo-hermitian, $H^\dagger = C HC$,    it still defines a unitary time evolution and has
real eigenvalues.  The non-interacting theory is actually a perfectly hermitian
description of particles and holes near the Fermi surface.  
  In this paper we further elaborated on this issue,  noting
that   the operator $C$ simply distinguishes between particles 
and holes.    Furthermore,  the kinematic constraints coming from the expansion
around the Fermi surface  require eigenstates of $H$ that are also eigenstates of $C$,
and for such states $H$ is actually hermitian.     Finally,  the model  is consistent with spin-statistics
since spin is a flavor symmetry and we are still quantizing spin $\smallhalf$ particles
as fermions.       

\bigskip

\n\bull    ~~   A small non-zero temperature can be introduced as a relativistic
mass term in the lagrangian.   Thus,  in euclidean space,  the  theory is $3$ dimensional
and spontaneous symmetry breaking is possible,  in accordance with 
the Mermin-Wagner   theorem.   

\bigskip
\n\bull  ~~ Our model gives a new  quantum field theoretic treatment of 
conventional s-wave superconductivity for attractive interactions in $3d$ and can be
extended down to $2d$.

\bigskip

\n\bull  ~~  The same model can be motivated  from the Hubbard model at half-filling.
Thus,  the model can interpolate between the nearly circular Fermi surface just
below and up to half-filling. 
    This analysis shows how to  vary the hole doping 
by varying the cut-offs,  or equivalently the inverse coupling,  and leads to a
calculable phase diagram.       The phase diagram has some universal geometrical
features that depend on the strength of the  coupling at short distances.  
In order to properly understand the phase diagram,  it was essential to 
recognize that hole doping was proportional to the inverse coupling and 
to implement certain  RG  scaling relations.     

\bigskip

\n\bull  ~~  It was necessary to derive a new kind of momentum-dependent gap equation
that incorporates the scattering of Cooper pairs near the Fermi surface.     
Only then can one see that there is an attractive d-wave channel  even if
the original interactions that led to AF order were repulsive.        This d-wave channel
is only attractive for $N<3$,   so it cannot be understood using large $N$ methods. 

\bigskip

\n\bull ~~ Although the $SO(5)$ symmetry helps to explain the existence of 
both AF and SC order since order parameters for both are present, 
   since the AF gap is s-wave and  the SC gap  is d-wave, 
 they are not simply related by symmetry.   In particular the d-wave gap equation
is second order in the coupling $g$ whereas the AF one is first order.   
What is related by $SO(5)$ symmetry to the AF phase is a conventional
s-wave SC obtained when one flips the sign of the interaction to make it
attractive.   

\bigskip

\n\bull ~~ The pseudogap is the region where the energy scales are comparable to
the scale of the dimensionfull coupling $g$,  but not low enough for the particles
to condense.  

\bigskip

Our theory certainly reproduces all of the qualitative features of high $T_c$
materials and we believe also explains the universal nodal Fermi velocity observed in 
\cite{FermiV}.    On the quantitative side,  
the  1-loop calculation of the optimal doping fraction is in good agreement with
experimental results, as is our estimate of $T_c$.       
It is beyond the scope of this article to present additional 
detailed comparisons  with the  many 
impressive experimental results obtained over the past 2 decades.

\section{Acknowledgments}

We wish to thank Seamus Davis, Jim Sethna,  Kyle Shen,  and Henry Tye  for discussions.  
 This work is
supported in part by the National Science Foundation.

\section{Appendix:  Derivation of the momentum dependent gap equation}

In order to consistently incorporate higher order scattering,
one can start with the so-called 1PI  effective action $S_{\rm eff}$
with the usual definition as the sum of 1-particle irreducible 
vertices\cite{Weinbergbook}.     Throughout this section we work in euclidean space. 
Since we are only interested in gaps that depend on the spatial component
$\kvec$ of $p = (\omega, \kvec)$,  one should start with a 4-particle term of
the form: 
\beq
\label{A.0}
S_{\rm eff} \vert_{\rm 4-particle} =  
\int dt  \int  d^d \xvec_1 \cdots  d^d \xvec_4  ~ 
\Gamma^{(4)} (x_1 , x_2 , x_3 , x_4) ~
\chi^-_\up (x_1 ) \chi^-_\down (x_2) \chi^+_\down (x_3) \chi^+_\up (x_4)
\eeq
where all $x_i$ are at the same time:  $x_i = (t, \xvec_i )$.   
However in order to streamline the derivation, and also to more clearly
express the result in terms of the usual momentum-space Green functions,
we treat space and time on equal footing and make the above reduction
at the end.   
The 4-particle term in our $N=2$ model then has the
form
\barray
\nonumber 
S_{\rm eff} \vert_{\rm 4-particle} &=&  
\int  d^D x_1 \cdots  d^D x_4  ~ 
\Gamma^{(4)} (x_1 , x_2 , x_3 , x_4) ~
\chi^-_\up (x_1 ) \chi^-_\down (x_2) \chi^+_\down (x_3) \chi^+_\up (x_4) 
\\ 
\nonumber 
&=&  \int (dp_1 ) \cdots (dp_4)  ~ \Gamma^{(4)} (p_1 , p_2 , p_3 , p_4) 
~ \chi^-_\up (p_1) \chi^-_\down (p_2) \chi^+_\down (p_3) \chi^+_\up (p_4) 
\earray
where $(dp) \equiv d^D p/(2\pi)^D$ 
and $\chi (p) $ is the Fourier transform of $\chi (x)$.  
The function $\Gamma^{(4)}$ is a 4-point correlation function of
$\chi$'s and by translational invariance has an overall $\delta$-function:
\beq
\label{A.2}
\Gamma^{(4)} (p_1 ,p_2 , p_3 , p_4 ) 
=  (2\pi)^D \delta^{(D)} ( p_1 + p_2 + p_3 + p_4 ) \( - 8 \pi^2 g + ....\)
\eeq
where we have included the tree-level contribution and $....$ represents
loop corrections.

To derive the gap equation we follow the auxiliary field method
of section VIII and introduce momentum dependence in the manner
described by Weinberg\cite{Weinberg,Weinbergbook}.   Introduce pair fields
$q^\pm (p_1 , p_2)$ and the auxiliary action
\barray
\label{A.3}
S_{\rm aux}  &=& \int (dp_1) \cdots (dp_4) \Gamma^{(4)} (p_1, p_2, p_3 , p_4)
\Bigl[-q^+ (p_1 , p_2) q^- (p_3 , p_4)
\\  \nonumber 
&~& ~~~~~~~~~~~~~~~~~~~ + q^- (p_1 , p_2) 
\chi^+_\up (p_3) \chi^+_\down (p_4)  
+ q^+ (p_1 , p_2) \chi^-_\down (p_3) \chi^-_\up (p_4) \Bigr]
\earray
The equations of motion for $q$ give 
$q^- = \chi^-_\down \chi^-_\up $ and $q^+ = \chi^+_\up \chi^+_\down$
and substituting back into $S_{\rm aux}$ recovers the correct
quartic interaction of $\chi$'s.   

We now specialize to Cooper pairs with total momentum zero:
\beq
\label{A.4}
q(p_1 , p_2 ) = (2\pi)^D \delta^{(D)} (p_1 + p_2)  \, q(p_1) 
\eeq
Define the kernel $G$  as follows:
\beq
\label{A.5}
\Gamma^{(4)} (p_1 , - p_1 , p_3 , p_4 ) = (2\pi)^D \delta^{(D)} (p_3 + p_4) 
\, G(p_1 , p_3 )
\eeq
Then the auxiliary action becomes
\barray
\label{A.6}
S_{\rm aux} &=&  \int (dp) (dp')  ~ G(p,p') 
\Bigl[   - V^{(D)} q^+ (p) q^- (p') 
+ q^- (p) \chi^+_\up (p') \chi^+_\down (-p') 
\\   \nonumber
&~& ~~~~~~~~~~~~~~~~~~~~~~~~~~~~~~~~~~~~~
+ q^+ (p) \chi^-_\down (p') \chi^-_\up (-p') 
\Bigl]
\earray
where $V^{(D)} = (2\pi)^D \delta^{(D)} (0) $ is the 
$D$-dimensional volume.   

The free field kinetic term is the following
\beq
\label{A.7}
S_{\rm free} = \int (dp) \sum_{\alpha = \up, \down} 
~ \chi^-_\alpha (p) (p^2 + m^2 ) \chi^+_\alpha (-p) 
\eeq
The gaussian functional integral over the $\chi$ fields can now
be performed to give an effective 
potential $V_{\rm eff} = S_{\rm eff}/V^{(D)}$. 
To describe  $V_{\rm eff}$ in a compact form it is convenient 
to define 
\beq
\label{A.8}
\qhat^\pm (p) =  \int (dp') G(p,p') \, q^\pm (p') 
\eeq
One then finds
\beq
\label{A.9}
V_{\rm eff} = - \int (dp)(dp') \, G^{-1} (p,p') ~
\qhat^+ (p') \qhat^- (p) - \inv{2}  \int (dp)  ~ 
\Tr \, \log \, A(p)
\eeq
where $A(p)$ is of the same form as in eq. (\ref{mean.8}) 
with $q^\pm \to - \qhat^\pm (p) $ and $s=0$.   
Above, $G^{-1}$ is the inverse of $G$ as an integral operator,
and not simply $1/G$.      

In order to understand how one recovers the constant gap equation,
let us pass to position space where
\beq
\label{A.10}
q(x_1 , x_2 ) = \int (dp) e^{i p \cdot (x_1 - x_2 )}  q(p) 
\eeq
For a constant kernel, it can be redefined to $G(p,p') = 1$,
and one sees that 
\beq
\label{A.11}
 q (x,x) = \int (dp) q(p)  = \qhat 
\eeq
thus a constant $\qhat (p)$ corresponds to a constant $q = q(x,x)$ in
position space.    Thus, where there is no time dependence
in the gap,  one can let $p = (\omega, \kvec)$ and define 
\beq
\label{A.12}
\qhat(\kvec) = \int (d\kvec') \, G(\kvec , \kvec') q(\kvec') 
\eeq
where now $G(\kvec, \kvec')$ is the same kernel but with the
time Fourier transform disregarded.    
In other words,  $G(\kvec, \kvec')$ is simply $G(p,p')$ 
with $p = (0, \kvec)$ and $p' = (0, \kvec')$.   

Finally,  $\delta_q V_{\rm eff} =0$ gives the gap equation
\beq
\label{A.13}
\qhat(\kvec) =  -  \int \frac{ d\omega \, d^d \kvec'}{(2\pi)^{d+1}} 
~  G(\kvec , \kvec' ) ~  \frac{ \qhat (\kvec' )}{(\omega^2 + \kvec'^2)^2 + 
\qhat (\kvec' )^2 }
\eeq
where $\qhat= \qhat^\pm$.  After relabeling $\qhat  \to q$ 
this is the equation (\ref{orb.1}).

\section{Appendix B:  relevant aspects of lattice fermions}

In this appendix we collect some known features of lattice 
fermion models that are referred to in the paper.  
All are contained in the reviews\cite{Affleck,Fradkin}.

Let $\rvec_i$ denote the positions of sites of a 
2-dimensional square lattice with lattice spacing $a$.  
Nearest neighbors $\rvec_i$ and $\rvec_j$ are related
by $\rvec_j = \rvec_i + \avec$ with 
$\avec \in \{ \avec_1 , .., \avec_4 \}$ 
where $\avec_1 = -\avec_3 = (a,0)$, $\avec_2 = -\avec_4 = (0,a)$.  
Introduce fermion operators $c_{\rvec_i , \alpha }$ where $\alpha=\up, \down$ 
represents spin and define the hamiltonian
\beq
\label{L.1}
H = -t \sum_{\rvec, \avec, \alpha}  \( \cdag_{\rvec,\alpha}
 c_{\rvec+ \avec,\alpha } \) 
\eeq
where
\beq
\label{L.1b}
\{ c_{\rvec, \alpha} , \cdag_{\rvec', \alpha'} \}  =
 \delta_{\rvec, \rvec'} \delta_{\alpha,\alpha'}
\eeq
The hamiltonian is hermitian due to $\sum_\avec = \sum_{-\avec}$.  

Introduce the momentum space expansion
\beq
\label{L.2}
c_{\rvec , \alpha} = \sum_{\kvec} e^{i\kvec\cdot \rvec} \, c_{\kvec, \alpha}
\eeq
Using $\sum_{\rvec} e^{i \kvec \cdot \rvec} = \delta_{\kvec, 0} $, 
the hamiltonian becomes
\beq
\label{L.3}
H =  \sum_{\kvec, \alpha} \vep_\kvec  \,  \cdag_{\kvec, \alpha}
 c_{\kvec, \alpha} 
\eeq
where
\beq
\label{latticeep}
\vep_\kvec =  -2t\( \cos k_x a  + \cos k_y a \)  
\eeq

Define the local spin operators:
\beq
\label{L.4b}
S^+_{\rvec}  = \inv{\sqrt{2}} \cdag_{\rvec\up} \, c_{\rvec\down}, 
~~~~~ S^-_{\rvec} = \inv{\sqrt{2}} \cdag_{\rvec\down} \, c_{\rvec\up},
~~~~~
S^{z}_\rvec = \inv{2} \( \cdag_{\rvec\up} \, c_{\rvec\up} - 
\cdag_{\rvec\down} \, c_{\rvec\down} \) 
\eeq
The conserved spin $SU(2)$ charges are 
$\vec{Q} = \sum_\rvec \vec{S}_\rvec$ and satisfy
the $SU(2)$ algebra with the convention in eq. (\ref{SC.3}).  

The Hubbard interaction is essentially unique up to 
shifts of the chemical potential as a consequence of
the Fermi statistics:
\beq
\label{Hubb.1}
H_{\rm int} = U \sum_\rvec ( n_{\rvec\up} - \smallhalf )(n_{\rvec\down} 
- \smallhalf ) 
\eeq
where $n_{\rvec \alpha} = \cdag_{\rvec\alpha} \, c_{\rvec\alpha} $
are local number operators.   The interaction can also 
be written as a spin-spin interaction since 
\beq
\label{Hubb.2}
\vec{S}_\rvec \cdot \vec{S}_\rvec = -\frac{3}{2} n_{\rvec \up} n_{\rvec\down}
+ \frac{3}{4} ( n_{\rvec\up} + n_{\rvec_\down} ) 
\eeq

In addition  to the above spin $SU(2)$ symmetry,  there is another 
commuting $SU(2)$ symmetry so that the largest symmetry of the
Hubbard model is $SU(2) \otimes SU(2) = SO(4)$\cite{YangZhang}.     The latter $SU(2)$
is intrinsic to the lattice and does not obviously have a
continuum limit.  
 
At half-filling there is one fermion per site,  which implies
$n_{\rvec \up} n_{\rvec\down} =0$ and 
$n_{\rvec\up} + n_{\rvec\down} =1$.   
Thus at half-filling one sees from eq. (\ref{Hubb.2}) 
\beq
\label{Hubbcon}
\vec{S}_\rvec \cdot \vec{S}_\rvec = \frac{3}{4}
\eeq
Equating the above with $ j(j+1)$, 
$j=1/2$ one sees that at half-filling the local spin operators
$\vec{S}$ form the 2-dimensional spin $\smallhalf$ representation of $SU(2)$. 
Away from half filling the constraint $(\vec{S}_\rvec)^2 = {\rm constant}$
needs to be relaxed.  

At large $U$ and half-filling,  the Hubbard model can be 
formulated as an effective spin $\smallhalf$ Heisenberg model\cite{Heisenberg,Anderson2}:
\beq
\label{Hubb.3}
H_{\rm eff} =  J \sum_{<i,j>} \( \vec{S}_{\rvec_i} \cdot 
\vec{S}_{\rvec_j}  - 1/4 \)
\eeq
where $J = 4 t^2/U$.   

In the continuum limit an effective theory for fluctuations
above the AF state is the non-linear $O(3)$ sigma model\cite{Haldane}.
One starts from a staggered configuration
\beq
\label{Hubb.4}
\vec{S}_\rvec = \pm \, \nvec_\rvec + \vec{\ell}_\rvec
\eeq
where $\pm$ is for even/odd sublattices.   In the continuum
limit the effective theory for the $\nvec$ field is
second-order in space and time derivatives with lagrangian density
\beq
\label{Hubb.5}
\CL =  \inv{2} \( \d_t \nvec \cdot \d_t \nvec -  \gradvec \nvec 
\cdot \gradvec \nvec \)  + \CL_{\rm top}
\eeq
where $\nvec^2$ is constrained due to eq. (\ref{Hubbcon})
\beq
\label{Hubb.6}
\nvec^2 = {\rm constant}
\eeq
The topological term has the form 
$\CL_{\rm top} \sim \nvec \cdot \d \nvec \times \d \nvec $.
In $1d$ the topological term is known to serve an important 
r\^ole  in determining the low energy fixed point\cite{Haldane}.
In $2d$ the topological term is not known to  play
an analogous r\^ole;  in fact it is an RG irrelevant operator
of dimension $7/2$ and we ignore it.

\end{document}